\documentclass{svjour3}                     
\smartqed  
\usepackage{graphicx}
\usepackage{appendix}
\usepackage{amsmath, amssymb}
\makeatletter
\def\maketag@@@#1{\hbox{\m@th\normalfont\normalsize#1}}
\makeatother

\usepackage{enumerate}
\usepackage{datetime2}
\usepackage{xcolor}
\usepackage{cancel}

\usepackage{mathtools}

\usepackage{subfig}

\usepackage{tikz}\usetikzlibrary{decorations.markings,patterns,calc}

\newcommand{\bA}{\mathbf{A}}
\newcommand{\bB}{\mathbf{B}}

\newcommand{\bI}{\mathbf{I}}

\newcommand{\bM}{\mathbf{M}}

\newcommand{\bO}{\mathbf{O}}

\newcommand{\bQ}{\mathbf{Q}}
\newcommand{\bR}{\mathbf{R}}
\newcommand{\bS}{\mathbf{S}}
\newcommand{\bT}{\mathbf{T}}
\newcommand{\bU}{\mathbf{U}}
\newcommand{\bV}{\mathbf{V}}
\newcommand{\bW}{\mathbf{W}}
\newcommand{\bX}{\mathbf{X}}
\newcommand{\bY}{\mathbf{Y}}
\newcommand{\bZ}{\mathbf{Z}}

\newcommand{\bm}{{\mathbf m}}

\newcommand{\bp}{{\mathbf p}}
\newcommand{\bq}{{\mathbf q}}
\newcommand{\br}{{\mathbf r}}

\newcommand{\bx}{{\mathbf x}}

\newcommand{\cI}{\mathcal{I}}
\newcommand{\cK}{\mathcal{K}}
\newcommand{\cP}{\mathcal{P}}

\newcommand{\cZ}{\mathcal{Z}}

\newcommand{\tq}{\tilde{q}}
\newcommand{\tQ}{\tilde{Q}}
\newcommand{\btq}{\mathbf{\tilde{q}}}

\newcommand{\btQ}{\mathbf{\tilde{Q}}}

\newcommand{\btX}{\mathbf{\tilde{X}}}
\newcommand{\talpha}{\tilde{\alpha}}
\newcommand{\tbeta}{\tilde{\beta}}

\renewcommand{\proof}{\noindent\textit{Proof}:\;}

\newcommand{\Tr}{\mathrm{Tr}\,}
\newcommand{\Pf}{\mathrm{Pf}\,}
\newcommand{\sgn}{\mathrm{sgn}}

\newcommand{\erf}{\mathrm{erf}}
\newcommand{\erfc}{\mathrm{erfc}}
\newcommand{\qdet}{\mathrm{qdet}}

\newcommand{\qqimp}{\qquad \Rightarrow \qquad}

\newcommand{\NMrs}[1]{\check{\mathsf{#1}}}

\makeatletter
\renewcommand*\env@matrix[1][*\c@MaxMatrixCols c]{%
  \hskip -\arraycolsep
  \let\@ifnextchar\new@ifnextchar
  \array{#1}}
\makeatother

\makeatletter
\newsavebox{\@brx}
\newcommand{\llangle}[1][]{\savebox{\@brx}{\(\m@th{#1\langle}\)}%
  \mathopen{\copy\@brx\kern-0.5\wd\@brx\usebox{\@brx}}}
\newcommand{\rrangle}[1][]{\savebox{\@brx}{\(\m@th{#1\rangle}\)}%
  \mathclose{\copy\@brx\kern-0.5\wd\@brx\usebox{\@brx}}}
\makeatother

\usepackage{hyperref}
\hypersetup{
bookmarksnumbered,
colorlinks=true,
linkcolor={red!50!black},
citecolor={blue!50!black},
urlcolor={blue!80!black},
}

\begin{document}

\title{Tracy-Widom distributions for the 
 Gaussian orthogonal and symplectic ensembles revisited: a skew-orthogonal polynomials approach
}

\titlerunning{TW distributions for the GOE and GSE revisited: a skew-orthogonal polynomials approach}        

\author{\author{Anthony Mays$^1$ \and Anita Ponsaing$^2$ \and Gr\'{e}gory Schehr$^3$} 
}

\authorrunning{A. Mays, A. Ponsaing and G. Schehr} 


\institute{A. Mays \at  School of Mathematics and Statistics, University of Melbourne, VIC 3010, Australia 
\\ \and A. Ponsaing \at ARC Centre of Excellence for Mathematical and Statistical Frontiers (ACEMS), School of Mathematical Sciences, University of Adelaide, SA 5000, Australia\\
\and G. Schehr \at  Universit\'e Paris-Saclay, CNRS, LPTMS, 91405, Orsay, France}

\date{}

\maketitle

\begin{abstract}
We study the distribution of the largest eigenvalue in the ``Pfaffian'' classical ensembles of random matrix theory, namely in the Gaussian orthogonal (GOE) and Gaussian symplectic (GSE) ensembles, using semi-classical skew-orthogonal polynomials, in analogue to the approach of Nadal and Majumdar (NM) for the Gaussian unitary ensemble (GUE). Generalizing the techniques of Adler, Forrester, Nagao and van Moerbeke, and using ``overlapping Pfaffian'' identities due to Knuth, we explicitly construct these semi-classical skew-orthogonal polynomials in terms of the semi-classical orthogonal polynomials studied by NM in the case of the GUE. With these polynomials we obtain expressions for the cumulative distribution functions of the largest eigenvalue in the GOE and the GSE. Further, by performing asymptotic analysis of these skew-orthogonal polynomials in the limit of large matrix size, we obtain an alternative derivation of the Tracy-Widom distributions for GOE and GSE. This asymptotic analysis relies on a certain Pfaffian identity, the proof of which employs the characterization of Pfaffians in terms of perfect matchings and link diagrams.

\end{abstract}

\section{Introduction}

Since their discovery more than 25 years ago, the Tracy-Widom (TW) distributions \cite{forrester1993spectrum,tracy1994level,TracWido1996} have become cornerstones of extreme value statistics of strongly correlated variables \cite{majumdar2020extreme}. While they were initially found as the limiting distributions describing the typical fluctuations 
of the largest eigenvalues of large random matrices belonging to the classical Gaussian ensembles of random matrix theory (RMT), namely the Gaussian orthogonal, unitary and symplectic ensembles (respectively denoted as GOE, GUE and GSE), they have since found a large number of applications (for a review see \cite{majumdar2007course}). Indeed, TW distributions have emerged in a variety of problems at the interface between statistical mechanics and mathematics, including the longest increasing subsequence of random permutations \cite{baik1999distribution}, directed polymers \cite{baik1999distribution,baik2000limiting,baik2018pfaffian} and related growth models \cite{prahofer2000universal,majumdar2004anisotropic,imamura2004fluctuations}, in the Kardar-Parisi-Zhang (KPZ) universality class in (1 + 1) dimensions as well as for the continuum (1+1)-dimensional KPZ equation \cite{sasamoto2010one,calabrese2010free,dotsenko2010bethe,amir2011probability,le2012kpz,gueudre2012directed,barraquand2020half}, sequence alignment problems \cite{majumdar2005exact}, height fluctuations of non-intersecting Brownian motions over a fixed time interval \cite{forrester2011non,liechty2012nonintersecting,nguyen2017non}, height fluctuations of non-intersecting interfaces in the presence of a long-range interaction induced by a substrate \cite{nadal2009nonintersecting}, or more recently in the context of trapped fermions \cite{dean2015finite,dean2016noninteracting,stephan2019free,dean2019noninteracting}, as well as in finance \cite{biroli2007top}. Remarkably, the TW distributions have been recently observed in experiments on nematic liquid crystals \cite{takeuchi2010universal,takeuchi2011growing} (for the GOE and GUE) as well as in experiments involving coupled fibre lasers (for the GOE), and in dissipative self-assembled systems \cite{makey2020universality} (for the~GUE).  

In the pioneering works on the largest eigenvalue in the classical ensembles of RMT \cite{forrester1993spectrum,tracy1994level,TracWido1996}, the authors used the powerful tools of determinantal (for GUE) or Pfaffian (for GOE and GSE) point processes. This naturally led to the expression
of these distributions in terms of a Fredholm determinant (for GUE) or a Fredholm Pfaffian (for GOE and GSE). Using  rather involved 
``operator theoretic" techniques \cite{tracy1994level,TracWido1996}, it was further shown how to relate these Fredholm determinants and Pfaffians 
to sets of partial differential equations. In the limit of large matrix size $N$, this eventually led to a fairly explicit expression of these distributions for GOE, GUE and GSE in terms of a special solution of a Painlev\'e~II equation (the so called Hastings-McLeod solution, see also below). 

More recently, an alternative derivation of the TW distribution for the GUE was proposed by Nadal and Majumdar in Ref. \cite{NadaMaju2011} using (semi-classical) orthogonal polynomials. The idea of the method is rather simple and also quite instructive since one sees how the Painlev\'e II equation emerges from the asymptotic analysis of the three-term recurrence relation satisfied by these orthogonal polynomials, which are some deformations of the standard Hermite polynomials, in the limit of large matrix size $N$. Furthermore, this approach was further extended in Ref. \cite{PerrSche2014} to compute the distribution of the first gap (between the first two eigenvalues), and more generally the statistics of near extreme eigenvalues in the GUE, which could be expressed in a rather compact form in terms of Painlev\'e transcendents, from which very precise asymptotics could be derived (see also \cite{witte2013joint} for yet another derivation of the statistics of the first gap in the GUE). It would thus be very useful to obtain such an alternative derivation of the TW distributions in the other classical ensembles, namely the GOE and the GSE. This would be particularly interesting in the case of GOE, since this would provide a very efficient method to compute the statistics of near-extreme eigenvalues for this ensemble, which is directly relevant to describe static \cite{monthus2013typical} and dynamical \cite{fyodorov2015large} properties of a well known mean-field spin-glass model, namely the spherical Sherrington-Kirkpatrick model. Up to now, the statistics of near-extreme eigenvalues in these ensembles have only been studied numerically \cite{perret2015density}. The goal of this paper is precisely to extend the method of Ref. \cite{NadaMaju2011} and provide an alternative derivation of the TW distributions in the GOE and the GSE, by developing an approach based on (semi-classical) skew-orthogonal polynomials. This is a first important step towards a precise and useful description of the statistics of near-extreme eigenvalues, e.g. the first gap between the two largest eigenvalues, in terms of Painlev\'e transcendents in these ensembles \cite{mays2020prep}.  

\section{Summary of main results}

In the following we consider Gaussian random matrices $\bM= [m_{ij}]$ belonging to the aforementioned classical ensembles of random matrices with real symmetric (GOE), complex Hermitian (GUE) or real quaternionic self-dual (GSE) entries respectively~\cite{Mehta2004,Forrester2010} (see also Appendix \ref{sec:classical}), characterised by a Dyson index $\beta = 1, 2$ and $4$ respectively. In these three cases, the probability measure associated to the matrix ensemble is given by\footnote{Note that, for $\beta = 4$, the $\Tr$ function needs to be interpreted as a \textit{quaternion trace} [see Eq. \eqref{e:qTr}].}
\begin{align}
\label{e:MatPDFs} \Pr(\bM) \propto e^{-\beta (\Tr \bM^2)/2} \;.
\end{align}
In what follows we denote by G$\beta$E these ensembles with $\beta=1$ for the GOE, $\beta=2$ for the GUE and $\beta=4$ for the GSE. By performing a change of variables from the matrix entries $m_{ij}$ to the eigenvalues and eigenvectors of ${\bM}$, one obtains the joint probability density function (JPDF) of the (real) eigenvalues $\lambda_1, \lambda_2, \cdots, \lambda_N$ in the G$\beta$E ensembles as (see \cite{Mehta2004,Forrester2010})
\begin{align}
\label{e:evJPDFs} \cP_{\beta} (\lambda_1, \dots, \lambda_N) &=\frac{1} {Z_{\beta,N}}\; \prod_{j=1}^N e^{- \beta \lambda_j^2 /2 } \prod_{j<k} |\lambda_k -\lambda_j|^{\beta},
\end{align}
where $Z_{\beta, N}$ is a normalization constant such that
\begin{align}
    \int_{-\infty}^{\infty} d\lambda_1 \cdots \int_{-\infty}^{\infty} d\lambda_N \cP_{\beta} (\lambda_1, \dots, \lambda_N) =1 \;
\end{align}
and is given explicitly by
\begin{align}
\label{e:ZbetaN} Z_{\beta, N} = \beta^{-\frac{N}{2} - \frac{N \beta}{4} (N-1)} (2\pi)^{\frac{N}{2}} \prod_{j=0}^{N-1} \frac{\Gamma \left( 1+(j+1) \frac{\beta}{2} \right)} {\Gamma \left( 1+\frac{\beta}{2} \right)} \;,
\end{align}
where $\Gamma(z)$ is the gamma function. We will compute the cumulative distribution function (CDF) of the largest eigenvalue, i.e. $F_{\beta, N} (y) \equiv \Pr (\lambda_{\max}^{(\beta)} <y)$, 
or equivalently, the probability that all eigenvalues are less than some upper bound $y$
\begin{align}
\label{d:CDF} F_{\beta, N} (y) \equiv \Pr (\lambda_{\max}^{(\beta)} <y) = N! \int_{-\infty}^y d\lambda_{1} \int_{\lambda_1}^y d \lambda_2 \cdots \int_{\lambda_{N-1}}^y d\lambda_N \mathcal{P}_{\beta} (\lambda_{1}, \dots,\lambda_N) \;,
\end{align}
where the factorial $N!$ comes from the fact that in Eq.~(\ref{d:CDF}), the eigenvalues are ordered, i.e. $\lambda_1 < \lambda_2 < \dots < \lambda_N \leq y$. (Note that this ordering is not required here, however it will be convenient later to work with ordered eigenvalues and therefore we impose the ordering from the beginning.) It is well known that the JPDF in Eq.~(\ref{e:evJPDFs}) can be interpreted as the Boltzmann weight of a one-dimensional gas of $N$ charged particles where $\lambda_i$ denotes the position of the $i$-th particle and $\beta$ the inverse temperature \cite{dyson1962statistical}. These particles 
interact via a repulsive logarithmic interaction while they are subjected to an external quadratic potential: this is the so-called log-gas. Hence the CDF $F_{\beta, N} (y)$ in Eq.~(\ref{d:CDF}) is the partition function of this log-gas in the presence of a hard wall at position $y$~\cite{majumdar2014top} --- such partition functions are called ``restricted partition functions'' in the following.  

\begin{figure}
\begin{center}
{\includegraphics[width=0.4\textwidth]{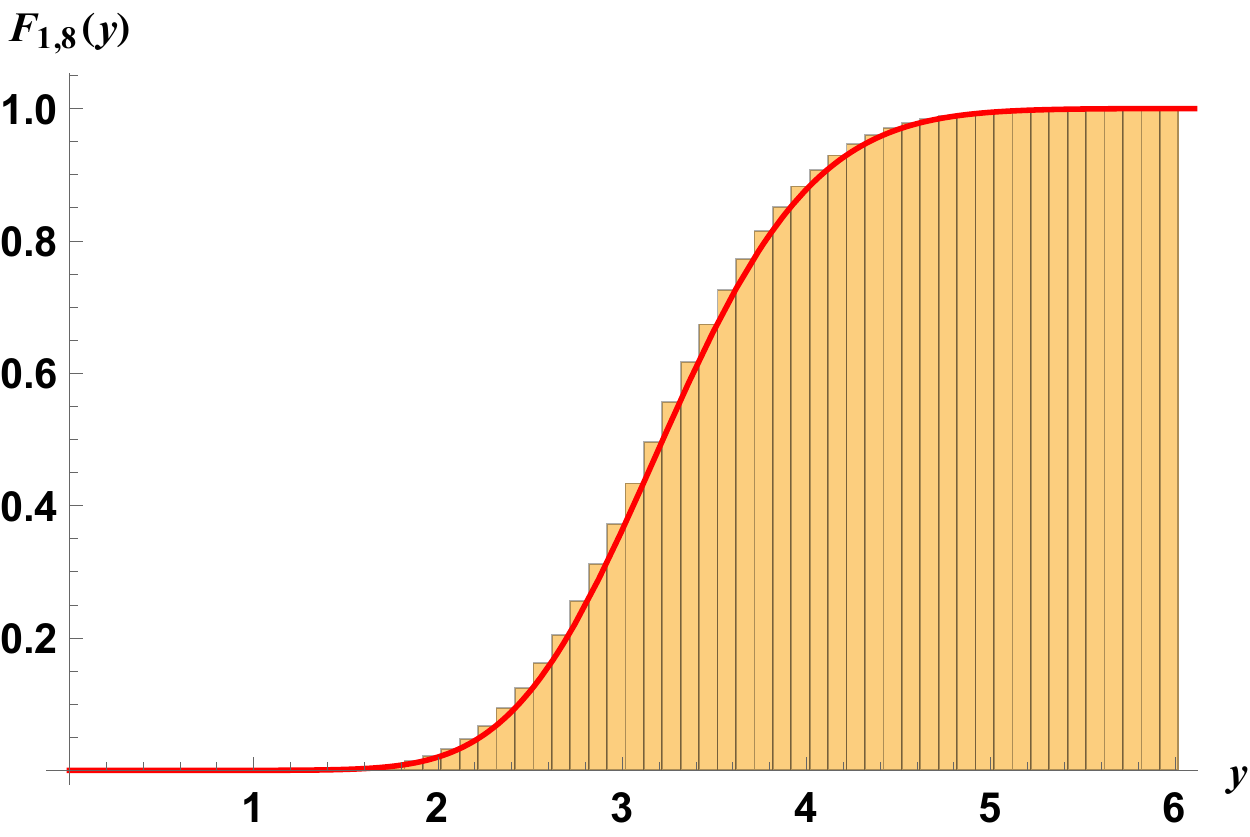}}  \qquad {\includegraphics[width=0.4\textwidth]{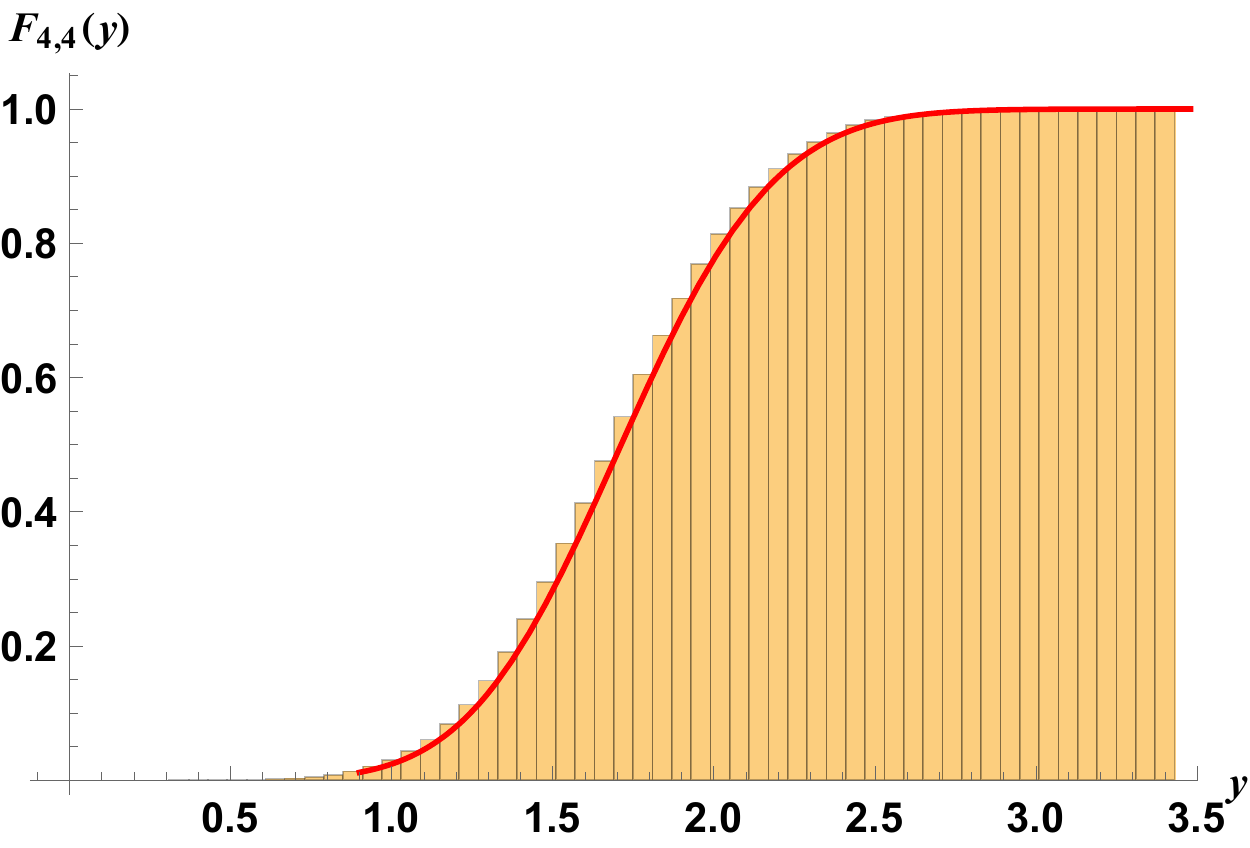}}
\caption{The histograms correspond to a numerical evaluation of the CDF of the largest eigenvalue sampled from $50,000$ matrices in the $\beta=1$ ensemble for $N=8$ (left panel) and in the $\beta=4$ ensemble for $N=4$ (right panel). The solid red line represents the exact result given, in the left panel, by Eq. (\ref{e:F1N}) and, in the right panel, by Eq. (\ref{e:F4N}).}\label{f:CDFs}
\end{center}
\end{figure}

To compute $F_{\beta, N} (y)$, it is useful to introduce sets of \textit{orthogonal} and \textit{skew-orthogonal} polynomials. Specifically, we define the $y$-dependent inner (or scalar) product for $\beta=2$
\begin{align}
    \label{d:IP2} (f,g)_{2}^y = \int_{-\infty}^y e^{-\lambda^2} f(\lambda) g(\lambda) d\lambda \;,
\end{align}
and the skew-inner products for $\beta=4$
\begin{align}
\nonumber \langle f , g \rangle_{4}^y &= \frac{1}{2} \int_{- \infty}^{y} dx \; e^{-2 x^2} \left[ f(x) g'(x)- g(x) f'(x) \right]\\
\label{d:IP4} &= \frac{1}{2} \int_{- \infty}^{y} dx \; e^{-x^2} \left[ f(x) \frac{d} {dx} \left( e^{-x^2} g(x)\right) - g(x) \frac{d} {dx} \left( e^{-x^2} f(x) \right) \right],
\end{align}
and for $\beta=1$
\begin{align}
\nonumber \langle f , g \rangle_{1}^y &= \frac1{2} \int_{-\infty}^{y} dx \; e^{-x^2/2} f(x) \int_{-\infty}^y d z \; e^{-z^2/2} g (z) \sgn (z- x)\\
\label{d:IP1} &= \frac1{2} \int_{-\infty}^{y} dx \; e^{-x^2/2} f(x) \int_{x}^{y} d z \; e^{-z^2/2} g (z) - \frac1{2} \int_{-\infty}^{y} dx \; e^{-x^2/2} f(x) \int_{-\infty}^{x} d z \; e^{-z^2/2} g (z).
\end{align}
Then for each ensemble, we seek a set of (monic) polynomials $\{ p_j (x,y) \}$ for $\beta = 2$, $\{Q_j (x, y) \}$ for $\beta = 4$, and $\{ R_j (x, y) \}$ for $\beta = 1$ (by increasing order of complexity, as we will see) with the orthogonality/skew-orthogonality properties
\begin{align}
\label{d:2orthog} (p_j, p_k)_2^y = h_j(y) \delta_{j,k},
\end{align}
\begin{align}
\nonumber \langle Q_{2j} , Q_{2k} \rangle_{4}^y&= \langle Q_{2j+1} , Q_{2k+1} \rangle_{4}^y= 0 \;, \\
\label{d:4sorthog} \langle Q_{2j} , Q_{2k+1} \rangle_{4}^y&= -\langle Q_{2k+1} , Q_{2j} \rangle_{4}^y = q_j(y) \delta_{j,k} \;,
\end{align}
and
\begin{align}
\nonumber \langle R_{2j} , R_{2k} \rangle_{1}^y&= \langle R_{2j+1} , R_{2k+1} \rangle_{1}^y= 0 \;, \\
\label{d:1sorthog} \langle R_{2j} , R_{2k+1} \rangle_{1}^y&=-\langle R_{2k+1} , R_{2j} \rangle_{1}^y = r_j (y) \delta_{j,k} \;,
\end{align}
where, to be explicit, the respective normalizations are
\begin{align}
\label{e:ynorm2} h_j(y) &:= (p_j, p_j)_2^y\\
\label{e:ynorm4} q_j(y) &:=\langle Q_{2j} , Q_{2j+1} \rangle_{4}^y\\
\label{e:ynorm1} r_j(y) &:= \langle R_{2j} , R_{2j+1} \rangle_{1}^y \;.
\end{align}
(Note that the orthogonal and skew-orthogonal polynomials depend on the parameter $y$, although we will often suppress the explicit notation of that dependence for brevity.) In fact, as for the case of the GUE \cite{NadaMaju2011}, the CDF $F_{\beta,N}(y)$ can be expressed only in terms of the norms $h_j(y)$, $q_j(y)$ and $r_j(y)$ for $\beta = 2, 4$ and $1$ respectively. For $\beta=2$, it was indeed shown in \cite{NadaMaju2011} that
\begin{align}
\label{e:F2N}F_{2, N} (y)  &= \prod_{j=0}^{N-1} \frac{h_j(y)} {h_j (\infty)} = \frac{2^{N(N-1)/2}}{\pi^{N/2}} \prod_{j=0}^{N-1} \frac{h_j(y)} {j!} \;.
\end{align}
In the present paper we show that for $\beta =1$ (and where $N$ is restricted to be even for simplicity), with the polynomials $R_j$ from \eqref{d:1sorthog}, we have
\begin{align}
\label{e:F1N} F_{1,N} (y) = \prod_{j=0}^{N/2 -1} \frac{r_j (y)} {r_j (\infty)} = \frac{2^{\frac{N}{2} \left( \frac{N}{2} -1 \right)}} {\pi^{N/4}} \prod_{j=0}^{N/2 -1} \frac{1} {(2j)!} \Pf \bV_{N-1},
\end{align}
where the matrix $\bV_m$, whose explicit expression is given in \eqref{d:Vm} below, contains the $\beta=2$ polynomials $p_{j} (x,y)$ and their normalizations $h_j(y)$. For $\beta=4$ we require a slightly modified (by a simple rescaling) skew-inner product with associated modified skew-orthogonal polynomials $\tilde{Q}_{j}$ and normalizations $\tilde{q}_j$  [see Eqs. (\ref{e:Rescaleqj}) and (\ref{e:RescaleQj})], which gives us
\begin{align}
\label{e:F4N} F_{4, N} (y) = \prod_{j=0}^{N-1} 2^{ -2j -\frac{1}{2}} \frac{\tilde{q}_j (\sqrt{2} y)} {q_j (\infty)} = \frac{2^{N^2}}{\pi^{N/2}} \prod_{j=0}^{N-1} \frac{1}{(2j +1)!} \Pf \bW_{2N-1}\Big|_{y \mapsto \sqrt{2} y},
\end{align}
where the matrix $\bW_m$ is given in \eqref{d:Wmat}, and again contains the $\beta=2$ polynomials $p_{j} (x,y)$ and their normalizations $h_j(y)$. In Fig. \ref{f:CDFs} we present a comparison between a numerical evaluation of these formulae \eqref{e:F1N} and \eqref{e:F4N} and a direct numerical computation of these CDFs by sampling GOE and GSE random matrices, showing very good agreement. We emphasize that the expressions on the right hand side of Eqs. \eqref{e:F1N} and \eqref{e:F4N} depend only on the $\beta=2$ orthogonal polynomials, and do not depend on the skew-orthogonal polynomials at all.

In the case of the GUE ($\beta = 2$), the orthogonal polynomials $p_k$ for the inner product in \eqref{d:IP2} have already been studied, first in \cite{NadaMaju2011} and later in \cite{PerrSche2014}. Here we call these polynomials the \textit{Nadal--Majumdar (NM) polynomials}. Interestingly, these NM polynomials naturally arise also in the study of the so called {\it level curvature distribution} at the soft edge of random Hermitian matrices \cite{fyodorov2011level}. Although they do not have a known closed formula, they satisfy the three-term recurrence relation [since they are orthogonal with respect to the inner product in (\ref{d:IP2})]
\begin{align}
\label{e:NM1} \lambda p_k (\lambda, y) &= p_{k+1} (\lambda, y) + \NMrs{S}_k (y) p_k (\lambda, y) + \NMrs{R}_k (y) p_{k-1} (\lambda, y)\\
\label{e:NM2} \NMrs{R}_{k}(y) &= \frac{h_k (y)} {h_{k-1}(y)}\\
\label{e:NM3} \NMrs{S}_k(y) & \neq 0,
\end{align}
where the last expression follows because the domain of integration in the inner product \eqref{d:IP2} is not symmetric. (Note that we have used the ``check'' and sans serif font to distinguish $\NMrs{R}_k$ from the $\beta=1$ polynomials $R_j$. We use the same style for $\NMrs{S}_k$ for consistency.) In the limit $y\to \infty$ the NM polynomials become the (monic, ``physicist's'') Hermite polynomials, i.e. \cite{NadaMaju2011} 
\begin{align}
\label{e:limHerms} p_j(\lambda, y) = \frac{1}{2^j} H_j (\lambda) + O \left( e^{-y^2} \right),
\end{align}
where the Hermite polynomials of index $j$, $H_j(x)$, are orthogonal with respect to the weight function $e^{-x^2}$, and the division by $2^j$ is here to ensure monicity. In fact, in this limit the inner product (\ref{d:IP2}) and skew-inner products \eqref{d:IP4}--\eqref{d:IP1} all reduce to their classical Gaussian counterparts, with norms \cite{Mehta2004,Forrester2010}
\begin{align}
\label{e:InfinityNorms} h_j (\infty)= \frac{\pi^{1/2}} {2^j} \Gamma (j+1), \quad q_j(\infty) = \frac{\pi^{1/2}} {2^{4j+\frac{3}{2}}} \Gamma (2j+2), \quad r_j(\infty) = \frac{\pi^{1/2}} {2^{2j}} \Gamma (2j+1) \;.
\end{align}
The corresponding classical skew-orthogonal polynomials are known, and recalled in Appendices \ref{a:limpolys4} and \ref{a:limpolys1}.

However, for finite $y$, there are no known statements analogous to \eqref{e:NM1}--\eqref{e:NM3} for $\beta=1$ and $4$ polynomials. Yet, as a first approach, we can iteratively use the skew-inner products \eqref{d:IP4} and \eqref{d:IP1} to construct these polynomials. An important property is that these polynomials are not unique, since skew-inner products are invariant under the polynomial transformation
\begin{align}
\label{e:oddsymm} \eta_{2j+1} \mapsto \eta_{2j+1} + c \; \eta_{2j}
\end{align}
where $c$ is any constant (and $\eta_k = Q_k$ or $R_k$), and therefore a set of skew-orthogonal polynomials is unique only up to this symmetry in the odd degree polynomials. By specifying the constant we employ this iterative process to construct the skew-orthogonal polynomials defined in Eqs. (\ref{d:4sorthog}) and (\ref{d:1sorthog}) in Appendix \ref{s:iterative}. However this method is not convenient for the asymptotic analysis of the quantities in \eqref{d:CDF}. Instead, in \cite{NadaMaju2011,PerrSche2014}, it was shown that the recurrence relations \eqref{e:NM1}--\eqref{e:NM3} can be exploited to obtain the asymptotic behaviors of the norms $h_j(y)$ and the polynomials $p_j (\lambda,y)$ themselves in the limit of large $N$ and large $y$. Here, we extend the approach developed in \cite{AdlevanMoer2002,AdleForrNagavanMoer2000} to obtain explicit expressions for the sets of semi-classical skew-orthogonal polynomials $\{ Q_j\}$ and $\{ R_j\}$ in the basis of the orthogonal polynomials $\{ p_j\}$ (the NM polynomials). This is the content of Proposition \ref{p:alpha4} (for the GSE) and Proposition \ref{p:alpha1} (for the GOE). Interestingly, the proofs of these results are achieved by using results on {\it overlapping Pfaffians}, studied by Knuth \cite{Knuth1996}. This is the first main technical contribution of this work. As a byproduct of our analysis, we also recover the classical skew-orthogonal polynomials as the $y\to \infty$ limit of our results here (see Appendices \ref{a:beta4} and \ref{a:beta1}).

We will then use this explicit construction, together with the asymptotic analysis of the polynomials $p_j(x, y)$, to compute the large $N$ asymptotic limit of $F_{1,N}$ and $F_{4,N}$. Indeed for the case of the GSE, we show that\footnote{Note that the factor $2^{-7/6}$ differs by a factor $2^{-1/6}$ from the result obtained in the original paper \cite{TracWido1996}. This mistake was actually noticed in \cite[p.47]{nadal2011matrices} --- see also \cite{borot2012right}. There, this factor was corrected by matching with known asymptotic results for large (positive and negative) arguments. Here, we obtain this correct factor $2^{-7/6}$ by a direct computation.}
\begin{eqnarray}\label{e:TW4_intro}
\lim_{N \to \infty} F_{4,N} \left( y = \sqrt{2N} + \frac{s}{2^{7/6} N^{1/6}} \right) = \exp{\left(-\frac{1}{2} \int_s^\infty (x-s) q^2(x) dx \right)} \cosh{\left(\frac{1}{2} \int_s^\infty q(x)dx \right)},
\end{eqnarray}
where $q(x)$ is the Hastings-McLeod solution of the Painlev\'e II equation, i.e.
\begin{eqnarray}\label{e:PII_intro}
q''(x) = x q(x) + 2 q^3(x) \;, {\rm with} \;\; q(x) \underset{x \to \infty}{\sim} {\rm Ai}(x) \;,
\end{eqnarray}
and ${\rm Ai}(x)$ is the standard Airy function. On the other hand, for the case of the GOE, we show that
\begin{align}
\label{e:TW1_intro} \lim_{N \to \infty} F_{1, N} \left( y = \sqrt{2N} + \frac{s}{\sqrt{2} N^{1/6}} \right) &= \exp\left( -\frac{1}{2} \int_s^{\infty} (x-s) q(x)^2 dx \right) \exp\left( - \frac{1}{2} \int_s^{\infty} q(x) dx \right),
\end{align}
with, again, $q(x)$ given in Eq. (\ref{e:PII_intro}). We thus recover the known expressions of the TW distributions for GSE and GOE \cite{TracWido1996}, by using here a completely different method. This is the second main achievement of the present paper. The key result used to obtain the TW distributions is an identity proved in Proposition \ref{lem:PfW} [see Eq. (\ref{e:expansion:Pf})] that allows us to obtain explicit expressions of the Pfaffians entering the expressions in Eqs. (\ref{e:F1N}) and (\ref{e:F4N}), which are then conveniently amenable to an asymptotic analysis in the limit of large $N$. The proof of this identity (\ref{e:expansion:Pf}) relies on the expression of a Pfaffian as a sum over perfect matchings recalled in \eqref{e:PfPerfMatch} of the Appendices --- this representation is used extensively throughout the present paper.

The paper is organized as follows.  In Section \ref{sec:PartFns} we use the polynomials $Q_j$ and $R_j$, defined in \eqref{d:4sorthog} and \eqref{d:1sorthog} respectively, to find Pfaffian expressions for restricted partition functions such as the CDFs $F_{1,N}(y)$ and $F_{4,N}(y)$ using standard techniques. In Section \ref{sec:SOPs} we construct explicitly these skew-orthogonal polynomials in terms of the NM polynomials $p_j (x,y)$ and their normalizations $h_j (y)$, finding in particular Pfaffian expressions for the coefficients and the normalizations $q_j(y)$ and $r_j(y)$. In Sections \ref{s:F4asympt} and \ref{s:F1asympt} we present the asymptotic analysis of $F_{4,N}(y)$ and $F_{1,N}(y)$ respectively, leading to the expressions given in Eqs. (\ref{e:TW4_intro}) and (\ref{e:TW1_intro}). Finally, Section \ref{s:conclusion} contains our conclusions and perspectives. Several technical details about the results presented in this paper have been left to the Appendices.  

\section{Restricted partition functions and generalizations}
\label{sec:PartFns}

In this section, we show how to compute restricted partition functions such as the CDFs $F_{\beta,N}(y)$ in Eq. (\ref{d:CDF}). We actually consider
slightly more general quantities defined as the following averages over the eigenvalue JPDFs for the GOE ($\beta=1$), GUE ($\beta=2$) and the GSE ($\beta =4$):
\begin{align}
\nonumber \hat{Z}_{\beta, N} [a, y] &= \left\langle \prod_{j=1}^N a(\lambda_j) \right\rangle_{\mathcal{P}_{\beta}}^y\\
\label{e:GenPF}&= \frac{1} {Z_{\beta,N}} \int_{-\infty}^y d\lambda_1 \cdots \int_{-\infty}^y d\lambda_N \prod_{j=1}^N a(\lambda_j) e^{- \beta \lambda_j /2 } \prod_{j<k} |\lambda_k -\lambda_j|^{\beta}.
\end{align}
Each of the $\hat{Z}_{\beta, N} [a, y]$ will be put into determinant/Pfaffian form --- the construction of the associated matrices will depend on its own set of monic polynomials. While these polynomials are in principle arbitrary, it is convenient to specify them to be the respective orthogonal/skew-orthogonal polynomials. If we think of the integral in \eqref{e:GenPF} as an average over a truncated version of the density \eqref{e:evJPDFs}, i.e.
\begin{eqnarray} \label{eq:janossy1}
\cP_{\beta} (\lambda_1, \dots, \lambda_N; y) := \cP_{\beta} (\lambda_1, \dots, \lambda_N) \chi_{(-\infty, y)} (\lambda_1, \dots, \lambda_N) \;,
\end{eqnarray}
where $\chi_{A} (\bx) =1$ if $\bx\in A^N$ and zero otherwise, then we are in the realm of Janossy densities \cite{Janossy1950} (see \cite{Soshnikov2004} for a clear introduction to the topic and references). In \cite{BoroSosh2003} the authors discussed ``determinantal'' Janossy densities (where the particle JPDF and $n$-point correlation functions can be written in terms of a determinant) and found the matrix kernel for the determinant. In \cite{Soshnikov2003} these results were extended to ``Pfaffian'' Janossy densities, that is, the author found the matrix kernel for Janossy JPDFs and $n$-point correlation functions that are expressed as Pfaffians. Our eigenvalue JPDFs \eqref{e:evJPDFs} have this determinantal ($\beta=2$) or Pfaffian ($\beta=1,4$) structure, and so the $n$-point correlations will also have determinantal/Pfaffian structure. We will explicitly construct these correlation functions in a future work, and use them to calculate gap probabilities and the density of states near the largest eigenvalue \cite{mays2020prep}.
Here, however, we restrict ourselves to the calculation of the averages \eqref{e:GenPF}, which gives us the CDF of the largest eigenvalue \eqref{d:CDF} via
\begin{align}
\label{e:FZ} F_{\beta, N} (y) = \hat{Z}_{\beta, N} [1,y] \;.
\end{align}
Below we treat the case $\beta = 2$, $\beta = 4$ and $\beta = 1$, again by increasing order of complexity. 

\subsection{$\beta=2$}

Although this is not needed for the $\beta=1,4$ cases, for completeness we also include the $\beta=2$ result, which can be obtained using the Vandermonde determinant identity (the procedure is a slight modification to that in \cite[\S 5.2.1]{Forrester2010})
\begin{align}
\label{e:GenPF2} \hat{Z}_{2,N} [a, y] = \frac{N!}{Z_{2 ,N}} \det \left[ \gamma_{j,k}^{(2)} [a, y] \right]_{j,k= 0, \dots, N-1},
\end{align}
where $Z_{2, N}$ is given in \eqref{e:ZbetaN} and
\begin{align}
\label{e:GenPF2b} \gamma_{j,k}^{(2)} [a, y] := \int_{-\infty}^y a(\lambda) e^{-x^2} p_j (\lambda, y) p_k (\lambda, y) d\lambda \;.
\end{align}
The polynomials $p_j$ in \eqref{e:GenPF2b} are the NM polynomials, i.e. the monic polynomials of degree $j$ that are orthogonal with respect to the inner product \eqref{d:IP2}. A consequence of this (in the limit $y\to \infty$) is the known result \cite{Mehta2004,Forrester2010}
\begin{align}
\label{e:Z2norms} Z_{2,N} = N! \prod_{j=0}^{N-1} h_j (\infty),
\end{align}
where $h_j (\infty)$ is given in \eqref{e:InfinityNorms}, which agrees with \eqref{e:ZbetaN}.

With $a(x)=1$ the integral $\gamma_{j,k}^{(2)}$ becomes the inner product \eqref{d:IP2}, so with the orthogonal polynomials $p_j$ we use \eqref{e:FZ} to obtain the known result \eqref{e:F2N}.

\subsection{$\beta=4$}

\begin{proposition}
The average \eqref{e:GenPF} for $\beta=4$ is
\begin{align}
\label{e:GenPF4} \hat{Z}_{4,N} [a, y] = \frac{N! 2^N}{Z_{4, N}}\; \Pf \left[ \gamma^{(4)}_{j,k}[a, y] \right]_{j,k= 0, \dots, 2N-1},
\end{align}
where
\begin{align}
\gamma^{(4)}_{j,k} [a ,y] := \frac{1}{2} \int_{- \infty}^{y} d\lambda \; a(\lambda) e^{-\lambda^2} \left[ Q_j(\lambda, y) \frac{d} {d\lambda} \left( e^{-\lambda^2} Q_k(\lambda, y)\right) - Q_k(\lambda, y) \frac{d} {d\lambda} \left( e^{-\lambda^2} Q_j (\lambda, y) \right) \right]
\end{align}
and the $Q_j$ are monic polynomials of degree $j$ that are skew-orthogonal with respect to the skew-inner product~\eqref{d:IP4}.
\end{proposition}

Using the theory of Section \ref{sec:SOPs} below, we can make a quick check of \eqref{e:GenPF4} by noting that when $a(x) = 1$ the matrix in \eqref{e:GenPF4} is of the form \eqref{d:sdiag}, and so from \eqref{e:Pfsdiag} the Pfaffian is given by the product $q_0 (y) q_1 (y) \cdots q_{N-1} (y)$. Then, in the limit $y\to \infty$, we recover the result analogous to \eqref{e:Z2norms} \cite{Mehta2004,Forrester2010}
\begin{align}
\hat{Z}_{4,N}[1,y]\Big|_{y\to \infty} =1 \qqimp Z_{4, N} = N! 2^N \prod_{j=0}^{N-1} q_j (\infty),
\end{align}
where $q_j (\infty)$ is given in \eqref{e:InfinityNorms}, and this agrees with \eqref{e:ZbetaN}.

\bigskip\proof This result is obtained using the same techniques as applied in \cite{TracWido1998,Mehta2004,Forrester2010}, but with a truncated domain of integration, and a correspondingly different set of polynomials. To keep this paper self-contained, we will go through the details. We start with the identity \cite{Mehta2004}
\begin{align}
\label{e:beta4Vdm} \prod_{1\leq j< k\leq N} (\lambda_j - \lambda_k)^4 = \det \left[ \begin{array}{c}
\lambda_j^{k-1}\\
(k-1) \lambda_j^{k-2}
\end{array}\right]_{j=1, \dots, N \atop k= 1, \dots, 2N},
\end{align}
and note that each even row is the derivative of the odd row immediately above it. Then in this matrix, for each column, by adding linear combinations of the columns to the left of that column (starting from the left-most column) we can create arbitrary monic polynomials, while preserving the derivative relationship between the even and odd rows. So for our purpose, we choose the polynomials to be the $Q_j$, which are skew-orthogonal with respect to the skew-inner product \eqref{d:IP4}, giving
\begin{align}
\nonumber \hat{Z}_{4, N} [a, y] &= \frac{1}{Z_{4, N}} \int_{-\infty}^y d\lambda_1 \cdots \int_{-\infty}^y d\lambda_N \prod_{j=1}^N a(\lambda_j) e^{- 2 \lambda_j } \det \left[ \begin{array}{cc}
Q_{2k-2} (\lambda_j) & Q_{2k-1} (\lambda_j)\\
Q_{2k-2}' (\lambda_j) & Q_{2k-1}' (\lambda_j)
\end{array}\right]_{j,k= 1, \dots, N}\\
\label{e:beta4Vdm2}&= \frac{1}{Z_{4, N}} \sum_{P\in S_{2N}} \varepsilon(P) \prod_{j=1}^N \int_{-\infty}^y d\lambda\; a(\lambda) e^{-2\lambda^2} Q_{P(2j-1) -1} (\lambda) Q_{P(2j) -1}' (\lambda),
\end{align}
where the second line follows from Laplace expansion of the determinant, and we apply the integrals to each matched pair of $Q$ and $Q'$. (Note that we suppress the dependence on $y$ for brevity.)

For each pair of indices on the $Q$ and $Q'$ in \eqref{e:beta4Vdm2}, we then match up each permutation with the corresponding permutation where that index pair is interchanged, hence picking up a $(-1)$, giving
\begin{align}
\nonumber &\hat{Z}_{4, N} [a, y] =\frac{1}{Z_{4, N}} \times\\
&\sum_{P\in S_{2N} \atop P(2j)> P(2j-1)} \varepsilon(P) \prod_{j=1}^N \int_{-\infty}^y d\lambda\; a(\lambda) e^{-2\lambda^2} \Big( Q_{P(2j-1) -1} (\lambda) Q_{P(2j) -1}' (\lambda) - Q_{P(2j) -1} (\lambda) Q_{P(2j-1) -1}' (\lambda) \Big),
\end{align}
where we need to restrict the sum to just those permutations obeying the rule $P(2j)> P(2j-1)$ for all $j$. Introducing a factor of $\frac{1}{2}$ for each integral (incurring a pre-factor of $2^N$), then using the definition of the Pfaffian recalled in \eqref{def:Pf} we obtain
\begin{align}
\hat{Z}_{4, N} [a, y] =\frac{2^N N!}{Z_{4, N}} \Pf \left[ \frac{1}{2} \int_{-\infty}^y d\lambda\; a(\lambda) e^{-2\lambda^2} \Big( Q_{j} (\lambda) Q_{k}' (\lambda) - Q_{k} (\lambda) Q_{j}' (\lambda) \Big)  \right]_{j,k=0, \dots, 2N-1}.
\end{align}
The equality between the first and second lines in \eqref{d:IP4} gives the result in \eqref{e:GenPF4}.

\hfill $\Box$

While \eqref{e:GenPF4}, with $a(x)=1$, gives us the CDF $F_{4,N}$, we will need the explicit forms of the polynomials $\{ Q_j \}$ before we can obtain the expression in \eqref{e:F4N}. This will be achieved below in Section \ref{sec:SOPs}.

\subsection{$\beta=1$, with $N$ even}

Recall that we have restricted $N$ to be even in this work. The parity of $N$ plays an important role since for the $\beta=1$ case we have the difficulty of the absolute value of the Vandermonde determinant in \eqref{e:evJPDFs}. To deal with it, we apply the method of integration over alternate variables, which was introduced by de Bruijn \cite{deBruijn1955} and applied to integrals similar to \eqref{e:GenPF} by Mehta \cite{Mehta2004}. However, this method is dependent on the parity of $N$, which can be seen when one pairs up the rows in \eqref{e:integAVs} below --- when $N$ is odd there would be one unpaired row, which needs to be specially dealt with. For simplicity, we will only work with the $N$ even case here, and the techniques for dealing with the $N$ odd case are contained in \cite{Mehta2004,Forrester2010,ForrMays2009,Mays2011thesis}.

\begin{proposition}
With $N$ even the average \eqref{e:GenPF} for $\beta=1$ is
\begin{align}
\label{e:GenPF1} \hat{Z}_{1,N} [a ,y] = \frac{2^{N/2} N!}{Z_{1,N}} \Pf \left[ \gamma^{(1)}_{j,k}[a ,y] \right]_{j,k= 0, \dots, N-1},
\end{align}
where
\begin{align}
\label{d:gamma1} \gamma^{(1)}_{j,k} [a ,y] := \frac1{2} \int_{-\infty}^{y} dx \; a(x) e^{-x^2/2} R_j (x, y) \int_{-\infty}^{y} d z \; a(z) e^{-z^2/2} R_k (z, y) \, \sgn(z-x)
\end{align}
and the $R_j$ are monic polynomials of degree $j$ that are skew-orthogonal with respect to the skew-inner product~\eqref{d:IP1}.
\end{proposition}

As with $\beta=4$ above, we can recover the known result \eqref{e:ZbetaN} \cite{Mehta2004,Forrester2010} with $a(x)=1, y\to \infty$ 
\begin{align}
\hat{Z}_{1,N} [1,y]\Big|_{y\to \infty} = 1 \qqimp Z_{1, N} = N! 2^{N/2} \prod_{j=0}^{N/2 -1} r_j (\infty),
\end{align}
where $r_j (\infty)$ is given in \eqref{e:InfinityNorms}.

\bigskip\proof As in the case of $\beta=4$ above, the techniques used here are found in \cite{TracWido1998,Mehta2004,Forrester2010} but we will delve into some of the details using the truncated integral for completeness.

We start by ordering the eigenvalues $-\infty < \lambda_1 <\cdot\cdot\cdot < \lambda_N < y$ (incurring a factor of $N!$) in \eqref{e:GenPF} so that we can remove the absolute value from the product of differences. Then we use the Vandermonde determinant expression (suppressing the polynomial dependence on $y$)
{\small
\begin{align}
\nonumber &\hat{Z}_{1,N} [a ,y]= \frac{N!}{Z_{1, N}} \int_{-\infty}^{y}d\lambda_N \int_{-\infty}^{\lambda_N}d\lambda_{N-1} \cdot\cdot\cdot \int_{-\infty}^{\lambda_2} d\lambda_1 \prod_{j=1}^N e^{-\lambda_j^2/2}\; a(\lambda_j)\prod_{1\leq j < k \leq N}(\lambda_k -\lambda_j)\\
\nonumber &=\frac{N!}{Z_{1, N}} \int_{-\infty}^{y}d\lambda_N \int_{-\infty}^{\lambda_N} d\lambda_{N-1} \cdot\cdot\cdot \int_{-\infty}^{\lambda_2} d\lambda_1 \det \left[ e^{-\lambda_j^2/2} a(\lambda_j) \lambda_j^{k-1} \right]_{j,k=1,...,N}\\
&=\frac{N!}{Z_{1, N}} \int_{-\infty}^{y} d\lambda_N \int_{-\infty}^{\lambda_N} d\lambda_{N-1} \cdot\cdot\cdot \int_{-\infty}^{\lambda_2} d\lambda_1 \det \left[ e^{-\lambda_j^2/2} a(\lambda_j) R_{k-1}(\lambda_j) \right]_{j,k=1,...,N},
\end{align}}where the third equality follows from elementary column operations. This is the same procedure that was applied to \eqref{e:beta4Vdm} in the $\beta=4$ case above, and it allows us to obtain any set of monic polynomials in the columns; for our purposes we specify the polynomials to be the $\{R_j\}$, which are skew-orthogonal with respect to the skew-inner product \eqref{d:IP1}.

Now we wish to apply the method of integration over alternate variables (mentioned above), and to prepare for that we change the order of the integrals, with even integrals on the left and odd integrals on the right
{\small\begin{align}
\nonumber &\hat{Z}_{1,N} [a ,y]=\frac{N!}{Z_{1, N}}\\
&\times \int_{-\infty}^{y}d\lambda_N \int_{-\infty}^{\lambda_N} d\lambda_{N-2} \cdot\cdot\cdot \int_{-\infty}^{\lambda_4} d\lambda_2 \int_{\lambda_{N-2}}^{\lambda_N} d\lambda_{N-1} \cdots \int_{\lambda_{2}}^{\lambda_{4}} d\lambda_{3} \int_{-\infty}^{\lambda_2} d\lambda_1 \det \left[ e^{-\lambda_j^2/2} a(\lambda_j) R_{k-1} (\lambda_j) \right]_{j,k=1,...,N}.
\end{align}}The purpose of this manipulation is that now in each odd integral (i.e. over the variables $\lambda_{2n-1}$) the only dependence of the corresponding variable is in the $(2n-1)$st row of the determinant, so the odd integrals can be applied to their respective rows:
{\small
\begin{align}
\label{e:integAVs} \hat{Z}_{1,N} [a ,y]&=\frac{N!}{Z_{1, N}} \int_{-\infty}^{y} d\lambda_N \int_{-\infty}^{\lambda_N}d\lambda_{N-2} \cdot\cdot\cdot \int_{-\infty}^{\lambda_4} d\lambda_2 \det \left[ \begin{array}{c}
\int_{-\infty }^{\lambda_{2j}} e^{-\lambda^2/2} a(\lambda) R_{k-1}(\lambda) d\lambda\\
e^{-\lambda_{2j}^2/2} a(\lambda_{2j}) R_{k-1}(\lambda_{2j})
\end{array}\right]_{j=1,...,N/2 \atop k=1,...,N},
\end{align}
}where we have added the first row to the third row, and the first and third rows to the fifth row, and so on, so all the integrals have lower terminal $-\infty$. (This sequence of steps is the \textit{method of integration over alternate variables}.)

We see that the determinant in \eqref{e:integAVs} is now symmetric in the variables $\lambda_2, \lambda_4,..., \lambda_N$, and so we can remove the ordering $\lambda_2 < \lambda_4 < ... < \lambda_N$ at the cost of dividing by $(N/2)!$. Taking the Laplace expansion of the determinant we find
\begin{align}
\hat{Z}_{1,N} [a ,y]= \frac{1}{Z_{1,N}} \frac{N!} {(N/2)!} \sum_{P\in S_N} \varepsilon(P) \prod_{j=1}^{N/2} \mu_{P(2j -1), P(2j)},
\end{align}
where
\begin{align}
\label{def:mu_GOE} \mu_{j,k}:=\int_{-\infty}^{y} dx\: e^{-x^2/2}\: a(x) \:R_{k-1}(x) \int_{-\infty}^x dz\: e^{-z^2/2}\: a(z)\: R_{j-1}(z),
\end{align}
and $\varepsilon(P)$ is the sign of the permutation $P$. By defining
\begin{align}
\gamma_{j,k}^{(1)} :=\frac{1}{2}(\mu_{j,k} -\mu_{k,j}),
\end{align}
incurring a factor of $2^{N/2}$, then we can restrict the sum to terms with $P(2j)> P(2j -1)$ for all $j$, giving
\begin{align}
\hat{Z}_{1,N} [a ,y]=\frac{1}{Z_{1,N}} 2^{N/2} \frac{N!} {(N/2)!} \sum_{P\in S_N \atop P(2j) > P(2j-1)} \varepsilon(P) \prod_{j=1}^{N/2} \gamma_{P(2j-1), P(2j)}^{(1)}.
\end{align}
Now using \eqref{def:Pf} we have the result in \eqref{e:GenPF1}--\eqref{d:gamma1} [where we cancel the factor of $(N/2)!$ to account for summing over distinct terms only].

\hfill $\Box$

As for $\beta=4$ above, we will need to first find the skew-orthogonal polynomials $\{ R_j \}$ before we can use \eqref{e:GenPF1} to obtain the expression for the CDF $F_{1,N}$ in \eqref{e:F1N}. This is precisely the aim of the next section.

\section{Explicit construction of the skew-orthogonal polynomials}
\label{sec:SOPs}

The averages \eqref{e:GenPF4} and \eqref{e:GenPF1} in Section \ref{sec:PartFns} above contain integrals over the respective skew-orthogonal polynomials $Q_j$ and $R_j$. The major advantage of using these polynomials can be seen if we first consider the case of $\beta=2$, from the expression \eqref{e:GenPF2}: we see that when $a(x)= 1$ the matrix in the determinant becomes $\left[ (p_j, p_k)_2^y \right]_{j,k=0, \dots, N-1}$, and so the determinant will be simply calculated if the polynomials $p_j$ are orthogonal with respect to the inner product \eqref{d:IP2} since the resulting matrix is diagonal. Indeed, this was the approach taken in \cite{NadaMaju2011,PerrSche2014}, where the orthogonal polynomials are the NM polynomials, which obey the relations \eqref{e:NM1}--\eqref{e:NM3}.

We will use the same approach for the $\beta=4$ and $\beta=1$ cases; that is we will construct the polynomials $Q_j$ and $R_j$ such that the matrices in \eqref{e:GenPF4} and \eqref{e:GenPF1} are of \textit{skew-diagonal form}\footnote{The term \textit{skew-diagonal} is used here in analogy with the term \textit{diagonal}, that is, the (non-trivial) skew-symmetric\newline ($\bM=-\bM^T$) analogue of a diagonal matrix.}
\begin{align}
\label{d:sdiag} \bS= \begin{bmatrix}
0&s_1& 0 &0 & \cdots& 0& 0\\
-s_1& 0& 0& 0 & \cdots& 0& 0\\
0& 0& 0& s_2& \cdots& 0& 0\\
0& 0& -s_2 & 0 & & \vdots& \vdots\\
\vdots& \vdots& \vdots& &\ddots\\
0& 0& 0& \cdots &&0& s_N\\
0& 0& 0& \cdots &&-s_N& 0
\end{bmatrix}.
\end{align}
The only non-zero elements of $\bS$ are in $2\times 2$ blocks $\begin{bmatrix} 0& s_j\\ -s_j &0 \end{bmatrix}$ on the diagonal, and we then have the simple result
\begin{align}
\label{e:Pfsdiag} \Pf \bS = \prod_{j=1}^N s_j.
\end{align}
In other words, we are looking for two sets of monic polynomials $\{Q_j\}, \{ R_j\}$ that satisfy the conditions in \eqref{d:4sorthog} and \eqref{d:1sorthog} respectively. Such polynomials are called \textit{skew-orthogonal polynomials}. Recall that these polynomials are only unique up to the transformation \eqref{e:oddsymm}, where $\eta_j= Q_j$ ($\beta=4$) and $\eta_j=R_j$ ($\beta=1$).

As discussed in Introduction we can, in principle, construct the polynomials iteratively using the conditions \eqref{d:4sorthog} and \eqref{d:1sorthog}, but this technique does not yield expressions that are amenable to asymptotic analysis. Nor is there a known closed form or recursive expression for these polynomials. Rather, we apply the method of \cite{AdlevanMoer2002,AdleForrNagavanMoer2000} where we aim to express our skew-orthogonal polynomials in the basis of the NM polynomials $\{ p_j \}$ from the analogous $\beta=2$ problem
\begin{align}
\label{e:SOPvsOP} \eta_j= p_j + \alpha_{j, j-1} p_{j-1} + \dots \alpha_{j, 0} p_0, \qquad \mbox{($\eta_j= Q_j$ for $\beta=4$, $\eta_j=R_j$ for $\beta=1$)}.
\end{align}
If we can find the coefficients $\alpha_{j,k}$ in \eqref{e:SOPvsOP} then we can use the properties of the polynomials $\{ p_j \}$ to obtain asymptotic results for the problems that we consider here. Note that \eqref{e:oddsymm} implies that we have freedom in the choice of the $\alpha_{2j+1, 2j}$ (that is, the second term in the odd-degree polynomials) in equation \eqref{e:SOPvsOP}, and we will typically choose $\alpha_{2j+1, 2j} =0$. [We will see that this choice is quite natural once we have the general formula for the coefficients, see \eqref{e:beta0}.] We find that the coefficients and the polynomial normalizations $q_j(y)$ and $r_j(y)$ are given as ratios of Pfaffians. To contrast this with the classical Gaussian case (in the limit $y\to \infty$) we have included the skew-orthogonal polynomials for $\beta=4$ in Appendix \ref{a:limpolys4} and for $\beta=1$ in Appendix \ref{a:limpolys1}.

The key to the method of \cite{AdleForrNagavanMoer2000} is an operator $A$, which acts thusly
\begin{align}
\label{d:Aop} A f [x] = e^{x^2/2} \frac{d}{dx} \left( e^{-x^2/2} f(x) \right).
\end{align}
We will also need the inverse operator
\begin{align}
\nonumber A^{-1} f[x] &= \frac{e^{x^2/2}} {2} \int_{-\infty}^{\infty} \sgn(x-z) e^{-z^2/2} f(z) dz\\
\label{d:Ainvop} &= \frac{e^{x^2/2}} {2} \left( \int_{-\infty}^{x} e^{-z^2/2} f(z) dz -  \int_{x}^{\infty} e^{-z^2/2} f(z) dz \right) \;.
\end{align}
That this is the inverse can be checked by explicitly calculating $A A^{-1} f[x]$ and $A^{-1}A f[x]$, and using the identity $\frac{d}{dx} \sgn (x-z) = 2 \delta (x-z)$ (where care is taken to use the \textit{distributional derivative}). The use of these operators will allow us to find relations between the $\beta=2$ inner product \eqref{d:IP2} and the $\beta=4,1$ skew-inner products \eqref{d:IP4} and \eqref{d:IP1}, and then to find the sought relations between the polynomials themselves.

Before proceeding, we point out for the interested reader that the original motivation for developing the technique in \cite{AdlevanMoer2002} was to relate the $\tau$-function solutions of the Toda lattice equations and the so-called ``Pfaff $\tau$-function'' solutions of the related Pfaff lattice. The Toda $\tau$-functions are matrix integrals that have determinantal expressions, but they also define polynomials that diagonalize the related matrix of inner products, which is essentially the matrix in \eqref{e:GenPF2}. One can then analogously define the Pfaff lattice, which has solutions given by Pfaff $\tau$-functions, which can be expressed in terms of Pfaffians (instead of determinants). Further, the polynomials defined by these Pfaff $\tau$-functions skew-diagonalize matrices of skew-inner products like \eqref{e:GenPF4} and \eqref{e:GenPF1}. The conversion between the Toda lattice equations and the Pfaff lattice equations is essentially the expression of the new (Pfaff) polynomials in the basis of the original (Toda) polynomials, and applying the matrix operations in \eqref{e:xwxt} and \eqref{e:xvxt}. We refer the reader to Ref.~\cite{AdlevanMoer2002} (and references therein) for more details. 

\subsection{$\beta=4$}
\label{s:SOPs4}

Recall that the goal is to find the coefficients $\alpha_{j,k}$ in \eqref{e:SOPvsOP} so that we can express the $Q_j$ in the basis of the $p_j$, which are the $\beta=2$ orthogonal polynomials [orthogonal with respect to the inner product \eqref{d:IP2}], and we will use the operator $A$ from \eqref{d:Aop}. This operator will allow us to develop both the $\beta=1$ and $\beta=4$ cases in the same framework, however we will need to define a slightly modified $\beta=4$ skew inner product, the skew-orthogonal polynomials of which are related to the $Q_j$ by a simple rescaling. The modified $\beta=4$ skew inner product is defined as
\begin{align}
\nonumber \llangle f , g \rrangle_{4}^y &:= \frac{1}{2} \int_{- \infty}^{y} dx \; e^{-x^2} \left[ f(x) g'(x)- g(x) f'(x) \right]\\
\label{d:modIP4}&= \frac{1}{2} \int_{- \infty}^{y} dx \; e^{-x^2/2} \left[ f(x) \frac{d} {dx} \left( e^{-x^2/2} g(x)\right) - g(x) \frac{d} {dx} \left( e^{-x^2/2} f(x) \right) \right],
\end{align}
which is the same as \eqref{d:IP4}, except that we have replaced $e^{-x^2} \mapsto e^{-x^2/2}$. We also define the associated monic skew-orthogonal polynomials $\{\tilde{Q}_j \}_{j=0,1, \dots}$ and normalizations $\{\tilde{q}_j \}_{j=0,1, \dots}$:
\begin{align}
\label{d:4modsorthoga} \llangle \tilde{Q}_{2j} , \tilde{Q}_{2k} \rrangle_{4}^y&= \llangle \tilde{Q}_{2j+1} , \tilde{Q}_{2k+1} \rrangle_{4}^y= 0,\\
\label{d:4modsorthogb} \llangle \tilde{Q}_{2j} , \tilde{Q}_{2k+1} \rrangle_{4}^y&= -\llangle \tilde{Q}_{2k+1} , \tilde{Q}_{2j} \rrangle_{4}^y = \tilde{q}_j(y) \delta_{j,k}.
\end{align}
Note that the use of the tilde $\tilde{~}$ here and elsewhere in this paper (which matches the notation in \cite{Forrester2010}) denotes that the quantity is related to this modified $\beta=4$ skew-inner product \eqref{d:modIP4}, rather than the standard skew-inner product \eqref{d:IP4}.

By performing a change of variables we have the relations
\begin{align}
\label{e:Rescaleqj} q_j (y) &= 2^{ -2j -\frac{1}{2}} \tilde{q}_j \left( \sqrt{2} y \right),\\
\label{e:RescaleQj} Q_k (x, y) &= 2^{-k /2} \tilde{Q}_k \left( \sqrt{2} x, \sqrt{2} y \right),
\end{align}
and so we can recover the polynomials that we are searching for. (Note that the factor of $2^{-k/2}$ ensures that the polynomials remain monic.) We can check these relations by generating the first few polynomials $\tQ_j$, as done for the $Q_j$ in Appendix \ref{s:iterative},
\begin{align}
\tilde{Q}_0 (\lambda, y)& = 1, \qquad \tilde{Q}_1 (\lambda, y) = \lambda, \qquad \tilde{Q}_2 (\lambda, y)= \lambda^2 +\tilde{b} \lambda + \frac{1 -y \tilde{b}} {2},\\
\tilde{Q}_3 (\lambda, y)&= \lambda^3 - 3 \frac{1- y \tilde{b}} {2} \lambda -\tilde{b} (1+ y^2)
\end{align}
[where we used \eqref{e:oddsymm} for $\tilde{Q}_3 (\lambda)$] and the corresponding normalizations 
\begin{align}
\tilde{q}_0 (y)&:= \llangle \tilde{Q}_0 , \tilde{Q}_1 \rrangle_{4}^y = \frac{\sqrt{\pi}}{4 } \;\erfc (-y) = \frac{e^{-y^2}} {2 \tilde{b}},\\
\tilde{q}_1 (y)&:= \llangle \tilde{Q}_2 , \tilde{Q}_3 \rrangle_{4}^y = \frac{1} {8} \left( 3 \sqrt{\pi}\, \erfc (- y) - e^{-y^2} y (9 +2 y^2) - e^{-y^2} ( 4 +y^2) \tilde{b} \right),
\end{align}
with the parameter $\tilde b$ given by
\begin{align}
\tilde{b}= \frac{2 e^{-y^2}} {\sqrt{\pi} (1+\erf ( y))}= \frac{2 e^{-y^2}} {\sqrt{\pi} \, \erfc (- y)}.
\end{align}

To use this modified skew-inner product we will introduce the operator $A$ from \eqref{d:Aop} into the $\beta=2$ inner product \eqref{d:IP2} and we have the properties (by integrating by parts)
\begin{align}
\label{e:fAg} (f, Ag)_{2}^y &= - (g, A f)_{2}^y + \Omega (f, g; y),\\
\label{e:fAf} (f, Af)_{2}^y &= \frac{\Omega (f, f; y)} {2},
\end{align}
where
\begin{align}
\label{d:Omega} \Omega (f, g; y) &:= \Big[ e^{-x^2} f(x) g(x) \Big]_{-\infty}^y = \lim_{x\to y} \Big( e^{-x^2} f(x) g(x) \Big) - \lim_{x\to -\infty} \Big( e^{-x^2} f(x) g(x) \Big).
\end{align}
Then we can write
\begin{align}
\nonumber \llangle f , g \rrangle_{4}^y &= \frac{1}{2} \Big( (f, Ag)_{2}^y - (g, Af)_{2}^y \Big)\\
\label{e:IP4vsIP2} &= (f, Ag)_{2}^y - \frac{\Omega (f, g; y)} {2}.
\end{align}

So we are searching for coefficients $\tilde{\alpha}_{j,k}$
\begin{align}
\label{e:4SOPvsOP} \tilde{Q}_{j} = p_j + \tilde{\alpha}_{j,j-1} \; p_{j-1} + \dots + \tilde{\alpha}_{j,1} \; p_1 + \tilde{\alpha}_{j,0}
\end{align}
such that the relations \eqref{d:4modsorthoga}--\eqref{d:4modsorthogb} hold, and we will use \eqref{e:IP4vsIP2}--\eqref{e:4SOPvsOP} to recast the problem in terms of the $\beta=2$ inner product and associated polynomials. Note that the coefficients $\talpha_{j,k}$ depend on $y$. Here we only present the results, with the detailed derivations in Appendix \ref{a:beta4}.

Define the skew-symmetric matrix
\begin{align}
\nonumber \bW_m &= \left[
\begin{array}{cccccc}
0 & w_{0,1} & w_{0,2} & w_{0,3}  & \cdots& w_{0,m}\\
-w_{0,1} & 0 & w_{1,2} & w_{1,3}  & \cdots& w_{1,m}\\
-w_{0,2} & -w_{1,2} & 0 & w_{2,3}  & \cdots& w_{2,m}\\
-w_{0,3} & -w_{1,3} & -w_{2,3} & 0  & & \\
\vdots & \vdots & \vdots & & \ddots & \\
-w_{0,m} & -w_{1,m} & -w_{2,m} &   & & 0
\end{array}
 \right]\\
\label{d:Wmat}&= \left[
\begin{array}{cccccc}
0 & h_{1} + \frac{\Omega_{0,1}}{2} & \frac{\Omega_{0,2}}{2} & \frac{\Omega_{0,3}}{2} & \cdots & \frac{\Omega_{0,m}}{2}\\
-h_{1} - \frac{\Omega_{0,1}}{2} & 0 & h_2 + \frac{\Omega_{1,2}}{2} & \frac{\Omega_{1,3}}{2} & \cdots & \frac{\Omega_{1,m}}{2}\\
-\frac{\Omega_{0,2}}{2} & -h_2 - \frac{\Omega_{1,2}}{2} & 0 & h_3 + \frac{\Omega_{2,3}}{2} & \cdots & \frac{\Omega_{2,m}}{2}\\
-\frac{\Omega_{0,3}}{2} & -\frac{\Omega_{1,3}}{2} & -h_3 - \frac{\Omega_{2,3}}{2} & 0 && \\
\vdots & \vdots & \vdots &  & \ddots & \\
-\frac{\Omega_{0,m}}{2} & -\frac{\Omega_{1,m}}{2} & -\frac{\Omega_{2,m}}{2} &  &  & 0 
\end{array}
 \right],
\end{align}
where
\begin{align}
w_{j,k}:= \delta_{j+1,k} h_k + \frac{1}{2}\Omega_{j,k}, \qquad \Omega_{j,k} := \Omega (p_j, p_k; y) = e^{-y^2} p_j (y, y) p_k (y, y).
\end{align}
Note that it will turn out that $\bW_m= \left[ \llangle p_j, p_k \rrangle_4^y \right]_{j,k=0, \dots, m}$, and so we can use the fact that $\tQ_j$ is a linear combination of the $p_j$'s to obtain information about the coefficients $\talpha_{j,k}$ in \eqref{e:4SOPvsOP}. Then we use a result of Knuth on overlapping Pfaffians \cite[(5.0)--(5.1)]{Knuth1996} (see also Appendix \ref{a:beta4} for more details) to obtain the following.
\begin{proposition}\label{p:alpha4}
Assuming
\begin{align}
\label{e:alpha04} \tilde{\alpha}_{j,j-1} (y) &=0, \qquad \mbox{$j$ odd},
\end{align}
then for $j\geq 2$
\begin{align}
\label{e:alphas4} \tilde{\alpha}_{j,k} (y)&= \left\{ \begin{array}{ll}
-\frac{{\displaystyle\Pf \bW_{j-1}^{(k\mapsto j)}}} {{\displaystyle\Pf \bW_{j-1}}}, &\quad  \mbox{$j$ even, $k\leq j-1$},\\
\\
-\frac{{\displaystyle\Pf \bW_{j-2}^{(k\mapsto j)}}} {{\displaystyle\Pf \bW_{j-2}}}, &\quad \mbox{$j$ odd, $k\leq j-2$},
\end{array}\right.
\end{align}
where $\bW_{m}^{(\eta \mapsto \nu)}$ is the matrix $\bW_{m}$ from \eqref{d:Wmat} with all occurrences of the index $\eta$ replaced by the index $\nu$, and
\begin{align}
\label{e:alphajj} \tilde{\alpha}_{j,j} (y) &=1, \qquad \forall j.
\end{align}

The normalizations are
\begin{align}
\label{e:qs} \tilde{q}_j (y)= \frac{\Pf \bW_{2j+1}} {\Pf \bW_{2j-1}},
\end{align}
with the convention
\begin{align}
\label{e:PfWm0} \Pf \bW_{-1} = 1.
\end{align}
\end{proposition}
The proofs of these results are contained in Appendix \ref{a:beta4}.

With Proposition \ref{p:alpha4}, we can obtain the skew-orthogonal polynomials for the skew-inner product \eqref{d:IP4} via \eqref{e:RescaleQj}
{\small \begin{align}
\label{e:Qpolys} Q_j (\lambda, y) = 2^{-k/2} \Big[ p_j ( \sqrt{2} \lambda, \sqrt{2} y ) + \talpha_{j, j-1} ( \sqrt{2} y )\, p_{j-1} ( \sqrt{2} \lambda, \sqrt{2} y ) + \dots + \talpha_{j,1} ( \sqrt{2} y )\, p_1 ( \sqrt{2} \lambda, \sqrt{2} y ) + \talpha_{j,0} ( \sqrt{2} y )\, \Big],
\end{align}}where the $p_j$ are the NM polynomials, and the normalizations are obtained from \eqref{e:qs} via \eqref{e:Rescaleqj}
\begin{align}
\label{e:Pfm2} q_j (y) = 2^{-2j- \frac{1}{2}} \frac{\Pf \bW_{2j+1}} {\Pf \bW_{2j-1}} \bigg|_{y\mapsto \sqrt{2} y}.
\end{align}
From \eqref{e:FZ} we know that the CDF of the largest eigenvalue is expressed in terms of the average \eqref{e:GenPF}, with the function $a(x) =1$. Also, with $a (x) =1$ we have $\gamma_{j,k}^{(4)}(1) = \langle Q_j, Q_k \rangle_{4}^y$. So using the skew-orthogonal polynomials \eqref{e:Qpolys}, the relations \eqref{d:4sorthog} tell us that the matrix in $\hat{Z}_{4,N} [1, y]$ is of the form \eqref{d:sdiag}, and so its Pfaffian is given by \eqref{e:Pfsdiag} with $s_j = q_j$. Thus, substitution of the normalization \eqref{e:Pfm2} into \eqref{e:GenPF4}, with $a(x)=1$, yields the result in \eqref{e:F4N}.

The expression for $\tilde{\alpha}_{j,k}$ can be seen to recover the classical Gaussian case (with $y\to \infty$), since in this limit the polynomials $p_j$ are the Hermite polynomials [from \eqref{e:limHerms}] and also that $\Omega_{j,k}=0$, so the matrix \eqref{d:Wmat} is then the same as that in \cite[Prop 6.2.1]{Forrester2010}. The derivation of the $\alpha_{j,k}$ in \eqref{e:LimPolys4} then proceeds identically.

\subsection{$\beta=1$, $N$ even}

As above, we want to express the skew-orthogonal polynomials $\{ R_j \}$ from \eqref{d:1sorthog} in terms of the polynomials $\{ p_j \}$ from \eqref{d:2orthog}. So we look for coefficients $\alpha_{j,k}$ such that
\begin{align}
\label{e:Rjalpha1} R_j = p_j + \alpha_{j, j-1} p_{j-1} + \dots + \alpha_{j, 1} p_1 + p_0,
\end{align}
and again these coefficients will depend on $y$. To make further progress, we use the operator $A^{-1}$ from \eqref{d:Ainvop}. First we note from \eqref{e:fAg} that
\begin{align}
\nonumber (f, A^{-1}g) = (A A^{-1}f, A^{-1}g) &= -(A^{-1}f, g) + \Omega(A^{-1}f, A^{-1} g; y)\\
\label{e:AinvSymm} &= -(g, A^{-1}f) - \Phi(f, g) - \Phi (g,f),
\end{align}
where
\begin{align}
\label{d:Phifg} \Phi(f, g)&:= \frac{1}{2} \int_{-\infty}^{y} e^{-z^2/2} f(z) dz \int_{y}^{\infty} e^{-z^2/2} g(z) dz.
\end{align}
Now we can re-write the skew-inner product \eqref{d:IP1} as
\begin{align}
\nonumber \langle f, g \rangle_{1}^y &= - \frac{1}{2} \Big( (f, A^{-1}g )_{2}^y - (g, A^{-1}f )_{2}^y + \Phi(f,g) - \Phi (g,f) \Big)\\
\label{e:IP1vsIP2} &= -(f, A^{-1}g)_{2}^y - \Phi(f, g).
\end{align}
From here we follow the same procedure as for $\beta=4$, but replacing the matrix $\bW_m$ in \eqref{d:Wmat} with the more complicated matrix
\begin{align}
\nonumber \bV_m &= \left[
\begin{array}{cccccc}
0 & v_{0,1} & v_{0,2} & v_{0,3}  & \cdots& v_{0,m}\\
-v_{0,1} & 0 & v_{1,2} & v_{1,3}  & \cdots& v_{1,m}\\
-v_{0,2} & -v_{1,2} & 0 & v_{2,3}  & \cdots& v_{2,m}\\
-v_{0,3} & -v_{1,3} & -v_{2,3} & 0  & & \\
\vdots & \vdots & \vdots & & \ddots & \\
-v_{0,m} & -v_{1,m} & -v_{2,m} &   & & 0
\end{array}
 \right]\\
\label{d:Vm} &= \begin{bmatrix}
0 & h_0 - X_{0,1} -\Phi_{0,1} & X_{2,0} + \Phi_{2,0} & X_{3,0} + \Phi_{3,0}& \dots \\
-h_0 +X_{0,1} +\Phi_{0,1} & 0 & h_1 - X_{1,2} -\Phi_{1,2} & X_{3,1} + \Phi_{3,1}& \dots \\
-X_{2,0} - \Phi_{2,0} & -h_1 +X_{1,2} +\Phi_{1,2} & 0 & h_2 - X_{2,3} -\Phi_{2,3}& \dots \\
-X_{3,0} - \Phi_{3,0} & -X_{3,1}- \Phi_{3,1} & -h_2 +X_{2,3} +\Phi_{2,3} & 0\\
\vdots & \vdots & \vdots 
\end{bmatrix},
\end{align}
where
\begin{align}
\label{d:Phijk} \Phi_{j,k} &:= \Phi(p_j, p_k) = \frac{1}{2} \int_{-\infty}^{y} e^{-z^2/2} p_j (z, y) dz \int_{y}^{\infty} e^{-z^2/2} p_k (z, y) dz,\\
\label{d:Xjk} X_{j,k} &:= \frac{1}{2} \left( \int_{-\infty}^y p_j (x, y) e^{-x^2/2}\; \erf\left( \frac{x}{\sqrt{2}} \right) dx \right) \int_{-\infty}^{\infty} e^{-z^2/2} p_k (z, y) dz.
\end{align}
Note that we have the equality
\begin{align}
\bV_m= \Big[ \langle p_j, p_k \rangle_{1}^y \Big]_{j,k=0, \dots, m}= -\Big[ (p_j, A^{-1} p_k)_{2}^y + \Phi(p_j, p_k) \Big]_{j,k=0, \dots, m}.
\end{align}
We now give expressions for the coefficients $\alpha_{j,k} (y)$ and normalizations $r_j(y)$ in terms of the matrix $\bV_m$. (We discuss the construction of the matrix in Appendix \ref{a:beta1}.)

\begin{proposition}\label{p:alpha1}
Assuming
\begin{align}
\label{e:alpha01} \alpha_{j,j-1} (y) &=0, \qquad \mbox{$j$ odd},
\end{align}
then for $j\geq 2$
\begin{align}
\label{e:alphas1b} \alpha_{j,k} (y)&= \left\{ \begin{array}{ll}
-\frac{{\displaystyle\Pf \bV_{j-1}^{(k\mapsto j)}}} {{\displaystyle\Pf \bV_{j-1}}}, &\quad \mbox{$j$ even, $k\leq j-1$},\\
\\
-\frac{{\displaystyle\Pf \bV_{j-2}^{(k\mapsto j)}}} {{\displaystyle\Pf \bV_{j-2}}}, &\quad \mbox{$j$ odd, $k\leq j-2$},
\end{array}\right.
\end{align}
where $\bV_{m}^{(\eta \mapsto \nu)}$ is the matrix $\bV_{m}$ with all occurrences of the index $\eta$ replaced by the index $\nu$, and
\begin{align}
\label{e:alphajj1} \alpha_{j,j} (y) &=1, \qquad \forall j.
\end{align}

The normalizations are
\begin{align}
\label{e:rs} r_j (y)= \frac{\Pf \bV_{2j+1}} {\Pf \bV_{2j-1}}
\end{align}
with the convention
\begin{align}
\label{e:PfVm0} \Pf \bV_{-1} = 1.
\end{align}
\end{proposition}
Note that the coefficients $\alpha_{j,k}$ depend on $y$. We give the proof of Proposition \ref{p:alpha1} in Appendix \ref{a:beta1}.

The coherence of the $\alpha_{j,k}$ in Proposition \ref{p:alpha1} with the $y\to \infty$ classical Gaussian result in \eqref{e:limHerms1} is not as straightforward as in the $\beta=4$ case above, and we go through the details in Appendix \ref{a:limpolys1}. The extra complications are  because the technique of \cite{AdleForrNagavanMoer2000} did not use an exact analogue of our matrix $\bV_m$ in \eqref{d:Vm}; they instead used some shrewd linear algebra to express the matrix $[(p_j, A^{-1} p_k)]$ in terms of the matrix $[(p_j, A p_k)]$ and some other matrices containing the polynomial normalizations $h_j$. This approach worked as it relied on inverting matrices that are (almost) diagonal, however the analogous step in our case (with finite $y$) involves inverting a full $N \times N$ matrix, and so it is infeasible here. At any rate, setting $y = \infty$, we see from \eqref{d:Phijk} that the matrix $\Phi (p_j, p_k) =0$, and from \eqref{e:limHerms} that the polynomials $p_j$ become the Hermite polynomials leading us to the simple expression \eqref{e:XjkLim} for the elements of $X_{j,k}$. Using these facts we recover the classical case in \eqref{e:limHerms1}, where $\alpha_{2j+1, 2j-1} = -j$ and is zero otherwise --- see Appendix \ref{a:limpolys1} for the details.

As with the $\beta=4$ case above, by use of the polynomials $R_j$ the matrix in $\hat{Z}_{1,N} [1,y]$ of \eqref{e:GenPF1} has the skew-diagonal structure in \eqref{d:sdiag} and so its Pfaffian is given by \eqref{e:Pfsdiag} with $s_j= r_j$. Substitution of \eqref{e:rs} into \eqref{e:GenPF1} gives the expression for the CDF of the largest eigenvalue in \eqref{e:F1N}.

\subsection{Skew-orthogonal polynomials for more general weight functions}

As mentioned after Eq. \eqref{e:GenPF}, the density functions in this paper are of the form \eqref{eq:janossy1}, which are a type of Janossy density, so a natural question is to ask if our methods can be applied more generally. We see from \eqref{e:IP4vsIP2} and \eqref{e:IP1vsIP2} that the key step involved in calculating the polynomial coefficients in Proposition \ref{p:alpha4} (for the GSE) and Proposition \ref{p:alpha1} (for the GOE) is writing the corresponding skew-inner product in terms of the GUE inner product. The quantity separating the procedure here from the classical case in \cite{AdleForrNagavanMoer2000} is $\Omega$ in \eqref{d:Omega}, which [via \eqref{e:AinvSymm}] also determines the quantity $\Phi$ in \eqref{d:Phifg}. This $\Omega$ function is particular to the Gaussian weight and the eigenvalue domain $(-\infty, y)$, however, from following the matrix manipulations in Appendices \ref{a:beta4} and \ref{a:beta1}, we can conclude that our method will work for Janossy densities over more general domains and for the other classical weight functions. Indeed, let
\begin{align}
w_{\beta} (x):= \left\{ \begin{array}{cc}
e^{-\beta x^2/2},& \mbox{(Gaussian)},\\
x^{a \beta/2 } e^{-\beta x/2},& \mbox{(Laguerre)},\\
(1-x)^{a \beta/2} (1+x)^{b \beta/2},& \mbox{(Jacobi)},\\
( 1+ x^2)^{-\beta (N-1)/2 +1},& \mbox{(Cauchy)},
\end{array}\right.
\end{align}
and define the inner product
\begin{align}
(f,g)_2^{Y} := \int_Y w_2 (x) f(x) g(x) dx
\end{align}
and the skew-inner products
\begin{align}
\label{f:GenIP4} \langle f, g \rangle_4^Y&:= \frac{1}{2} \int_Y w_4 (x) \left[ f(x) g'(x)- g(x) f'(x) \right] dx,\\
\label{f:GenIP1} \langle f , g \rangle_{1}^Y &:= \frac{1}{2} \int_Y w_1 (x) f(x) \int_Y  w_1 (z) g (z) \sgn (z- x)\, dz\, dx,
\end{align}
where $Y\subset U_w$, with $U_w$ the maximal domain for the weight function $w_\beta$. Then we can define a new $\Omega$ and apply the procedures in Appendices \ref{a:beta4} and \ref{a:beta1} to obtain the skew-orthogonal polynomials in terms of the orthogonal polynomials. (Of course, explicitly calculating the $\beta=1$ and $\beta=4$ polynomials using this method relies on knowing the orthogonal polynomials for the corresponding $\beta=2$ problem; a non-trivial hurdle.) Note that in the case of Hermitian matrix models (i.e. with $\beta = 2$), such orthogonal polynomials have been studied in the context of the counting statistics of eigenvalues in these ensembles (see e.g. \cite{cao2014continuous,witte2012variance}).

For the purposes of illustration assume $Y:= (y_1, y_2) \subset \mathbb{R}$, then we replace \eqref{d:Omega} and \eqref{d:Phifg} by
\begin{align}
\Omega (f, g; Y) &:= \lim_{x\to y_2} \Big( w_2(x) f(x) g(x) \Big) - \lim_{x\to y_1} \Big( w_2(x) f(x) g(x) \Big)\\
\Phi (f,g; Y)&:= \frac{1}{2} \int_{y_1}^{y_2} w_1(z) f(z) dz \left( \int_{y_2}^{\infty} w_1(z) g(z) dz - \int_{-\infty}^{y_1} w_1(z) g(z) dz \right),
\end{align}
and then Propositions \ref{p:alpha4} and \ref{p:alpha1} hold, with the matrices $\bW_m$ and $\bV_m$ modified accordingly.

\section{Asymptotic analysis of the CDF of the largest eigenvalue for $F_{4,N}$ for large $N$}
\label{s:F4asympt}

In this section we show that our formula for the CDF $F_{4,N} (y)$ is amenable to an asymptotic analysis, in the large $N$ limit, which allows us to obtain an alternative derivation of the Tracy-Widom formula for $\beta = 4$ \cite{TracWido1996}. Indeed, we will show below that, from the expression in (\ref{e:F4N}), we can obtain
\begin{eqnarray}\label{e:TW4}
\lim_{N \to \infty} F_{4,N} \left( y = \sqrt{2N} + \frac{s}{2^{7/6}}N^{-1/6} \right) = \exp{\left(-\frac{1}{2} \int_s^\infty (x-s) q^2(x) dx \right)} \cosh{\left(\frac{1}{2} \int_s^\infty q(x)dx \right)},
\end{eqnarray}
where $q(x)$ is the Hastings-McLeod solution of the Painlev\'e II equation as in \eqref{e:PII_intro}.

%

To show this result (\ref{e:TW4}), starting from our expression in (\ref{e:F4N}), we will first provide an explicit expression for $\Pf \bW_{2N-1}$, where the matrix $\bW_{m}$ is defined in (\ref{d:Wmat}). It is convenient first to define
\begin{eqnarray}\label{d:M1}
M_{i_1, i_2}(y) = \frac{1}{2}\dfrac{\prod_{m={i_1}}^{i_2-1} h_{2m+2} (y)}{\prod_{m=i_1}^{i_2} h_{2m+1} (y)} p_{2i_1}(y, y) p_{2i_2+1}(y, y) e^{-y^2} \;, \;\;\;\; i_2 \geq i_1 \;,
\end{eqnarray}
where we recall that the $p_k$'s are the NM orthogonal polynomials (\ref{e:NM1})--(\ref{e:NM3}) while the $h_k$'s are their corresponding norms (\ref{e:ynorm2}). (We will often suppress the explicit dependence on $y$ for concision.) In view of the asymptotic analysis, it is useful to rewrite \eqref{d:M1} as
 \begin{eqnarray}\label{d:M2}
M_{i_1, i_2}(y) = \frac{1}{2} \frac{1}{\sqrt{\NMrs{R}_{2i_2+1} (y)}} \prod_{m=i_1}^{m=i_2-1} \sqrt{\frac{\NMrs{R}_{2m+2} (y)}{\NMrs{R}_{2m+1} (y)}} \, \psi_{2i_1}(y, y) \psi_{2i_2+1}(y, y) \;,  \;\;\;\; i_2 \geq i_1 \;
\end{eqnarray}
 in terms of $\NMrs{R}_m = h_m/h_{m-1}$ from \eqref{e:NM2} and the so-called ``wave functions'' $\psi_k(x, y)$ given by
 \begin{eqnarray}\label{d:psi}
 \psi_k(x, y) = \frac{p_k(x, y)}{\sqrt{h_k (y)}} e^{-\frac{x^2}{2}} \;.
 \end{eqnarray}
With these definitions we find the following convenient expression for the Pfaffian in \eqref{e:F4N}.

\begin{proposition}\label{lem:PfW}
With $M_{j,k}$ defined in \eqref{d:M1} and $\bW_m$ from \eqref{d:Wmat} we have (suppressing the explicit dependence on $y$)
\begin{align}
\label{e:expansion:Pf} \Pf \bW_{2N-1}(y) &= \left(\prod_{j=0}^{N-1} h_{2j+1} \right) \left(1 + \sum_{p=1}^N \; \sum_{\cI_{2p}} M_{i_1, i_2} M_{i_3,i_4} \cdots M_{i_{2p-1}, i_{2p}} \right) \;,
\end{align}
where we have used the notation
\begin{align}
\label{d:I2p} \cI_{2p}: 0\leq i_1 \leq i_2 < i_3 \leq i_4 < i_5 \leq i_6 < \dots <i_{2p-1}\leq i_{2p} \leq N-1
\end{align}
for the terminals on the second sum. (Note that the indices in the sum obey both strict and non-strict inequalities in the sequence $i_{2j-1}\leq i_{2j} < i_{2j+1} \leq i_{2j+2}$.)
\end{proposition}

\proof First we define
\begin{align}
\label{e:sigmaj} \sigma_{j} := \frac{p_{j} (y, y)}{\sqrt{2}} e^{-y^2/2}
\end{align}
so that
\begin{align}
\label{d:welts} w_{j,k} = h_k \delta_{j+1,k} + \sigma_j \sigma_k.
\end{align}

Here we use the expression for the Pfaffian in \eqref{e:PfPerfMatch}, where the sum is over all perfect matchings on $2N$ sites $\{ 0, 1, \dots, 2N-1\}$ and so
\begin{align}
\label{e:PfWPMs} \Pf \bW_{2N-1}(y) = \sum_{\mu\in M_{2N}} \varepsilon (\mu)\; w_{i_1, j_1} w_{i_2, j_2}\cdot\cdot\cdot w_{i_N, j_{N}},
\end{align}
where we recall that the perfect matchings $\mu\in M_{2N}$ are represented by link diagrams as in Figure \ref{f:PMeg}.

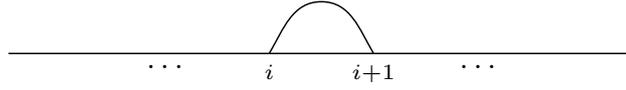
\begin{figure}
\begin{center}
\resizebox{0.5\textwidth}{!}{\begin{tikzpicture}[baseline=(current  bounding  box.center)] %
\draw (-0.5,0)--(5.5,0); %
\node[below] at (1,0) {${\dots}$}; %
\node[below] at (2,0) {${\scriptstyle i}$}; %
\node[below] at (3,0) {${\scriptstyle i+1}$}; %
\node[below] at (4,0) {${\dots}$}; %
\draw[smooth] (2,0) to[out=60,in=180] (2.5,0.5) to[out=0,in=120] (3,0); %
\end{tikzpicture}}
\end{center}
\caption{\label{f:LLs}A ``little link'' from site $i$ to $i+1$ corresponding to the factor $h_{i+1} + \sigma_i \sigma_{i+1}$.}
\end{figure}

We see from \eqref{d:welts} that the summand will include a factor of $h_{i+1} + \sigma_i \sigma_{i+1}$ if and only if the link diagram of the perfect matching includes a ``little link'' from site $i$ to $i+1$ (see Figure \ref{f:LLs}), and otherwise every factor is of the form $\sigma_i \sigma_{j}$.
We denote by $M_{2N; i_j, \dots, i_k}$ the set of perfect matchings of $2N$ sites with ``little links'' $(i_j, i_j+1), \dots, (i_k, i_k +1)$ and no others. Then \eqref{e:PfWPMs} becomes
\begin{align}
\nonumber \Pf \bW_{2N-1}(y) &= \sum_{\mu\in M_{2N}} \varepsilon (\mu)\; \Big( \sigma_{0} \dots \sigma_{2N-1} \Big)\\
\nonumber &+ \sum_{i_1=0}^{2N-2} \sum_{\mu\in M_{2N; i_1}} \varepsilon (\mu)\; \Big( \sigma_{0} \dots \sigma_{i_1 -1} \Big) h_{i_1+1} \Big( \sigma_{i_1+2} \dots \sigma_{2N-1} \Big) \\
\nonumber &+ \sum_{i_1=0}^{2N-2} \sum_{i_2=i_1+2}^{2N-2} \sum_{\mu\in M_{2N; i_1, i_2}} \varepsilon (\mu)\; \Big( \sigma_{0} \dots \sigma_{i_1 -1} \Big) h_{i_1+1} \Big( \sigma_{i_1 +2} \dots \sigma_{i_2 -1}\Big) h_{i_2+1} \Big( \sigma_{i_2+2} \dots \sigma_{2N-1} \Big) \\
\nonumber &\vdots\\
\label{e:PfW1} &+ \sum_{i_1=0}^{2N-2} \sum_{i_2=i_1+2}^{2N-2} \dots \sum_{i_N=i_{N-1}+2}^{2N-2} \sum_{\mu\in M_{2N; i_1, \dots i_{N-1}}} \varepsilon (\mu)\; h_{i_1+1} \dots h_{i_{N-1}+1},
\end{align}
where we see that each $h_{j+1}$ replaces a pair $\sigma_j \sigma_{j+1}$ in the summand.

In \eqref{e:PfW1} all summands are now independent of the matching $\mu$, except for the factor of $\varepsilon(\mu)$, and so we factor these out and want to show that
\begin{align}
\label{e:SignSum} \sum_{\mu\in M_{2N; I}} \varepsilon (\mu) = 1
\end{align}
for any set of indices $I$. From \eqref{e:NumPMs} we have that there is an even number of perfect matchings, excluding the identity perfect matching $\{ (0,1), (2,3), \dots, (2N-2, 2N-1) \}$, which has ``little links'' at all sites. We can pair these non-identities in such a way that for each perfect matching with a sign of $(+1)$ there is a partner with sign $(-1)$, and so the sum in \eqref{e:SignSum} will have contribution of zero from these terms, leaving just the identity matching. We do this pairing according to the following algorithm.

Any link diagram of a non-identity matching will have at least one non-identity link [i.e.~a link not of the form ($2j$, $2j + 1$)], and will match one of the two forms in Figure \ref{f:LinkPairing1}, where $2j$ is the left site of the left-most non-identity link (and thus $2j+1$ is by necessity also part of a non-identity link). Every non-identity perfect matching $\mu$ of the form in Figure \ref{f:LinkPairing1}~(a) can be paired with a $\hat{\mu}$ of the form in Figure \ref{f:LinkPairing1}~(b), where the link patterns are identical except at the four sites $\{2j, 2j+1, 2s, 2t+1\}$. The extra crossing in $\hat{\mu}$ then implies that $\varepsilon( \mu) =-\varepsilon( \hat{\mu})$, so
\begin{align}
\sum_{\mu\in M_{2N;I}} \varepsilon(\mu) = \varepsilon(\text{identity}) = 1.
\end{align}
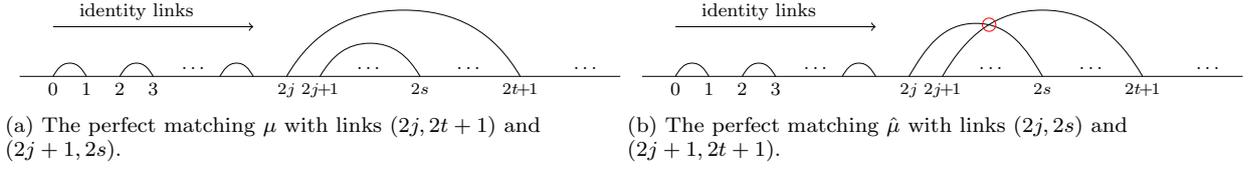
\begin{figure}
\begin{center}
\subfloat[The perfect matching $\mu$ with links $(2j, 2t+1)$ and {\protect\newline} $(2j+1, 2s)$.]{
\resizebox{0.49\textwidth}{!}{
\begin{tikzpicture}[baseline=(current  bounding  box.center)] %
\draw (-0.5,0)--(8.5,0); %
\node[below] at (0,0) {\small $0$}; %
\node[below] at (0.5,0) {\small $1$}; %
\node[below] at (1,0) {\small $2$}; %
\node[below] at (1.5,0) {\small $3$}; %
\node[above] at (2.125,0) {\dots}; %
\node[below] at (3.5,0) {\scriptsize $2j$}; %
\node[below] at (4,0) {\scriptsize $2j\!\!+\!\!1$}; %
\node[above] at (4.75,0) {\dots}; %
\node[below] at (5.5,0) {\scriptsize $2s$}; %
\node[above] at (6.25,0) {\dots}; %
\node[below] at (7,0) {\scriptsize $2t\!\!+\!\!1$}; %
\draw[smooth] (0,0) to[out=60,in=180] (0.25,0.2) to[out=0,in=120] (0.5,0); %
\node[above] at (8,0) {\dots}; %
\draw[smooth] (1,0) to[out=60,in=180] (1.25,0.2) to[out=0,in=120] (1.5,0); %
\draw[smooth] (2.5,0) to[out=60,in=180] (2.75,0.2) to[out=0,in=120] (3,0); %
\draw[smooth] (3.5,0) to[out=60,in=180] (5.25,1) to[out=0,in=120] (7,0); %
\draw[smooth] (4,0) to[out=60,in=180] (4.75,0.5) to[out=0,in=120] (5.5,0); %
\node[above] at (1.25,0.75) {\small identity links};
\draw[->] (0,0.75) -- (3,0.75);
\end{tikzpicture}
}}
\subfloat[The perfect matching $\hat{\mu}$ with links $(2j, 2s)$ and {\protect\newline} $(2j+1, 2t+1)$.]{
\resizebox{0.49\textwidth}{!}{
\begin{tikzpicture}[baseline=(current  bounding  box.center)] %
\draw (-0.5,0)--(8.5,0); %
\node[below] at (0,0) {\small $0$}; %
\node[below] at (0.5,0) {\small $1$}; %
\node[below] at (1,0) {\small $2$}; %
\node[below] at (1.5,0) {\small $3$}; %
\node[above] at (2.125,0) {\dots}; %
\node[below] at (3.5,0) {\scriptsize $2j$}; %
\node[below] at (4,0) {\scriptsize $2j\!\!+\!\!1$}; %
\node[above] at (4.75,0) {\dots}; %
\node[below] at (5.5,0) {\scriptsize $2s$}; %
\node[above] at (6.25,0) {\dots}; %
\node[below] at (7,0) {\scriptsize $2t\!\!+\!\!1$}; %
\draw[smooth] (0,0) to[out=60,in=180] (0.25,0.2) to[out=0,in=120] (0.5,0); %
\node[above] at (8,0) {\dots}; %
\draw[smooth] (1,0) to[out=60,in=180] (1.25,0.2) to[out=0,in=120] (1.5,0); %
\draw[smooth] (2.5,0) to[out=60,in=180] (2.75,0.2) to[out=0,in=120] (3,0); %
\draw[smooth] (3.5,0) to[out=60,in=180] (4.5,0.8) to[out=0,in=120] (5.5,0); %
\draw[smooth] (4,0) to[out=60,in=180] (5.5,1) to[out=0,in=120] (7,0); %
\draw[red] (4.7,0.78) circle (0.1cm);
\node[above] at (1.25,0.75) {\small identity links};
\draw[->] (0,0.75) -- (3,0.75);
\end{tikzpicture}
}}
\end{center}
\caption{\label{f:LinkPairing1}The two possible configurations of the first non-identity links (from the left), one with a crossing and the other without. We convert between them by switching the end points, picking up a factor of $(-1)$.}
\end{figure}

This leaves us with
\begin{align}
\nonumber \Pf \bW_{2N-1}(y) &= \Big( \sigma_{0} \dots \sigma_{2N-1} \Big)+ \sum_{i_1=0}^{2N-2} \; \Big( \sigma_{0} \dots \sigma_{i_1 -1} \Big) h_{i_1+1} \Big( \sigma_{i_1+2} \dots \sigma_{2N-1} \Big) \\
\nonumber &+ \sum_{i_1=0}^{2N-2} \sum_{i_2=i_1+2}^{2N-2} \; \Big( \sigma_{0} \dots \sigma_{i_1 -1} \Big) h_{i_1+1} \Big( \sigma_{i_1 +2} \dots \sigma_{i_2 -1}\Big) h_{i_2+1} \Big( \sigma_{i_2+2} \dots \sigma_{2N-1} \Big) \\
\nonumber &\vdots\\
\label{e:PfW2} &+ h_{1} h_{3} \dots h_{2N -1},
\end{align}
where we have replaced the bottom line in \eqref{e:PfW1} by the product over the odd indexed $h_j$, which is the only term in that sum (since there is only one way to replace all pairs of $\sigma_j \sigma_{j+1}$).

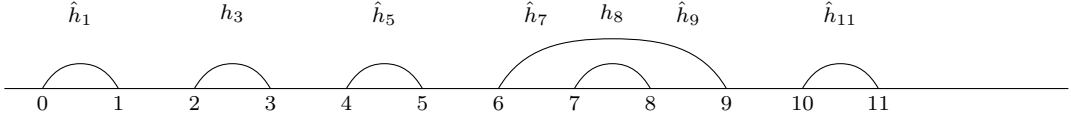
\begin{figure}
\begin{center}
\begin{tikzpicture}[baseline=(current  bounding  box.center)] %
\draw (-0.5,0)--(13.5,0); %
\node[below] at (0,0) {\small $0$}; %
\node at (0.5,1) {$\hat{h}_1$}; %
\node[below] at (1,0) {\small $1$}; %
\node[below] at (2,0) {\small $2$}; %
\node at (2.5,1) {$h_3$}; %
\node[below] at (3,0) {\small $3$}; %
\node[below] at (4,0) {\small $4$}; %
\node at (4.5,1) {$\hat{h}_5$}; %
\node[below] at (5,0) {\small $5$}; %
\node[below] at (6,0) {\small $6$}; %
\node at (6.5,1) {$\hat{h}_7$}; %
\node[below] at (7,0) {\small $7$}; %
\node at (7.5,1) {$h_8$}; %
\node[below] at (8,0) {\small $8$}; %
\node at (8.5,1) {$\hat{h}_9$}; %
\node[below] at (9,0) {\small $9$}; %
\node[below] at (10,0) {\small $10$}; %
\node at (10.5,1) {\small $\hat{h}_{11}$}; %
\node[below] at (11,0) {\small $11$}; %
\draw[smooth] (0,0) to[out=60,in=180] (0.5,0.33) to[out=0,in=120] (1,0); %
\draw[smooth] (2,0) to[out=60,in=180] (2.5,0.33) to[out=0,in=120] (3,0); %
\draw[smooth] (4,0) to[out=60,in=180] (4.5,0.33) to[out=0,in=120] (5,0); %
\draw[smooth] (6,0) to[out=60,in=180] (7.5,0.66) to[out=0,in=120] (9,0); %
\draw[smooth] (7,0) to[out=60,in=180] (7.5,0.33) to[out=0,in=120] (8,0); %
\draw[smooth] (10,0) to[out=60,in=180] (10.5,0.33) to[out=0,in=120] (11,0); %
\end{tikzpicture}
\end{center}
\caption{\label{f:LinkDiagEg1}The link diagram corresponding to the term in \eqref{e:PfTermEg}, showing which links contribute factors of $h$. The little links can contribute factors of $\sigma$ or $h$, while long links can only contribute factors of $\sigma$. The missing factors of $h_{2j+1}$ are denoted by a ``hat''.}
\end{figure}

We now just need to match up the expression in \eqref{e:PfW2} with products of $M_{i_1, i_2}$ from \eqref{d:M1} --- first we note
\begin{align}
\label{d:Msigma} M_{i_1, i_2}(y) = \sigma_{2i_1}(y)  \frac{h_{2i_1+2}}{h_{2i_1 +1}} \frac{h_{2i_1+4}} {h_{2i_1 +3}} \cdots \frac{h_{2i_2}} {h_{2i_2 -1}} \frac{\sigma_{2i_2+1}(y)} {h_{2i_2 +1}} \;, \;\;\;\; i_2 \geq i_1 \;.
\end{align}
In each term of \eqref{e:PfW2}, we start from the left with $\sigma_0$ and pair up each even $\sigma_{2j}$ with the nearest odd $\sigma_{2k+1}$ to its right ($k\geq j$). This will be easiest to see if we start with an example of one of the terms in \eqref{e:PfW2}, such as the term
\begin{align}
\label{e:PfTermEg} \Big( \sigma_{0} \dots \sigma_{i_1 -1} \Big) h_{i_1+1} \Big( \sigma_{i_1 +2} \dots \sigma_{i_2 -1}\Big) h_{i_2+1} \Big( \sigma_{i_2+2} \dots \sigma_{2N-1} \Big) = \sigma_0 \sigma_1 h_3 \sigma_4 \sigma_5 \sigma_6 h_8 \sigma_9 \sigma_{10} \sigma_{11}
\end{align}
with $N=6, i_1=2, i_2=7$, where the corresponding link diagram is drawn in Figure \ref{f:LinkDiagEg1}. In the diagram, we have included the labels of the $h_j$ which are present ($h_3$ and $h_8$) and, for convenience, the odd $h_j$ which are missing using a ``hat'' ($\hat{h}_1, \hat{h}_5, \hat{h}_7, \hat{h}_9$ and $\hat{h}_{11}$).
Dividing through by $h_1 h_3 \cdots h_{11}$ we obtain
\begin{align}
\label{e:sigmaMeg} \frac{\sigma_0 \sigma_1} {h_1} \frac{\sigma_4 \sigma_5} {h_5} \frac{\sigma_6 h_8 \sigma_9}{h_7 h_9} \frac{\sigma_{10} \sigma_{11}} {h_{11}}= M_{0,0} M_{2,2} M_{3,4} M_{5,5}
\end{align}
using \eqref{d:Msigma}. We see that all the possible combinations of $M_{i_1,i_2} M_{i_3,i_4} M_{i_5,i_6} M_{i_7,i_8}$, with $0\leq i_1\leq i_2< i_3\leq i_4< i_5\leq i_6< i_7\leq i_8 \leq 5$, will appear in the second line of \eqref{e:PfW2} --- each of the odd $h_{2j+1}$ that appear are cancelled on division by $h_1 h_3 \cdots h_{11}$ and all the even $h_{2k}$ are subsumed into the $M_{i,j}$ containing the surrounding $\sigma$'s.

In general, for any $M_{i_1, i_2}$ we will have a factor $\frac{\sigma_{2i_1} \sigma_{2i_2 +1}} {h_{2i_2 +1}}$ for each neighbouring even--odd pair of $\sigma$'s and the $h$ corresponding to the right hand edge, and this pair will be accompanied by a factor of $\frac{h_{2j}} {h_{2j-1}}$ for each missing pair $\sigma_{2j-1} \sigma_{2j}$ in the interval $i_1 < j < i_2$, which is the expression in \eqref{d:Msigma}. Then, to obtain \eqref{e:expansion:Pf}, we rewrite \eqref{e:PfW2} as a sum over the number of $\sigma$ pairs in each term and divide through by $h_1 h_3 \cdots h_{2N-1}$. Finally, we rewrite the products of $\sigma$'s and $h$'s as in \eqref{e:sigmaMeg} and the indices obey the rule in~\eqref{d:I2p}.

\hfill $\Box$

We recall that the $h_k$'s as well as the $M_{i_1, i_2}$ depend explicitly on $y$.  The expression for the Pfaffian in (\ref{e:expansion:Pf}) is quite convenient to analyse the large $N$ limit of $F_{4,N}(y)$ which thus reads [see Eq. (\ref{e:F4N})]
\begin{eqnarray} \label{e:F4N2}
F_{4,N}(y) =\frac{2^{N^2}}{\pi^{N/2}} \prod_{j=0}^{N-1} \frac{1}{(2j +1)!} \left(\prod_{j=0}^{N-1} h_{2j+1} \right) \left(1 + \sum_{p=1}^N \; \sum_{\cI_{2p}} M_{i_1, i_2} M_{i_3,i_4} \cdots M_{i_{2p-1}, i_{2p}} \right) \Bigg|_{y \mapsto \sqrt{2} y} \;.
\end{eqnarray}
Let us first check from this formula (\ref{e:F4N2}) that $\lim_{y \to \infty}F_{4,N}(y) = 1$. From (\ref{d:M1}), and the knowledge from \eqref{e:InfinityNorms} that when $y \to \infty$, the norms $h_j$'s converge to the norms of the Hermite polynomial of degree $j$, i.e.
\begin{eqnarray}\label{e:norm2lim}
h_j (\infty) = \lim_{y \to \infty} h_{j}(y) = \sqrt{\pi} \frac{j!}{2^j},
\end{eqnarray}
it is rather clear that  
\begin{eqnarray}\label{e:norm1}
\lim_{y \to \infty} \left(1 + \sum_{p=1}^N \; \sum_{\cI_{2p}} M_{i_1, i_2} M_{i_3,i_4} \cdots M_{i_{2p-1}, i_{2p}} \right)\ = 1\;. 
\end{eqnarray}
We can also use \eqref{e:norm2lim} to obtain
\begin{eqnarray}\label{e:norm3}
\lim_{y \to \infty} \prod_{j=0}^{N-1} h_{2j+1}(y) = {\pi}^{N/2} \frac{\prod_{j=0}^{N-1} (2j+1)!}{2^{\sum_{j=0}^{N-1}(2j+1)}} = \pi^{N/2} \frac{\prod_{j=0}^{N-1} (2j+1)!}{2^{N^2}} \;,
\end{eqnarray}
which implies, by combining (\ref{e:F4N2}), (\ref{e:norm1}) and (\ref{e:norm3}), that
\begin{eqnarray}\label{e:norm4}
\lim_{y \to \infty} F_{4,N}(y) = 1 \;,
\end{eqnarray}
as it should. 

We now proceed to obtain the scaled limit \eqref{e:TW4}. Let us start by analyzing the first factors of $F_{4,N}(y)$ in (\ref{e:F4N2}) and define
\begin{eqnarray}\label{d:ZN}
{\cal Z}_N = \frac{2^{N^2}}{\pi^{N/2}} \prod_{j=0}^{N-1} \frac{1}{(2j +1)!} \left(\prod_{j=0}^{N-1} h_{2j+1} \right)  \;.
\end{eqnarray}
It is easy to check that
\begin{eqnarray} \label{e:ZN1}
\frac{{\cal Z}_{N-1} {\cal Z}_{N+1}}{{\cal Z}_N^2} = \frac{1}{N(N+1/2)} \frac{h_{2N+1}}{h_{2N-1}} = \frac{\NMrs{R}_{2N+1} \NMrs{R}_{2N}}{N(N+1/2)}  \;.
\end{eqnarray}
Let us assume the asymptotic scaling behavior 
\begin{eqnarray}\label{e:scaling}
\ln {\cal Z}_N(y) \mathop{\rightarrow}\limits_{N\to \infty} f(2^{7/6} N^{1/6} (y-\sqrt{2N})) \;,
\end{eqnarray} 
with some function $f$, independent of $N$, yet to be determined. Assuming this scaling behavior (\ref{e:scaling}), and setting $y = \sqrt{2N} + (s/2^{7/6}) N^{-1/6}$, the left hand side of Eq. (\ref{e:ZN1}) becomes
\begin{eqnarray}\label{e:ZN2}
\ln {\cal Z}_{N-1} + \ln {\cal Z}_{N+1} - 2  \ln {\cal Z}_{N} = 2^{4/3} f''(s) N^{-2/3} + {o}(N^{-2/3}) \;.
\end{eqnarray}

Let us now analyse the right hand side of (\ref{e:ZN1}) in the large $N$ limit, where, from \cite{NadaMaju2011}, we have the asymptotic behavior
\begin{eqnarray} \label{e:RNasympt1}
\NMrs{R}_N\left(\sqrt{2N} + \frac{x}{\sqrt{2}} N^{-1/6}\right)  = \frac{N}{2} \left(1 - N^{-2/3} q^2(x) + {o}(N^{-{2/3}}) \right) \;,
\end{eqnarray}
where $q(s)$ is defined in \eqref{e:PII_intro}. This implies, setting again $y = \sqrt{2N} + (s/2^{7/6}) N^{-1/6}$, that 
\begin{eqnarray}\label{e:RNasympt2}
\NMrs{R}_{2N}(\sqrt{2}\, y) = N \left(1 - (2N)^{-2/3} q^2(s) + o(N^{-2/3}) \right) \;.
\end{eqnarray}
Hence the logarithm of the right hand side of Eq. (\ref{e:ZN1}) reads
\begin{eqnarray}\label{e:RNasympt3}
\ln\left( \frac{\NMrs{R}_{2N+1}(\sqrt{2}\, y) \NMrs{R}_{2N}(\sqrt{2}\, y)}{N(N+1/2)} \right) = - 2^{1/3} N^{-2/3} q^2(s) + o(N^{-2/3}) \;.
\end{eqnarray}
Taking the logarithm of the relation in (\ref{e:ZN1}) and equating the leading terms, of order ${\cal O}(N^{-2/3})$ on both sides, one finds
\begin{eqnarray}\label{e:RNasympt4}
f''(s) = -\frac{1}{2} q^2(s) \;.
\end{eqnarray}
Integrating twice this relation (\ref{e:RNasympt4}), using that $\lim_{s \to \infty} f'(s) = 0$ [since the probability density function $F'_{4,N}(y) \to 0$ as $y \to \infty$] as well as $\lim_{s \to \infty} f(s) = 0$ [since $F_{4,N}(y) \to 1$ as $y \to \infty$, see Eq. (\ref{e:norm4})], one obtains
\begin{eqnarray}\label{e:RNasympt5}
f(s) = -\frac{1}{2} \int_s^{\infty} (x- s) q^2(x) dx \;.
\end{eqnarray}
Therefore, recalling (\ref{e:scaling}) one obtains
\begin{eqnarray}\label{e:RNasympt6}
\lim_{N\to \infty} {\cal Z}_N \left(y = \sqrt{2N} + (s/2^{7/6})N^{-1/6}\right) \Big|_{y\mapsto \sqrt{2} y}= \exp{\left[-\frac{1}{2} \int_s^{\infty} (x- s) q^2(x) dx\right] } \;,
\end{eqnarray}
which gives the first factor of the Tracy-Widom distribution for $\beta=4$ [see Eq. (\ref{e:TW4})].

We now analyse the large $N$ behavior of the second factor in the expression of the Pfaffian in Eq. (\ref{e:expansion:Pf}). For this purpose, we will take advantage of the analysis performed in \cite{PerrSche2014}. In fact, one can show that, in the large $N$ limit, the multiple sums in Eq. (\ref{e:expansion:Pf}) are dominated by the region where $i_1, i_2, \cdots, i_{2p}$ are close to $N$. For later convenience, we reverse the order of the indices in the product of $M_{i_1, i_2}$ by looking for $M_{N- k_1, N- k_2}$, then from the results obtained in \cite{PerrSche2014} for the asymptotic forms of the ``wave functions'' in (\ref{d:psi})
\begin{align}
\label{e:LargePsi} \psi_N (y, y) \mathop{\sim}\limits_{N\to \infty} 2^{1/4} N^{-1/12} q \left( \sqrt{2} N^{1/6} (y- \sqrt{2N} ) \right)
\end{align}
and using $y\mapsto \sqrt{2} y = 2\sqrt{N} + \frac{s}{2^{2/3} N^{1/6}}$ we have
\begin{align}
\label{e:largePsi2} \psi_{2(N- k)} (y, y) &\sim 2^{1/6} N^{-1/12} q\left( s+ 2^{2/3} \frac{k}{N^{1/3}} \right).
\end{align}
Using \eqref{e:RNasympt2} for the pre-factors in \eqref{d:M2} one gets
\begin{align}
\label{e:Masympt2a} M_{N-k_1, N-k_2}\left(y = 2\sqrt{N} + \frac{s}{2^{2/3} N^{1/6}} \right) \sim \frac{1}{(2N)^{2/3}} q\left(s + 2^{2/3} \frac{k_1}{N^{1/3}}\right) q\left(s + 2^{2/3} \frac{k_2}{N^{1/3}}\right)  \;,
\end{align}
which we will be the useful form in the following. Indeed, performing first the change of variables $i_j = N - k_j$ in the second factor of Eq. (\ref{e:F4N2}) and then using (\ref{e:Masympt2a}) one finds, at leading order for large $N$, setting again $y = \sqrt{2N} + (s/2^{7/6}) N^{-1/6}$, 
\begin{align}
\label{e:Masympt3} \sum_{\cI_{2p}} M_{i_1, i_2} M_{i_3,i_4} \cdots M_{i_{2p-1}, i_{2p}} \Big|_{y\mapsto \sqrt{2} y} &\sim \frac{1}{(2N)^{\frac{2p}{3}}}\sum_{\cK_{2p}} q\left(s + 2^{\frac{2}{3}} \frac{k_1}{N^{\frac{1}{3}}} \right) \cdots q\left(s + 2^{\frac{2}{3}} \frac{k_{2p}} {N^{\frac{1}{3}}} \right) \;,
\end{align}
where, similar to \eqref{d:I2p}, we denote
\begin{align}
\label{e:K2p} \cK_{2p}: N \geq k_1 \geq k_2 > k_3 \geq k_4 > k_5 \geq k_6> \dots > k_{2p -1} \geq k_{2p} \geq 1\; .
\end{align}
In the limit $N \to \infty$ the discrete sums over the $k_j$'s become integrals. Performing the change of variables $v_j = 2^{2/3} k_j/N^{1/3}$ one finds
\begin{eqnarray}\label{e:Masympt4a}
 &&\;\;\; \sum_{\cI_{2p} } M_{i_1, i_2} M_{i_3,i_4} \cdots M_{i_{2p-1}, i_{2p}} \nonumber \\
 &\sim& \frac{1}{2^{2p}} \int_0^{\infty} dv_{2p}\int_{v_{2p}}^{\infty} dv_{2p-1} \cdots \int_{v_2}^{\infty} dv_1  \, q(s+v_1) \cdots q(s+v_{2p-1}) q(s+v_{2p}) \;.
\end{eqnarray}
Since the integrand in (\ref{e:Masympt4a}) is completely symmetric under the permutation of the variables $v_i$'s, the nested integral can actually simply be written as
\begin{eqnarray}\label{e:Masympt5}
 &&\;\;\; \sum_{\cI_{2p} } M_{i_1, i_2} M_{i_3,i_4} \cdots M_{i_{2p-1}, i_{2p}} \sim \frac{1}{(2p)!} \left(\frac{1}{2} \int_s^\infty dx \, q(x)\right)^{2p} \;.
\end{eqnarray}
Finally, summing over $p$ in Eq. (\ref{e:F4N2}), one obtains
\begin{eqnarray}\label{e:Masympt6a}
 \;\;\; \lim_{N \to \infty} \left[1 + \sum_{p=1}^N \; \sum_{\cI_{2p} } M_{i_1, i_2} M_{i_3,i_4} \cdots M_{i_{2p-1}, i_{2p}}\right]  &=& \sum_{p=0}^\infty \frac{1}{(2p)!} \left(\frac{1}{2} \int_s^\infty dx \, q(x)\right)^{2p} \nonumber \\
 &=& \cosh\left(\frac{1}{2}  \int_s^\infty dx \, q(x) \right) \;.
\end{eqnarray}
Combining Eqs. (\ref{e:F4N2}), (\ref{d:ZN}), (\ref{e:RNasympt6}) and (\ref{e:Masympt6a}), one obtains the desired expression given in (\ref{e:TW4}) for the $\beta=4$ Tracy-Widom distribution.   

\section{Asymptotic analysis of the CDF of the largest eigenvalue for $F_{1,N}$ for large $N$}
\label{s:F1asympt}

We now show that starting with \eqref{e:F1N}, we can obtain the limiting formula for $\beta=1$ \cite{TracWido1996}
\begin{align}
\label{e:TW1} \lim_{N \to \infty} F_{1, N} \left( y = \sqrt{2N} + \frac{s}{\sqrt{2} N^{1/6}} \right) &= \exp\left( -\frac{1}{2} \int_s^{\infty} (x-s) q(x)^2 dx \right) \exp\left( - \frac{1}{2} \int_s^{\infty} q(x) dx \right),
\end{align}
where we proceed in much the same way as in Section \ref{s:F4asympt} above for $\beta=4$.

From the definitions in \eqref{d:Phijk} and \eqref{d:Xjk} we have
{\small\begin{align}
\nonumber X_{j,k} + \Phi_{j,k} &= \frac{1}{2} \left( \int_{-\infty}^y e^{-\frac{x^2}{2}} p_j (x, y) \; \erf\left( \frac{x}{\sqrt{2}} \right) dx \right) \int_{-\infty}^{\infty} e^{-\frac{x^2}{2}} p_k (x, y) dx + \frac{1}{2} \int_{-\infty}^{y} e^{-\frac{x^2}{2}} p_j (x, y) dx \int_{y}^{\infty} e^{-\frac{x^2}{2}} p_k (x, y) dx\\
\label{e:Ljk1} &= \frac{1}{2} \left( \int_{-\infty}^y e^{-\frac{x^2}{2}} p_j (x, y) \; \erfc \left( - \frac{x}{\sqrt{2}} \right) dx \right) P_k (\infty, y) - \frac{1}{2} P_{j} (y, y) P_{k} (y, y),
\end{align}}
where we have introduced the notation
\begin{align}
P_j(x, y) := \int_{-\infty}^x e^{-z^2/2} p_{j} (z, y) dz, \qquad P_j(\infty, y) := \int_{-\infty}^{\infty} e^{-z^2/2} p_{j} (z, y) dz.
\end{align}
For later use we also similarly define
\begin{align}
\label{d:CapPsi} \Psi_j(\infty, y) := \int_{-\infty}^{\infty} \psi_j (z, y) dz. 
\end{align}
Using the identities \eqref{e:Ointeg} and \eqref{e:Einteg} to perform the integrals in \eqref{e:Ljk1} gives
\begin{align}
\label{e:Pjxy} P_j(x, y)&= \frac{1}{2} \erfc \left( - \frac{x}{\sqrt{2}} \right) P_j (\infty, y) - e^{-x^2/2}p_{j-1} (x, y) + e^{-x^2/2} \mathrm{LoP}_{j-2}\\
\nonumber \int_{-\infty}^x e^{-\frac{z^2}{2}} p_j (z, y) \; \erfc \left( - \frac{z}{\sqrt{2}} \right) dz &= - e^{-\frac{x^2}{2}} \erfc \left( - \frac{x}{\sqrt{2}} \right) p_{j-1} (x, y) + \frac{1}{4} \erfc \left( - \frac{x}{\sqrt{2}} \right)^2 P_j (\infty, y)\\
\label{e:erfcinteg} &+ e^{-\frac{x^2}{2}} \mathrm{LoP}_{j-2} + e^{-x^2} \mathrm{LoP}_{j-2}
\end{align}
where the notation $\mathrm{LoP}_{j-2}$ denotes ``lower-order polynomials'' up to degree $j-2$, that is, some combination of $p_0 (x, y), p_1 (x,y), \dots, p_{j-2} (x,y)$.

Noting that $\erfc(x) \in (0,2)$ (so it is bounded) and recalling the $O \left( e^{-y^2} \right)$ corrections in \eqref{e:limHerms}, we substitute \eqref{e:Pjxy} and \eqref{e:erfcinteg} into \eqref{e:Ljk1} to give (at leading order for large $y$)
\begin{align}
\label{e:Guess2} X_{j,k} + \Phi_{j,k} &\mathop{\sim}\limits_{y \to \infty} \frac{1}{2} e^{-\frac{y^2}{2}} \Big( p_{k-1} (y, y) P_j (\infty, y)- p_{j-1} (y, y) P_k (\infty, y) \Big).
\end{align}
Keeping just the leading order polynomial (meaning that we use only the larger of $j$ or $k$), we substitute \eqref{e:Guess2} into \eqref{d:Vm} to obtain
\begin{align}
\label{e:Vm1} \bV_m &\sim \begin{bmatrix}
0 & h_0 -\frac{e^{-y^2/2}}{2} P_{0}\, p_{0} & -\frac{e^{-y^2/2}}{2} P_{0}\, p_{1} & -\frac{e^{-y^2/2}}{2} P_{0}\, p_{2} & \dots \\
-h_0 +\frac{e^{-y^2/2}}{2} P_{0}\, p_{0} & 0 & h_1 -\frac{e^{-y^2/2}}{2} P_{1}\, p_{1} & -\frac{e^{-y^2/2}}{2} P_{1} \,p_{2} & \dots \\
\frac{e^{-y^2/2}}{2} P_{0}\, p_{1} & -h_1 +\frac{e^{-y^2/2}}{2} P_{1}\, p_{1} & 0 & h_2 -\frac{e^{-y^2/2}}{2} P_{2}\, p_{2} & \dots \\
\frac{e^{-y^2/2}}{2} P_{0}\, p_{2} & \frac{e^{-y^2/2}}{2} P_{1}\, p_{2} & -h_2 +\frac{e^{-y^2/2}}{2} P_{2}\, p_{2} & 0\\
\vdots & \vdots & \vdots &
\end{bmatrix},
\end{align}
where $P_0 = P_0 (\infty, y)$ and we have suppressed all the function arguments to save space. Looking at the matrix in \eqref{e:Vm1}, we see that it has identical structure to \eqref{d:Wmat} if we make the following replacements
\begin{align}
N& \mapsto \frac{N}{2}, &h_{k}(y) &\mapsto h_{j-1} (y), &p_{j}(y,y) &\mapsto -P_{j} (\infty, y), &p_{k}(y,y) &\mapsto p_{k-1} (y,y),
\end{align}
where we use $j$ to denote the row index and $k$ for the column index. This allows us to use the Pfaffian identity in \eqref{e:expansion:Pf} to conclude that for large $y$ (recalling that $N$ is even)
\begin{align}
\label{e:F1lim1} F_{1,N}(y) &\sim \frac{2^{\frac{N}{2} \left( \frac{N}{2} -1 \right)}} {\pi^{N/4}} \prod_{j=0}^{N/2 -1} \frac{1} {(2j)!} \left(\prod_{j=0}^{\frac{N}{2} -1} h_{2j} (y) \right) \left(1 + \sum_{p=1}^{N/2} \; \sum_{\cI_{2p}} T_{i_1, i_2} T_{i_3,i_4} \cdots T_{i_{2p-1}, i_{2p}} \right),\\
&\cI_{2p}: 0\leq i_1 \leq i_2 < i_3 \leq i_4 < i_5 \leq i_6 < \dots <i_{2p-1}\leq i_{2p} \leq \frac{N}{2}-1
\end{align}
with
\begin{align}
\nonumber T_{i_1, i_2}(y) &= -\frac{1}{2}\dfrac{\prod_{m={i_1}}^{i_2-1} h_{2m+1} (y)}{\prod_{m=i_1}^{i_2} h_{2m} (y)} P_{2i_1}(\infty, y) p_{2i_2}(y, y) e^{-y^2/2} \;, \;\;\;\; i_2 \geq i_1,\\
\label{e:TPfelt} &= -\frac{1}{2} \prod_{m=i_1+1}^{m=i_2} \sqrt{\frac{\NMrs{R}_{2m-1} (y)}{\NMrs{R}_{2m} (y)}} \Psi_{2i_1}(\infty, y) \psi_{2i_2}(y,y) \;, \;\;\;\; i_2 \geq i_1,
\end{align}
where $\Psi_j (\infty, y)$ is from \eqref{d:CapPsi} and $\NMrs{R}_k(y) = h_k (y) /h_{k-1}(y)$ is from \eqref{e:NM2}. In terms of the proof of Proposition \ref{lem:PfW}, only superficial modifications are needed, with the main change here being that a ``little link'' from site $j$ to site $j+1$ (as in Figure \ref{f:LLs}) now corresponds to a factor of $h_j \delta_{j-1,j} - \frac{P_j (\infty, y) p_j (y,y)}{2} e^{-y^2/2}$.

Denoting the prefactor in \eqref{e:F1lim1} by
\begin{align}
\cZ_N:= \frac{2^{\frac{N}{2} \left( \frac{N}{2} -1 \right)}} {\pi^{N/4}} \prod_{j=0}^{N/2 -1} \frac{1} {(2j)!} \left(\prod_{j=0}^{\frac{N}{2} -1} h_{2j} (y) \right)
\end{align}
then
\begin{align}
\frac{\cZ_{N-2} \cZ_{N+2}} {\cZ_N^2} = \frac{4}{N(N-1)} \frac{h_N (y)}{h_{N-2} (y)} = \frac{4 \NMrs{R}_N \NMrs{R}_{N-1}}{N(N-1)} 
\end{align}
and so using \eqref{e:RNasympt1}
\begin{align}
\label{e:log1} \ln \left( \frac{4 \NMrs{R}_N \NMrs{R}_{N-1}}{N(N-1)} \right) \sim - 2 \frac{q(s)^2} {N^{2/3}}+ o \left( N^{-2/3} \right).
\end{align}
Now, analogously to \eqref{e:scaling} we assume
\begin{eqnarray}\label{e:beta1scaling}
\ln {\cal Z}_N(y) \mathop{\rightarrow}\limits_{N\to \infty} f(\sqrt{2} N^{1/6} (y-\sqrt{2N})) \;,
\end{eqnarray}
and then with $y= \sqrt{2N} + \frac{s}{\sqrt{2} N^{1/6}}$ we have
\begin{eqnarray}
\label{e:log2} \ln {\cal Z}_{N-2} + \ln {\cal Z}_{N+2} - 2  \ln {\cal Z}_{N} = \frac{4}{N^{2/3}} f''(s) + {o} \left( N^{-2/3} \right) \;.
\end{eqnarray}
Equating \eqref{e:log1} and \eqref{e:log2} we have (to leading order)
\begin{align}
f''(s) = - \frac{1}{2} q^2(s)
\end{align}
identically with the $\beta=4$ case in \eqref{e:RNasympt5}. Therefore, we have
\begin{eqnarray}
\lim_{N\to \infty} {\cal Z}_N \left(y = \sqrt{2N} + \frac{s}{\sqrt{2} N^{1/6}} \right)= \exp{\left[-\frac{1}{2} \int_s^{\infty} (x- s) q^2(x) dx\right] } \;,
\end{eqnarray}
which is the first factor in \eqref{e:TW1}.

For the right-most factor in \eqref{e:F1lim1} we use the known asymptotic behaviour \eqref{e:LargePsi}, with $y= \sqrt{2N} + \frac{s}{\sqrt{2} N^{1/6}}$, to find
\begin{align}
\psi_{N- 2k} (y, y) \sim 2^{1/4} N^{-1/12} q\left( s+ 2 \frac{k}{N^{1/3}} \right).
\end{align}
For $\Psi_{N -2k} (\infty, y)$ in \eqref{d:CapPsi}, we recall that this is an integral over the entire domain, so the integral will be dominated by the behaviour of the integrand in the bulk regions. From \eqref{e:limHerms} and \eqref{e:InfinityNorms} we have the large $y$ behaviour
\begin{align}
\psi_{N} (x, y) &\mathop{\sim}\limits_{y\to \infty} \frac{H_N (x)} {\pi^{1/4} 2^{N/2} \sqrt{\Gamma (N+1)}} e^{-x^2/2}
\end{align}
and using the recursion relations \cite[Chapter 22]{AbraSteg1972}
\begin{align}
\label{e:Hermids} H_{j+1} (x) = 2x H_j - H'_j (x), \qquad H'_j(x) = 2j H_{j-1} (x)
\end{align}
we can show
\begin{align}
\label{e:Hermint} \int_{-\infty}^{\infty} e^{-\frac{z^2}{2}} H_j (z) dz &= \left\{ \begin{array}{cl}
2^{j+ 1/2} \Gamma \left( \frac{j+1}{2} \right),& \mbox{$j$ even},\\
0,& \mbox{$j$ odd},
\end{array}\right.
\end{align}
where the second line follows since Hermite polynomials of odd degree are odd functions. So we have (recalling that $N$ is even)
\begin{align}
\label{e:PsiLim} \Psi_{N- 2k} (\infty, y) &\mathop{\sim}\limits_{y\to \infty} \frac{2^{3/4}} {N^{1/4}},
\end{align}
and with \eqref{e:RNasympt1} we obtain
\begin{align}
T_{\frac{N}{2} -k_1, \frac{N}{2} -k_2} \left( y= \sqrt{2N} + \frac{s}{\sqrt{2} N^{1/6}} \right) \sim - \frac{1}{N^{1/3}} q\left( s+ 2 \frac{k_2}{N^{1/3}} \right).
\end{align}
Then we find ourselves at the analogue of \eqref{e:Masympt3}
\begin{align}
\sum_{\cI_{2p}} T_{i_1, i_2} T_{i_3,i_4} \cdots T_{i_{2p-1}, i_{2p}} &\sim \frac{(-1)^p}{N^{\frac{p}{3}}}\sum_{\cK_{p}} q\left(s + 2 \frac{k_2}{N^{\frac{1}{3}}} \right) q\left(s + 2 \frac{k_4}{N^{\frac{1}{3}}} \right) \cdots q\left(s + 2 \frac{k_{2p}} {N^{\frac{1}{3}}} \right) \;,
\end{align}
where we denote
\begin{align}
\label{e:Kp} \cK_{p}: \frac{N}{2} \geq k_2 > k_4 > k_6> \dots > k_{2p -2} > k_{2p} \geq 1\; .
\end{align}
[Note that, in contrast to $\mathcal{K}_{2p}$ in \eqref{e:K2p}, $\cK_{p}$ contains only the even indices. In the $\beta=4$ case, both even and odd indices contributed factors of $q$, as can be seen in \eqref{e:Masympt2a}. But here the odd indices are attached to the integrals $\Psi_{N-2k_{\mathrm{odd}}}$, and only contribute factors of $N$ and $2$ as per \eqref{e:PsiLim}.] Changing variables $v_j = 2k_j /N^{1/3}$ gives
\begin{align}
\nonumber &\sum_{\cI_{2p}} T_{i_1, i_2} T_{i_3,i_4} \cdots T_{i_{2p-1}, i_{2p}} \\
\nonumber & \sim \frac{(-1)^p}{2^{p}} \int_0^{\infty} q(s+v_{2p}) dv_{2p }\int_{v_{2p}}^{\infty} q(s+v_{2p-2}) dv_{2p-2} \cdots \int_{v_6}^{\infty} q(s+v_4) dv_4 \int_{v_4}^{\infty} q(s+v_2)  dv_2 \\
&= \frac{(-1)^p}{p!} \left(\frac{1}{2} \int_s^\infty dx \, q(x)\right)^{p},
\end{align}
where, in the final line, we have removed the ordering from the integration variables since the integrand is symmetric in the $v_j$'s. Summing over $p$ and taking the limit we have
\begin{align}
\nonumber \lim_{N \to \infty} \left[1 + \sum_{p=1}^{N/2} \sum_{\cI_{2p} } T_{i_1, i_2} T_{i_3,i_4} \cdots T_{i_{2p-1}, i_{2p}}\right]  &= \sum_{p=0}^\infty \frac{1}{p!} \left(-\frac{1}{2} \int_s^\infty dx \, q(x)\right)^{p}\\
 &=\exp\left(-\frac{1}{2}  \int_s^\infty dx \, q(x) \right),
\end{align}
which is the second factor in \eqref{e:TW1}.

Lastly, we note that the Pfaffian identity \eqref{e:expansion:Pf} that we used here for $\beta=1$ and for $\beta=4$ above will hold more generally, for all anti-symmetric matrices of the form
\begin{align}
\bM= \bT + \bB,
\end{align}
where $\bT$ has upper triangular elements $\delta_{j+1,k} t_{j,k}$ and $\bB= [b_{j,k}]_{j,k=1, \dots, 2N}$ has upper-triangular entries $b_{j,k}= f_j g_k$ for some functions $f$ and $g$. In which case,
\begin{align}
\Pf \bM &= \left(\prod_{j=0}^{N-1} t_{2j+1} \right) \left(1 + \sum_{p=1}^N \; \sum_{\cI_{2p}} L_{i_1, i_2} L_{i_3,i_4} \cdots L_{i_{2p-1}, i_{2p}} \right) \;,
\end{align}
where
\begin{eqnarray}
L_{i_1, i_2} = \dfrac{\prod_{m={i_1}}^{i_2-1} t_{2m+2}}{\prod_{m=i_1}^{i_2} t_{2m+1}} f_{2i_1}\, g_{2i_2+1}, \;\;\;\; i_2 \geq i_1 \;,
\end{eqnarray}
with the summation indices $\cI_{2p}$ defined in \eqref{d:I2p}.

\section{Conclusions and perspectives}\label{s:conclusion}

In this paper, we have revisited the computation of the cumulative distribution function 
of the largest eigenvalue in the classical ensembles of RMT, namely the GOE and the GSE, using the
techniques of skew-orthogonal polynomials, thus extending the approach of Nadal and Majumdar \cite{NadaMaju2011}
developed for the GUE. By adapting the method of Refs. \cite{AdleForrNagavanMoer2000,AdlevanMoer2002}, we have
constructed explicitly these (semi-classical) skew-orthogonal polynomials in terms of the so-called ``Nadal-Majumdar''
orthogonal polynomials introduced in the case of the GUE. This construction involves some non-trivial Pfaffians,
which we have related to ``overlapping Pfaffians'', studied originally by Knuth \cite{Knuth1996}. We were then able to
carry out the asymptotic analysis of these skew-orthogonal polynomials and of their norms to obtain 
the well known Tracy-Widom distributions, using a method which is quite different from the original
one \cite{TracWido1996}, and also different from the more recent one obtained via the so-called stochastic
Airy operator \cite{bloemendal2013limits}. This relied on a certain Pfaffian identity, the most general statement of which is given at the end of Section \ref{s:F1asympt}.

As discussed in Section \ref{sec:PartFns}, it is known that ``Pfaffian'' Janossy densities (of which our $\beta=1$ and $\beta=4$ densities are examples) have $n$-point correlation functions given by Pfaffians. These correlation functions can be calculated via standard techniques (see \cite{Mehta2004,Forrester2010}) --- these calculations will be presented in a follow-up work \cite{mays2020prep}. By using the skew-orthogonal polynomials constructed in the present work, this will allow us to analyze the density of states near the largest eigenvalue and the statistics of the gap between the two largest eigenvalues in the GSE and the GOE. These quantities are particularly interesting in the challenging case of GOE since they naturally enter into the computation of physical observables in the spherical Sherrington-Kirkpatrick model of mean-field spin glasses \cite{fyodorov2015large}.

\begin{acknowledgements}
A.M. would like to thank Michael Wheeler, Peter Forrester and Shi-Hao Li for helpful discussions. A.M. and A.P. are supported by the Australian Research Council (ARC) Centre of Excellence for Mathematical and Statistical Frontiers (ACEMS), ARC Grant No. CE140100049. A.M. thanks LPTMS for their hospitality during a visit supported by CNRS.
\end{acknowledgements}


\newpage

\begin{appendices}
\appendixpage

\renewcommand{\theequation}{\thesection.\arabic{equation}}
\setcounter{equation}{0}
\numberwithin{proposition}{section}
\numberwithin{definition}{section}
\numberwithin{figure}{section}

\section{Reminder on the classical ensembles of RMT: GOE, GUE and GSE} \label{sec:classical}

For self-consistency, we recall here the definition of the classical ensembles of RMT studied in this paper: 
\begin{itemize}
    \item{The \textit{Gaussian Orthogonal Ensemble} (GOE) is the set of $N\times N$ real symmetric matrices
    \begin{align}
        \label{d:GOE} \bM = \frac{\bY+ \bY^T}{2},
    \end{align}
    where $\bY$ contains standard normally distributed elements $y_{j,k} \sim \mathcal{N}[0,1]$ resulting in the matrix PDF proportional to $e^{- (\Tr \bM^2)/2}$ which is invariant under orthogonal conjugation $\bM \mapsto \bO^T \bM \bO$.}
    
    \item{The \textit{Gaussian Unitary Ensemble} (GUE) is the set of complex Hermitian matrices
    \begin{align}
        \bM = \frac{\bY + \bY^{\dagger}}{2}
    \end{align}
    with real independent Gaussian components $y_{j,k} \sim \mathcal{N}[0,\frac{1}{\sqrt{2}}] + i \mathcal{N}[0,\frac{1}{\sqrt{2}}]$ giving a matrix PDF proportional to $e^{- \Tr \bM^2}$ which is invariant under unitary conjugation $\bM \mapsto \bU^{\dagger} \bM \bU$.}
    
    \item{The \textit{Gaussian Symplectic Ensemble} (GSE) is defined similarly for normally distributed quaternionic entries. We provide in Appendix \ref{app:Quats} some definitions related to quaternions, however this will not be required for understanding the current work, as we use the equivalent $2\times 2$ representation of quaternions
    \begin{align}
    \label{d:quat}\begin{bmatrix}
    a_1+ ib_1 &a_2 +i b_2\\
    -a_2+ ib_2 &a_1 -i b_1
    \end{bmatrix} \qquad (a_1, a_2, b_1, b_2 \in \mathbb{R}).
    \end{align}
    The ensemble is then the set of $2N\times 2N$ matrices,
    \begin{align}
    \label{d:GSE}    \bM= \frac{\bY+ \bY^{\dagger}} {2},
    \end{align}
    where each $2\times 2$ block of $\bY$ is of the form \eqref{d:quat} with each independent real component normally distributed $a_1, b_1, a_2, b_2 \sim \mathcal{N}[0, \frac{1}{2} ]$. The matrix PDF is then proportional to $e^{- \Tr \bM^2}$, which is invariant under symplectic conjugation, that is conjugation by a unitary matrix $\bM \mapsto \bU^{\dagger} \bM \bU$, with the restriction that
    \begin{align}
        \bU \bZ_{N} \bU^T = \pm \bZ_{N} \;,
    \end{align}
    where
    \begin{align}
\label{d:Zmat} \bZ_N:=
\begin{bmatrix}
0& 1\\
-1& 0
\end{bmatrix} \otimes \bI_N =
\begin{bmatrix}
0& 1& 0 & 0&0& &\\
-1& 0& 0 & 0&0& &\dots\\
0&0&0&1& 0 &&\\
0&0&-1&0& 0& &\\
&\vdots &&&&&\ddots
\end{bmatrix},
\end{align}
and $\bI_{N}$ is the $N\times N$ identity matrix.
    }

\end{itemize}

\section{Quaternions}\label{app:Quats}

Here we provide a brief overview of some definitions related to quaternions. A \textit{quaternion} is typically written in the form
\begin{align}
q= q_0 + iq_1 +j q_2 +kq_3, \qquad q_0, q_1, q_2, q_3 \in \mathbb{R},
\end{align}
where $i, j$ and $k$ are the quaternionic generalization of the imaginary unit and obey Hamilton's famous bridge equation
\begin{align}
i^2 = j^2 = k^2 = ijk = -1,
\end{align}
which defines their algebraic behaviour. (Note that we restrict the coefficients $q_j$ to be real --- these are called \textit{real quaternions} by other authors \cite{Mehta2004,Forrester2010}, to contrast with the more general case where the coefficients are complex. We have no need of the more general case in this work.) A more convenient representation of the same algebra is given by mapping the quaternions to the $2\times 2$ complex matrices
\begin{align}
\label{d:quat2x2} q= \begin{bmatrix}
a && b\\
-\bar{b} && \bar{a}
\end{bmatrix},
\end{align}
where $a:= q_0 + iq_1 \in \mathbb{C}$ and $b= q_2+ i q_3 \in \mathbb{C}$. This representation is equivalent to a linear combination of the Pauli spin matrices (see, for example, \cite{Forrester2010}).

The analogue of complex conjugation for quaternions is
\begin{align}
q^*= q_0 - iq_1 -j q_2 -kq_3 \qquad \longleftrightarrow \qquad q^{\dagger} = \begin{bmatrix}
\bar{a} & & -b\\
\bar{b} & & a
\end{bmatrix},
\end{align}
where we see that $q^*$ is the same as the Hermitian conjugate of the $2\times 2$ matrix representation. A matrix of quaternionic entries is said to be \textit{self-dual} if
\begin{align}
\bQ := [q_{j,k}] = [q^*_{k,j}] =: \bQ^*,
\end{align}
or equivalently, if the matrix of $2\times 2$ quaternionic blocks \eqref{d:quat2x2} is Hermitian. The Gaussian Symplectic Ensemble in \eqref{d:GSE} is then equivalently defined as the set of $2N \times 2N$ Hermitian matrices $[q_{j,k}]$ with entries
\begin{align}
q_{j,j} = \begin{bmatrix}
x_{j,j}& 0\\
0 & x_{j,j}
\end{bmatrix}, \qquad q_{j,k} = \begin{bmatrix}
z_{j,k}& w_{j,k}\\
-\bar{w}_{j,k} & \bar{z}_{j,k}
\end{bmatrix}, \qquad (k>j)
\end{align}
with
\begin{align}
x_{j,j} \mathop{\sim}\limits_d \mathcal{N}\left[ 0, \frac{1}{2} \right] \in \mathbb{R} \quad \mbox{and} \quad z_{j,k}, w_{j,k} \mathop{\sim}\limits_d \mathcal{N}\left[ 0, \frac{1}{2\sqrt{2}} \right] + i \mathcal{N}\left[ 0, \frac{1}{2\sqrt{2}} \right] \in \mathbb{C}.
\end{align}

In \eqref{e:MatPDFs} we write the matrix PDF for $\beta=1,2$ and $4$, however for $\beta=4$ this requires the use of the \textit{quaternion trace}, which for a quaternionic matrix $\hat{\bQ}_{N\times N}$, is
\begin{align}
\label{e:qTr} \mathrm{qTr}\; \hat{\bQ}_{N\times N} = \sum_{j=1}^N (q_0)_{j,j} = \frac{1}{2} \Tr \bQ_{2N\times 2N},
\end{align}
where in the second equality this is the usual matrix trace and $\bQ_{2N\times 2N}$ is the equivalent matrix with entries given by the $2\times 2$ matrices \eqref{d:quat2x2}. A related concept is the \textit{quaternion determinant}, which is defined for self-dual quaternion matrices by
\begin{align}
\label{def:qdet} \qdet\; \hat{\bQ}_{N\times N} :=\sum_{P\in S_N}(-1)^{N-|c(P)|}\prod_{(ab\cdots s)\in c(P)} \; (q_{ab}q_{bc}\cdots q_{sa})_{0},
\end{align}
where $c(P)$ is the set of cycles of the permutation $P$, and the subscript ${(\dots)}_{0}$ denotes that one takes the scalar part $q_{0}$ of the resulting quaternion. As with the quaternion trace in \eqref{e:qTr}, there is a relationship between the quaternion determinant and the usual determinant, given by
\begin{align}
\label{e:qdet} \left( \qdet\; \hat{\bQ}_{N\times N} \right)^2 = \det \bQ_{2N\times 2N},
\end{align}
where again the matrix on the right is the equivalent complex matrix made of the $2\times 2$ blocks \eqref{d:quat2x2}.

\section{Pfaffians}\label{a:Pfaffs}
\setcounter{equation}{0}

Pfaffians are very closely related to quaternion determinants \eqref{def:qdet}, however they do not require any of the quaternionic technicalities, so we prefer to use Pfaffians in this work. A brief historical survey on the topic is provided in \cite[\S 6]{Knuth1996}

\begin{definition} [Pfaffian]
Let $\bM=[m_{j, k}]_{j,k =1,... ,2N}$, where $m_{j, k}=-m_{k, j}$, so that $\bM$ is an anti-symmetric matrix of even size. Then the \textit{Pfaffian} of $\bM$ is defined by
\begin{align}
\nonumber \mathrm{Pf}\; \bM&=\sum^*_{P\in S_{2N} \atop P(2j)>P(2j -1)} \varepsilon (P) m_{P(1),P(2)} m_{P(3),P(4)}\cdot\cdot\cdot m_{P(2N-1), P(2N)}\\
\nonumber &=\frac{1}{N!} \sum_{P\in S_{2N} \atop P(2j) >P(2j -1)} \varepsilon (P) m_{P(1),P(2)} m_{P(3),P(4)}\cdot\cdot\cdot m_{P(2N-1), P(2N)}\\
\label{def:Pf} &=\frac{1}{2^N N!}\sum_{P\in S_{2N}} \varepsilon (P) m_{P(1), P(2)} m_{P(3), P(4)}\cdot\cdot\cdot m_{P(2N-1),P(2N)},
\end{align}
where $S_{2N}$ is the group of permutations of $2N$ letters and $\varepsilon (P)$ is the signature of the permutation $P$. The * above the first sum indicates that the sum is over distinct terms only (that is, all permutations of the pairs of indices are regarded as identical).
\end{definition}

Note that in the second equality of (\ref{def:Pf}) the factors of $2$ are associated with the restriction $P(2j)>P(2j -1)$ while the factorial is associated with counting only distinct terms [$N!$ is the number of ways of arranging the $N$ pairs of indices $ P(2l-1), P(2l)$]. Pfaffians can be calculated via a version of Laplace expansion, however the Pfaffian minors $\bM^{(j,k)}$ that one needs to calculate are obtained by blocking out both the $j$th and $k$th row and the $j$th and $k$th column.

The definition of a Pfaffian is very close to that of a determinant, and for the matrix $\bM$ (antisymmetric of size $2N \times 2N$), they are related by
\begin{align}
\label{e:PfDet2a} (\Pf \bM )^2 = \det \bM.
\end{align}
The clear similarity between \eqref{e:PfDet2a} and \eqref{e:qdet} highlights the equivalent nature of quaternion determinants and Pfaffians; they are specifically connected via the matrix $\bZ_N$ in \eqref{d:Zmat}, where we note that
\begin{align}
\label{e:detZ} \Pf (\bZ_{N}) = \det (\bZ_{N}) = 1.
\end{align}
With $\bZ_N$ we have
\begin{align}
\label{e:Pfqdet1} \Pf( \bM ) &= \qdet(\bM \bZ_{N} ) = \qdet( \bZ_{N} \bM),\\
\label{e:Pfqdet2} \Pf( \bM ) &= (-1)^N \qdet(\bM \bZ_{N}^T ) = (-1)^N \qdet( \bZ_{N}^T \bM).
\end{align}

We will also have need of the identity \cite{deBruijn1955}
\begin{align}
\label{e:PfBAB} \Pf (\bB \bM \bB^T)= \det (\bB) \Pf (\bM),
\end{align}
where $\bB$ is a general $2N \times 2N$ matrix.

\subsection{Pfaffians and elementary row/column operations}

Given the similarity between Pfaffians and determinants, it is not surprising that Pfaffians have similar behaviour to determinants, particularly for elementary row and column operations.

Recall the determinant identity
\begin{align}
\label{e:scaledet} \alpha \det \begin{bmatrix}
\bm_1& \bm_2 & \dots & \bm_N
\end{bmatrix} = \det \begin{bmatrix}
\bm_1& \dots & \bm_{j-1} & \alpha \bm_j & \bm_{j+1} & \bm_N
\end{bmatrix},
\end{align}
for a general $N\times N$ matrix, where $j$ is any integer from $1$ up to $N$. That is, the determinant can be scaled by scaling any column (or row) of the matrix. There is an equivalent identity for Pfaffians, however when we scale a column/row we also scale its corresponding row/column by the same factor. Explicitly, with $\bM$ as above, we have
\begin{align}
\label{e:scalePf} \alpha \; \Pf \bM &= \Pf \begin{bmatrix}[ccc|c|ccc]
    {} &&& \alpha m_{1,j} & {} \\
    {} & * && \vdots & &*& \\
    {} &&& \alpha m_{j-1,j} & {} \\
    \hline
    -\alpha m_{1,j} & \dots & -\alpha m_{j-1,j} &  0& \alpha m_{j,  j+1} & \dots & \alpha m_{j, 2N}\\
    \hline
    &&&- \alpha m_{j, j+1}&&&\\
    &* && \vdots&& * &\\
    &&&- \alpha m_{j, 2N}&&&
\end{bmatrix},
\end{align}
where the ``$*$'' represents that the remaining matrix elements are unchanged. Note this row and column scaling preserves the anti-symmetry of the matrix. If we take $\alpha \mapsto \sqrt{\alpha}$ then it can be seen that this scaling is consistent with \eqref{e:PfDet2a} and \eqref{e:scaledet}.

Similarly, we have an analogue of the identity
\begin{align}
\det \begin{bmatrix}
\bm_1& \bm_2 & \dots & \bm_N
\end{bmatrix} = \det \begin{bmatrix}
\bm_1& \dots & \bm_{j-1} & \bm_j +\alpha \bm_k & \bm_{j+1} & \bm_N
\end{bmatrix}
\end{align}
where the determinant is unchanged by adding to any column/row a scalar multiple of any other column/row. The Pfaffian analogue is obtained by adding a multiple of a column (or row) to another column (or row), and adding the same multiple of the same row (or column) to the matching row (or column),
\begin{align}
\label{e:Pfrowadd} \Pf \bM = \Pf \widehat{\bM}_{j,k; \alpha},
\end{align}
where the matrix $\widehat{\bM}_{j,k; \alpha}$ is identical to $\bM$ except for column and row $j$, which equal
\begin{align}
Col(j)\mapsto Col(j) + \alpha Col(k), \qquad Row(j)\mapsto Row(j) + \alpha Row(k)
\end{align}
Note that anti-symmetry is preserved, and we again see that this is consistent with \eqref{e:PfDet2a}.

Lastly, we have
\begin{align}
\det \begin{bmatrix}
\bm_1& \bm_2 & \dots & \bm_N
\end{bmatrix} = - \det \begin{bmatrix}
\bm_1& \dots & \bm_{j-1} & \bm_k & \bm_{j+1} \dots & \bm_{k-1} & \bm_j & \bm_{k+1} & \dots & \bm_N,
\end{bmatrix}
\end{align}
where determinants pick up a factor of $(-1)$ for each column/row swap. The analogous result for Pfaffians is more complicated
\begin{align}
\label{e:Pfrowswap} \Pf \bM = - \Pf \widehat{\bM}_{j \leftrightarrow k},
\end{align}
where the matrix $\widehat{\bM}_{j \leftrightarrow k}$ is identical to $\bM$ except that
\begin{align}
Col(j) \leftrightarrow Col(k) \quad \mbox{and} \quad Row(j) \leftrightarrow Row(k),
\end{align}
where the swaps happen in succession. This again preserves anti-symmetry and is consistent with \eqref{e:PfDet2a}.

\subsection{Pfaffians and perfect matchings}\label{a:PfsPMs}

In order to prove Proposition \ref{lem:PfW} we will use an expression equivalent to \eqref{def:Pf} in terms of perfect matchings and link patterns. Expressions for Pfaffians in terms of perfect matchings have been known for a long time, and they are discussed in many places --- we refer to \cite{Knuth1996,Rote2001}.

\begin{figure}
\begin{center}
\begin{tikzpicture}[baseline=(current  bounding  box.center)] %
\draw (-0.5,0)--(5.5,0); %
\node[below] at (0,0) {$1$}; %
\node[below] at (1,0) {$2$}; %
\node[below] at (2,0) {$3$}; %
\node[below] at (3,0) {$4$}; %
\node[below] at (4,0) {$5$}; %
\node[below] at (5,0) {$6$}; %
\draw[smooth] (0,0) to[out=60,in=180] (2,0.75) to[out=0,in=120] (4,0); %
\draw[smooth] (1,0) to[out=60,in=180] (1.5,0.33) to[out=0,in=120] (2,0); %
\draw[smooth] (3,0) to[out=60,in=180] (4,0.5) to[out=0,in=120] (5,0); %
\end{tikzpicture}
\end{center}
\caption{The link pattern for the perfect matching $\mu= \{(2,3), (5,1), (4, 6)\} = \{ (1, 5), (2,3), (4, 6)\}$ on sites $\{ 1,2,3, 4, 5, 6\}$ from \eqref{e:PMeg2}. \label{f:PMeg}}
\end{figure}
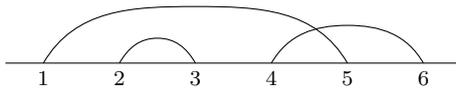

A \textit{perfect matching} $\mu$ is a set of links between $2N$ sites, where each site is connected to exactly one other site. Diagrammatically, this is expressed as a \textit{link diagram}, and most easily seen via an example: let
\begin{align}
\label{e:PMeg2} \mu= \{(2,3), (5,1), (4, 6)\}
\end{align}
and the link diagram is given in Figure \ref{f:PMeg}. The sign $\varepsilon(\mu)$ of the perfect matching is given by$(-1)^{\# \chi}$, where $\# \chi$ is the number of crossings in the link pattern --- for the example in \eqref{e:PMeg2} we have $\varepsilon(\mu) = (-1)^1$. We denote the set of all perfect matchings on $2N$ sites by $M_{2N}$, and the number of perfect matchings is
\begin{align}
\label{e:NumPMs} |M_{2N}| = (2N -1)!! = (2N -1) \cdot (2N-3) \cdots (3) \cdot (1),
\end{align}
since there are $2N-1$ sites for the first site to pair with, then $2N-3$ sites for the second site to pair with, etc.
(Note that usually a perfect matching is defined as a set of edges on a graph such that every vertex is included exactly once. However this characterization will not be useful for us, and for a complete graph it is equivalent to the definition we use in terms of link patterns.)

The connection to Pfaffians comes from the fact that there is a bijection from $M_{2N}$ to a subset of $S_{2N}$, the set of permutations of $\{1, \dots, 2N \}$. The bijection is found by taking a perfect matching $\mu$ and ordering the components of each pair such that $\mu= \{ (\mu_{j,L}, \mu_{j,R}) \}_{j= 1, \dots, N}$, where $\mu_{j,L}$ and  $\mu_{j,R}$ are respectively the left and right terminals of link $j$. (In graph parlance, this creates a \textit{directed link pattern}, where all links point from, say, left to right.) Then we institute an ordering between the pairs according to some scheme (say, that $\mu_{j,L}< \mu_{j+1 ,L}$), which results in a unique representative ordered set of pairs for each perfect matching. Then, by removing the pairing, we obtain a unique $s\in S_{2N}$. For the example in \eqref{e:PMeg2} we find
\begin{align}
\label{e:PMeg3} M_{6} \ni \{(2,3), (5,1), (4, 6)\} = \{ (1, 5), (2,3), (4, 6)\}  \mapsto (1,5,2,3,4,6) \in S_6.
\end{align}
The reverse mapping $S_{2N} \supset \hat{S}_{2N} \to M_{2N}$ is clear: $s\in \hat{S}_{2N}$ is a permutation of $1, \dots, 2N$ such that $s(2j-1)< s(2j)$ and $s(2j-1)< s(2k-1)$ for $j<k$. 

In order for this mapping to make sense, we need $\varepsilon(\mu)= \varepsilon(s)$, that is the number of crossings in the perfect matching $\mu$ must be the same as the sign of the permutation $s$, given by $(-1)^\tau$ where $\tau$ is the number of transpositions required to return $s$ to the identity permutation. This can be shown by first noting that the identity permutation gives a link pattern with no crossings, and then that a crossing can always be removed by a single transposition, while a link pattern with no crossings can be transformed to the identity by an even number of transpositions.

The conditions defining $\hat{S}_{2N}$ are the same restrictions on $S_{2N}$ as those implied by the first line of \eqref{def:Pf}, and so we have the following equivalent expression for the Pfaffian
\begin{align}
\label{e:PfPerfMatch} \Pf \bM = \sum_{\mu\in M_{2N}} \varepsilon (\mu)\; m_{i_1, j_1} m_{i_2, j_2}\cdot\cdot\cdot m_{i_N, j_{N}},
\end{align}
where $M_{2N}$ is the set of all perfect matchings $\mu= \{ (i_1, j_1), \dots, (i_N, j_N) \}$ on $2N$ sites, and $\varepsilon (\mu)$ is the sign of the perfect matching, or equivalently, the sign of the corresponding permutation.

\section{Iterative construction of the first few skew-orthogonal polynomials} \label{s:iterative}

In this Appendix, we iteratively construct the first few skew-orthogonal polynomials defined in Eqs.~(\ref{d:4sorthog}) and (\ref{d:1sorthog}).  

First for $\beta=4$, by monicity, we must have $Q_0 (\lambda) = 1$ and by \eqref{e:oddsymm} we can assume that $Q_1 (\lambda) = \lambda$, then we use the skew-inner product relations \eqref{d:IP4} to iteratively solve for the higher degree polynomials, so the first four skew-orthogonal polynomials are
\begin{align}
Q_0 (\lambda, y)& = 1, \qquad Q_1 (\lambda, y) = \lambda, \qquad Q_2 (\lambda, y)= \lambda^2 +b \lambda + \frac{1 -2yb} {4},\\
Q_3 (\lambda, y)&= \lambda^3 - 3 \frac{1- 2y b} {4} \lambda -b \frac{1+2 y^2} {2}
\end{align}
[where we used \eqref{e:oddsymm} for $Q_3 (\lambda, y)$] with normalizations
\begin{align}
q_0 (y)&:= \langle Q_0 , Q_1 \rangle_{4}^y = \frac{\sqrt{\pi}}{4 \sqrt{2}} \;\erfc (-\sqrt{2} y) = \frac{e^{-2y^2}} {4 b},\\
q_1 (y)&:= \langle Q_2 , Q_3 \rangle_{4}^y = \frac{1} {64} \left( 3 \sqrt{2 \pi} \erfc (- \sqrt{2} y) -2 e^{-2 y^2} y (9 +4 y^2) - 4 e^{-2 y^2} ( 2 +y^2) b \right),
\end{align}
where
\begin{align}
b= \frac{\sqrt{2} e^{-2y^2}} {\sqrt{\pi} (1+\erf (\sqrt{2} y))}= \frac{\sqrt{2} e^{-2y^2}} {\sqrt{\pi} \erfc (- \sqrt{2} y)}.
\end{align}

For $\beta=1$, again by monicity and \eqref{e:oddsymm} we have, $R_0(\lambda) =1, R_1(\lambda) = \lambda$ and using the relations \eqref{d:IP1} we can obtain the first four polynomials
\begin{align}
R_0(\lambda, y)& =1, \qquad R_1(\lambda) = \lambda,\\
R_2 (\lambda, y)&= \lambda^2 + \lambda \frac{c}{\sqrt{\pi}} \left(2 e^{-y^2 /2} + \sqrt{2 \pi} y \: \erfc (-y/ \sqrt{2}) \right) + c e^{y^2/2} \erfc (-y) -1,\\
R_3 (\lambda, y)&= \lambda^3 +\lambda c \left( \frac{2 ye^{-y^2/2}} {\sqrt{\pi}} - \frac{2}{c} - e^{y^2/2} \: \erfc (-y) + y^2 \sqrt{2 }\: \erfc (-y/ \sqrt{2} ) \right) - \frac{2 c e^{-y^2/2}} {\sqrt{\pi}}
\end{align}
with
\begin{align}
c= \left( 2e^{y^2/2} \erfc(-y) - \sqrt{2} \erfc(-y/ \sqrt{2}) \right)^{-1}
\end{align}
and
\begin{align}
r_0(y) &=\frac{\sqrt{\pi}} {2} \left( \erfc (-y) - \frac{e^{-y^2/2}} {\sqrt{2}} \erfc (- y/ \sqrt{2}) \right) = \frac{\sqrt{\pi} e^{-y^2/2}}{4 c}\\
\nonumber r_1 (y) &= \frac{\sqrt{\pi }}{8}  \text{erfc}(-y) -\frac{y e^{-y^2}}{4} - c \left( \frac{e^{-\frac{3y^2}{2}}}{\sqrt{\pi }} + \frac{y^2 \sqrt{\pi}}{2 \sqrt{2}} \erfc (-y) \erfc \left(-\frac{y}{\sqrt{2}}\right)\right.\\
&\left. +\frac{y e^{-\frac{y^2}{2}}}{2} \erfc (-y) +\frac{y e^{-y^2} \erfc \left(-\frac{y}{\sqrt{2}}\right)}{\sqrt{2}}- \frac{\sqrt{\pi} e^{\frac{y^2}{2}} }{4} \erfc (-y)^2 \right).
\end{align}

\section{Skew-orthogonal polynomials for $\beta=4$}
\label{a:beta4}
\setcounter{equation}{0}

For the ease of the reader, we try to use the same notation as in \cite[\S 6.2 \& \S 6.4]{Forrester2010}, where the case $y= \infty$ is discussed in detail. Also note that all the quantities in this section depend on $y$, however we will suppress the explicit notation of such, to save space.

The goal is to write the $\beta=4$ skew-orthogonal polynomials $\{ Q_j \}$, defined by \eqref{d:IP4} and \eqref{d:4sorthog}, in terms of the polynomials orthogonal with respect to the inner product \eqref{d:IP2}, the NM polynomials $p_j$ which obey the relations \eqref{e:NM1}--\eqref{e:NM3}. However, as discussed in Section \ref{s:SOPs4} we will instead use the modified skew-inner product \eqref{d:modIP4}, and look for polynomials $\{\tilde{Q}_j \}$ that obey the relations \eqref{d:4modsorthoga} and \eqref{d:4modsorthogb}, up to the invariance \eqref{e:oddsymm}. Since the orthogonal polynomials form a complete set we can find coefficients $\tilde{\alpha}_{j,k}$ such that
\begin{align}
\label{e:pQ} \tilde{Q}_j = p_j + \tilde{\alpha}_{j,j-1}p_{j-1} + \dots + \tilde{\alpha}_{j,1}p_1 + \tilde{\alpha}_{j,0} p_0, \qquad \tilde{\alpha}_{j,k}\in \mathbb{C}.
\end{align}
Recall that the tilde $\tilde{~}$ means that the quantity is associated with this modified skew-inner product. From monicity and \eqref{e:oddsymm} we have
\begin{align}
\label{e:alpharules} \tilde{\alpha}_{j,j}=1, \qquad \tilde{\alpha}_{2j+1, 2j}=0.
\end{align}

We can write \eqref{e:pQ} in the matrix form
\begin{align}
\btQ = \btX \bp
\end{align}
where
\begin{align}
\label{d:tQp} \btQ= \left[ \begin{array}{c}
\tilde{Q}_0 \\
\tilde{Q}_1\\
\vdots
\end{array}
\right], \qquad \bp= \left[ \begin{array}{c}
p_0 \\
p_1\\
\vdots
\end{array}
\right]
\end{align}

\begin{align}
\label{d:Xmat} \btX = \left[
\begin{array}{cccccc}
1 & 0 & 0 & 0 & 0 & \cdots\\
0 & 1 & 0 & 0 & 0 & \cdots\\
\tilde{\alpha}_{2, 0} & \tilde{\alpha}_{2, 1} & 1 & 0 & 0 & \cdots\\
\tilde{\alpha}_{3, 0} & \tilde{\alpha}_{3, 1} & 0 & 1 & 0 & \cdots\\
\tilde{\alpha}_{4, 0} & \tilde{\alpha}_{4, 1} & \tilde{\alpha}_{4, 2} & \tilde{\alpha}_{4, 3} & 1 & \\
\vdots & \vdots & \vdots & \vdots &  & \ddots
\end{array}
 \right].
\end{align}
For the calculation, we will find it more convenient to work with the equation
\begin{align}
\label{e:xinvtQ} \bp= \btX^{-1} \btQ.
\end{align}
Since the skew-orthogonal polynomials will also form a complete set, we know that $\btX$ is invertible and we denote
\begin{align}
\label{d:Xinvmat} \btX^{-1}= \left[
\begin{array}{cccccc}
1 & 0 & 0 & 0 & 0 & \cdots\\
\tilde{\beta}_{2,0} & 1 & 0 & 0 & 0 & \cdots\\
\tilde{\beta}_{2, 0} & \tilde{\beta}_{2, 1} & 1 & 0 & 0 & \cdots\\
\beta_{3, 0} & \beta_{3, 1} & \beta_{3, 2} & 1 & 0 & \cdots\\
\tilde{\beta}_{4, 0} & \tilde{\beta}_{4, 1} & \tilde{\beta}_{4, 2} & \tilde{\beta}_{4, 3} & 1 & \\
\vdots & \vdots & \vdots & \vdots &  & \ddots
\end{array}
 \right]= \left[
\begin{array}{cccccc}
1 & 0 & 0 & 0 & 0 & \cdots\\
0 & 1 & 0 & 0 & 0 & \cdots\\
\tilde{\beta}_{2, 0} & \tilde{\beta}_{2, 1} & 1 & 0 & 0 & \cdots\\
\beta_{3, 0} & \beta_{3, 1} & 0 & 1 & 0 & \cdots\\
\tilde{\beta}_{4, 0} & \tilde{\beta}_{4, 1} & \tilde{\beta}_{4, 2} & \tilde{\beta}_{4, 3} & 1 & \\
\vdots & \vdots & \vdots & \vdots &  & \ddots
\end{array}
\right]
\end{align}
where we have used the assumptions analogous to \eqref{e:alpharules}
\begin{align}
\label{e:betarules} \tbeta_{j,j}=1, \qquad \tbeta_{2j+1, 2j}=0.
\end{align}
So instead of looking for the coefficients in \eqref{e:pQ} we will solve for the coefficients $\tbeta_{j,k}$ in
\begin{align}
\label{e:Qp} p_j = \tQ_j + \tbeta_{j,j-1} \tQ_{j-1} + \dots + \tbeta_{j,1} \tQ_1 + \tbeta_{j,0} \tQ_0, \qquad \tbeta_{j,k}\in \mathbb{C},
\end{align}
and then hope to invert the relations to recover the $\talpha_{j,k}$. We also define the matrix of inner products
\begin{align}
\btq = \Big[ \llangle \tQ_j, \tQ_k \rrangle_{4}^y \Big]= \left[
\begin{array}{cccccc}
0 & \tq_0 & 0 & 0 & 0 & \cdots\\
-\tq_0 & 0 & 0 & 0 & 0 & \cdots\\
0 & 0 & 0 & \tq_1 & 0 & \cdots\\
0 & 0 & -\tq_1 & 0 & 0 & \cdots\\
0 & 0 & 0 & 0 & 0 & \\
\vdots & \vdots & \vdots & \vdots &  & \ddots
\end{array}
 \right].
\end{align}

Using \eqref{e:IP4vsIP2} we can write the modified $\beta=4$ skew-inner product in terms of the $\beta=2$ inner product, with the inclusion of the operator $A$ defined in \eqref{d:Aop}. To make use of this we first note that if $f_k$ is any monic polynomial of degree $k$ then we have
\begin{align}
Af_k[x] &= -\left( xf_k (x) - f_k' (x) \right) = -\left( p_{k+1} (x) + \sum_{j=0}^{k-1} c_j p_{j} (x) \right),
\end{align}
where we have decomposed $xf_k (x) - f_k' (x)$ into a sum over the (monic) orthogonal polynomials $p_j$, with coefficients $c_j$. Combining this fact with \eqref{e:fAg}, \eqref{e:fAf} and the normalization of the $p_j$ from \eqref{e:ynorm2} we have the matrix
\begin{align}
\nonumber \bA&:= \left[ (p_j, A p_k )_{2}^y \right]_{j, k =0, ..., N -1}\\
&= \left[
\begin{array}{ccccccc}
\frac{\Omega_{0, 0}} {2} & h_1+ \Omega_{0, 1} & \Omega_{0, 2} & \Omega_{0, 3} & \Omega_{0, 4} & \Omega_{0, 5} &\cdots\\
-h_1 & \frac{\Omega_{1, 1}} {2} & h_2+ \Omega_{1, 2} & \Omega_{1, 3} & \Omega_{1, 4} & \Omega_{1, 5} &\cdots\\
0 & -h_2 & \frac{\Omega_{2, 2}} {2} & h_3+ \Omega_{2, 3} & \Omega_{2, 4} & \Omega_{2, 5} &\cdots\\
0 & 0 & -h_3 & \frac{\Omega_{3, 3}} {2} & h_4+ \Omega_{3, 4} & \Omega_{3, 5} &\cdots\\
0 & 0 & 0 & -h_4 & \frac{\Omega_{4, 4}} {2} & h_5+ \Omega_{4, 5} &\\
\vdots & \vdots & \vdots & \vdots & \ddots & \ddots & \ddots
\end{array}\right].
\end{align}

So now we can write
\begin{align}
\nonumber \mathbf{\tilde{q}}&:= \left[ \llangle \tilde{Q}_j, \tilde{Q}_k \rrangle_{4}^y \right]= \llangle[\Big]{\btQ \btQ^T}\rrangle[\Big]_{4}^y = \llangle[\Big]{\btX \bp \bp^T \btX^T}\rrangle[\Big]_{4}^y\\
\nonumber & = \btX \llangle[\Big]{\bp \bp^T}\rrangle[\Big]_{4}^y \btX^T = \btX \Big( \bA - \frac{1}{2} \left[\Omega_{j,k} \right] \Big) \btX^T\\
\label{e:xwxt}&= \btX \bW \btX^T,
\end{align}
where $\bW= \bA - \frac{1}{2} \left[\Omega_{j,k} \right]$ is the anti-symmetric matrix in \eqref{d:Wmat}. (Note that for a matrix $\bM$ the notation $\llangle \bM \rrangle_4^y$ implies that the average is applied elementwise to the matrix.) Rearranging \eqref{e:xwxt}
\begin{align}
\label{e:xinvq} \btX^{-1} \btq (\btX^{-1} )^T = \bW,
\end{align}
and expanding out the left hand side we get
\begin{align}
\nonumber \Big[ \btX^{-1} \btq (\btX^{-1})^T \Big]_{j,k} &= \sum_{m = 0, 1, \dots, j \atop n = 0, 1, \dots, k} \tbeta_{j,m} \tilde{\bq}_{m,n} \tbeta_{k,n}\\
\nonumber &=\sum_{m\, \mathrm{even}} \tbeta_{j, m} \tilde{\bq}_{m,m+1} \tbeta_{k,m+1}+ \sum_{m\, \mathrm{odd}} \tbeta_{j, m} \tilde{\bq}_{m,m-1} \tbeta_{k,m-1}\\
&=\sum_{m\, \mathrm{even}} \tbeta_{j, m} \tq_{m/2} \tbeta_{k,m+1} -\sum_{m\, \mathrm{odd}} \tbeta_{j, m} \tq_{(m-1)/2} \tbeta_{k,m-1},
\end{align}
noting that this is a finite sum since all $\beta_{\mu, \nu}$ are zero when $\nu> \mu$. So we have the set of equations
\begin{align}
\label{e:betaeqns} 0&= \sum_{m \, \mathrm{even}} \tq_{m/2} \left( \tbeta_{j,m} \tbeta_{k, m+1}- \tbeta_{j,m+1} \tbeta_{k, m} \right) - w_{j,k}
\end{align}
and we are now in a position to solve for the normalizations $\tq_j$ and the coefficients $\tbeta_{j,k}$.

\subsection{Expressions for $\tq_j$}

Let the matrices in \eqref{e:xinvq} be of size $2n \times 2n$. Then, taking the Pfaffian we get
\begin{align}
\Pf \bW = \Pf (\btX^{-1} \btq (\btX^T)^{-1})= \det (\btX^{-1}) \Pf \btq = \Pf \btq \;,
\end{align}
where we used the Pfaffian identity \eqref{e:PfBAB} for the second equality, and the fact that $\btX^{-1}$ is a triangular matrix with $1$s on the diagonal for the third equality.

Because $\btq$ is a skew-diagonal matrix, as in \eqref{d:sdiag}, we have
\begin{align}
\label{e:prodqj} \Pf \btq = \prod_{j=0}^{n-1} \tq_j = \Pf \bW_{2n-1}.
\end{align}
Beginning with $n=1$ and iterating, we obtain \eqref{e:qs}, with the convention \eqref{e:PfWm0}.

\subsection{Expressions for $\tbeta_{j,k}$}

Let $k$ be even, then the last term in the sum of \eqref{e:betaeqns} is $-\tbeta_{j,k+1} \tq_{k/2}$ (when $m= k$), and so solving for this $\tbeta$ we obtain
\begin{align}
\label{e:betake} \tbeta_{j,k+1} = \frac{1}{\tq_{k/2}} \left[ \sum_{m=0, \atop m\, \mathrm{even}}^{k-2} \tq_{m/2} \left(  \tbeta_{j,m} \tbeta_{k, m+1} - \tbeta_{j,m+1} \tbeta_{k, m}\right) -w_{j,k} \right], \qquad \mbox{$k$ even}
\end{align}
For $k$ odd the last term (when $m=k-1$) is $\tq_{(k-1)/2}( \tbeta_{j, k-1}  - \tbeta_{j,k} \tbeta_{k, k-1})$, but recall from \eqref{e:betarules} that (when $k$ is odd) we have set $\tbeta_{k, k-1} =0$ [using \eqref{e:oddsymm}], so we obtain
\begin{align}
\label{e:betako} \tbeta_{j, k-1} = -\frac{1}{\tq_{(k-1)/2}} \left[ \sum_{m=0, \atop m\, \mathrm{even}}^{k-3}  \tq_{m/2} \left(  \tbeta_{j,m} \tbeta_{k, m+1} - \tbeta_{j,m+1} \tbeta_{k, m}\right) -w_{j,k} \right], \qquad \mbox{$k$ odd}.
\end{align}

From these two expressions we see that each $\tbeta_{j,2k}$ and $\tbeta_{j, 2k+1}$ only depends on the $\tbeta$s in the same row, and in columns $0,1,\dots 2k-1$. This allows us to inductively solve for the $\tbeta$: first we solve for $\tbeta_{j,0}, \tbeta_{j,1}$, then $\tbeta_{j,2}, \tbeta_{j,3}$, etc.

It is this decoupling of the $\tbeta$ equations that are the reason for working with $\btX^{-1}$ instead of $\btX$.

\begin{proposition}\label{p:betas4}
\begin{align}
\label{e:betas} \tbeta_{j,k}&= \left\{ \begin{array}{cl}
\frac{{\displaystyle\Pf \bW_{k+1}^{(k\mapsto j)}}} {{\displaystyle\Pf \bW_{k+1}}},&\quad \mbox{$k$ even},\\
\frac{{\displaystyle\Pf \bW_{k}^{(k\mapsto j)}}} {{\displaystyle\Pf \bW_{k}}},&\quad \mbox{$k$ odd},
\end{array}
\right.
\end{align}
where $\bW_{\mu}^{(\eta \mapsto \nu)}$ is the matrix $\bW_{\mu}$ from \eqref{d:Wmat} with all occurrences of the index $\eta$ replaced by the index $\nu$.
\end{proposition}

\proof As mentioned above, we will employ an inductive proof. We need both even and odd base cases. Expanding out \eqref{e:betaeqns} with $k=0$ we have
\begin{align}
&0= -\tbeta_{j,1} \tq_0 \tbeta_{0,0} - w_{j,0}\\
&\Rightarrow \tbeta_{j,1}= -\frac{w_{j,0}} {\tq_0}= \frac{w_{0,j}} {w_{0,1}} = \frac{{\displaystyle\Pf \bW_{1}^{(1\mapsto j)}}} {{\displaystyle\Pf \bW_{1}}}.
\end{align}
Similarly, [recalling that $\tbeta_{1,0}=0$ from \eqref{e:betarules}] with $k=1$, we get
\begin{align}
&0= \tbeta_{j,0}\tq_0 \tbeta_{1,1}-  \tbeta_{j,1} \tq_0 \tbeta_{1,0} - w_{j,1}\\
&\Rightarrow \tbeta_{j,0}= \frac{w_{j,1}} {\tq_0} = \frac{{\displaystyle\Pf \bW_{1}^{(0\mapsto j)}}} {{\displaystyle\Pf \bW_{1}}}.
\end{align}

Now we move to the inductive step. For convenience, here we restrict to $k$ even. Assume that we have \eqref{e:betas} for all $\tbeta_{j,0}, \tbeta_{j,1}, \dots, \tbeta_{j,k-1}$ and we substitute \eqref{e:betas} and \eqref{e:qs} into \eqref{e:betake} to get
\begin{align}
\label{e:beta1} \tbeta_{j,k+1} = \frac{\Pf \bW_{k-1}} {\Pf \bW_{k+1}} \left[ \sum_{m=0, \atop m\, \mathrm{even}}^{k-2} \left( \frac{{\displaystyle\Pf \bW_{m+1}^{(m\mapsto j)}}} {\Pf \bW_{m-1}} \frac{{\displaystyle\Pf \bW_{m+1}^{(m+1 \mapsto k)}}} {{\displaystyle\Pf \bW_{m+1}}} - \frac{{\displaystyle\Pf \bW_{m+1}^{(m+1 \mapsto j)}}} {{\displaystyle\Pf \bW_{m+1}}} \frac{{\displaystyle\Pf \bW_{m+1}^{(m\mapsto k)}}} {\Pf \bW_{m-1}} \right) -w_{j,k} \right].
\end{align}
Using \cite[(1.1)]{Knuth1996} we obtain
\begin{align}
\nonumber &\Pf \bW_{m+1}^{(m\mapsto j)} \Pf \bW_{m+1}^{(m+1\mapsto k)}- \Pf \bW_{m+1}^{(m+1\mapsto j)} \Pf \bW_{m+1}^{(m\mapsto k)}\\
\label{e:KnuthPf1} &= \Pf \bW_{m-1} \Pf \bW_{m+3}^{(m+2\mapsto k, m+3\mapsto j)}- \Pf \bW_{m+1}\Pf \bW_{m+1}^{(m\mapsto k,m+1\mapsto j)}.
\end{align}
The notation in \cite{Knuth1996} is quite different to that used here, so we briefly outline how \eqref{e:KnuthPf1} follows from \cite[(1.1)]{Knuth1996}, which we quote here, rearranged for convenience
\begin{align}
\label{e:KnuthPf2} - f[\alpha x z] f[\alpha w y] + f[\alpha w z] f[\alpha x y] &= f[\alpha] f[\alpha w x y z]- f[\alpha w x] f[\alpha y z]
\end{align}
where $w,x,y,z \in \mathbb{Z}$ are matrix indices and $\alpha \in \mathbb{Z}^p$ is an ordered set of indices. For index sets $\alpha_1\in \mathbb{Z}^p, \alpha_2 \in \mathbb{Z}^q$ the product $\alpha_1 \alpha_2 \in \mathbb{Z}^{p+q}$ is the concatenation of the index sets. The function $f[\alpha]$ is then the Pfaffian of the matrix $\big[ f[jk] \big]$ with index set $\alpha$, i.e.
\begin{align}
\label{e:KnuthPf3} f[\alpha] = \Pf \Big[ f[jk] \Big]_{j,k \in \alpha}
\end{align}
defined recursively, where for a pair of indices $f[jk]$ is the matrix element, and
\begin{align}
\label{e:KnuthPf4} f[jk]=-f[kj]
\end{align}
since Pfaffian matrices are anti-symmetric. So then to match \eqref{e:KnuthPf2} with \eqref{e:KnuthPf1} we take
\begin{align}
\alpha=\{ 0, 1, \dots, m-1\}, \quad w=\{m\}, \quad x= \{m+1\}, \quad y=\{k\}, \quad z= \{j\},
\end{align}
and apply \eqref{e:KnuthPf4} to rearrange the indices as needed.

Substituting \eqref{e:KnuthPf1} into \eqref{e:beta1} we obtain
\begin{align}
\tbeta_{j,k+1} &= \frac{\Pf \bW_{k-1}} {\Pf \bW_{k+1}} \left[ \sum_{m=0, \atop m\, \mathrm{even}}^{k-2} \frac{{\displaystyle\Pf \bW_{m+3}^{(m+2 \mapsto k, m+3\mapsto j)}}} {{\displaystyle\Pf \bW_{m+1}}} - \frac{{\displaystyle\Pf \bW_{m+1}^{(m\mapsto k, m+1\mapsto j)}}} {\Pf \bW_{m-1}} -w_{j,k} \right],
\end{align}
which is a telescoping sum, leaving
\begin{align}
\tbeta_{j,k+1}&= \frac{\Pf \bW_{k-1}} {\Pf \bW_{k+1}} \left[ \frac{{\displaystyle\Pf \bW_{k+ 1}^{(k+1 \mapsto j)}}} {{\displaystyle\Pf \bW_{k-1}}} - \frac{{\displaystyle\Pf \bW_{1}^{(0\mapsto k, 1\mapsto j)}}} {\Pf \bW_{-1}} -w_{j,k} \right]\\
&= \frac{{\displaystyle\Pf \bW_{k+ 1}^{(k+1 \mapsto j)}}} {{\displaystyle\Pf \bW_{k+1}}}
\end{align}
since ${\displaystyle\Pf \bW_{1}^{(0\mapsto k, 1\mapsto j)}} = w_{k,j}= -w_{j,k}$, and we also used the convention \eqref{e:PfWm0}.

For the odd case, one proceeds from \eqref{e:betako} in a similar fashion.

\hfill $\Box$

Note from \eqref{e:betas} that
\begin{align}
\tbeta_{2n+1, 2n} = \frac{{\displaystyle\Pf \bW_{2n+1}^{(2n\mapsto 2n+1)}}} {{\displaystyle\Pf \bW_{2n+1}}},
\end{align}
and since $w_{n,n}=0$ for all $n$, then the Pfaffian in the numerator has two identical columns (the right-most) and two identical rows (the bottom-most), which implies
\begin{align}
\label{e:beta0} {\displaystyle\Pf \bW_{2n+1}^{(2n\mapsto 2n+1)}}=0 \qqimp \tbeta_{2n+1, 2n} =0.
\end{align}
Also, we clearly have
\begin{align}
\tbeta_{j,j}&= \left\{ \begin{array}{cl}
\frac{{\displaystyle\Pf \bW_{j+1}^{(j\mapsto j)}}} {{\displaystyle\Pf \bW_{j+1}}},&\quad \mbox{$j$ even},\\
\frac{{\displaystyle\Pf \bW_{j}^{(j\mapsto j)}}} {{\displaystyle\Pf \bW_{j}}},&\quad \mbox{$j$ odd},
\end{array}\right\} \quad= 1
\end{align}
so we recover \eqref{e:betarules}.

\subsection{Expressions for $\talpha_{j,k}$ in Proposition \ref{p:alpha4}} \label{a:Proofbeta4}

From the matrix product
\begin{align}
\btX^{-1} \btX =\bI
\end{align}
we have
\begin{align}
\label{e:alphas3} \talpha_{j,k}= - \sum_{m=k}^{j-1} \tbeta_{j,m} \talpha_{m,k} 
\end{align}
for $j>k$. Using this and the expressions for the $\tbeta_{j,k}$ in \eqref{e:betas} we can find expressions for the $\talpha_{j,k}$.

\vspace{10pt}\noindent\textit{Proof of Proposition \ref{p:alpha4}}:\; From \eqref{e:Qp} we have
\begin{align}
\nonumber \tQ_j&= p_j - \sum_{k=0}^{j-1} \tbeta_{j, k} \tQ_k\\
\label{e:tqpj} &=p_j - \tbeta_{j,j-1} \tQ_{j-1} - \sum_{k=0}^{j-2} \tbeta_{j, k} \tQ_k,
\end{align}
so $\talpha_{j,j} = \tbeta_{j,j}=1$. Then, with \eqref{e:4SOPvsOP}, this also implies
\begin{align}
\tbeta_{j,j-1} \tQ_{j-1} = \tbeta_{j,j-1} \left( p_{j-1} + \talpha_{j-1, j-2} p_{j-2}+ \dots \right),
\end{align}
and thus
\begin{align}
\tQ_j&= p_j -\tbeta_{j,j-1} p_{j-1} - \mbox{lower degree polynomials}.
\end{align}
So we have that
\begin{align}
\label{e:talphatbeta} \talpha_{j,j-1}= -\tbeta_{j, j-1},
\end{align}
which is equal to zero when $j$ is odd by \eqref{e:betarules}. Now we have consistency with both \eqref{e:alpha04} and \eqref{e:alphajj}.

For \eqref{e:alphas4} we will use an inductive proof similar to that used in Proposition \ref{p:betas4}. We see from \eqref{e:alphas3} that each $\talpha_{j,k}$ only depends on the $\tbeta$'s (which are known) and the $\talpha$'s above it in the same column of the matrix $\btX$ [in \eqref{d:Xmat}]. From \eqref{e:talphatbeta} we have
\begin{align}
\label{e:alphaPf1} \talpha_{j,j-1} = - \tbeta_{j, j-1}= \left\{ \begin{array}{cl}
- \frac{{\displaystyle\Pf \bW_{j-1}^{(j-1\mapsto j)}}} {{\displaystyle\Pf \bW_{j-1}}}, &\quad \mbox{$j$ even},\\
0, &\quad \mbox{$j$ odd},
\end{array}\right.
\end{align}
and from \eqref{e:alphas3}
\begin{align}
\label{e:alphaPf2} \talpha_{j,j-2} = -\tbeta_{j,j-2}\talpha_{j-2, j-2} - \tbeta_{j, j-1} \talpha_{j-1, j-2}= -\tbeta_{j,j-2} = \left\{ \begin{array}{ll}
- \frac{{\displaystyle\Pf \bW_{j-1}^{(j-1\mapsto j)}}} {{\displaystyle\Pf \bW_{j-1}}}, &\quad \mbox{$j$ even},\\
- \frac{{\displaystyle\Pf \bW_{j-2}^{(j-2\mapsto j)}}} {{\displaystyle\Pf \bW_{j-2}}}, &\quad \mbox{$j$ odd},
\end{array}\right.
\end{align}
since one of $\tbeta_{j, j-1}$ or $\talpha_{j-1, j-2}$ must be zero by \eqref{e:betarules} or \eqref{e:alphaPf1}. The equations \eqref{e:alphaPf1} and \eqref{e:alphaPf2} give us expressions for all $\talpha$s on the first and second lower diagonals of $\btX$. So for any column $k$, there is a row $j$ for which all the $\talpha_{j- m, k}$ above it are known, so we have our base cases.

Now for the inductive step, we expand \eqref{e:alphas3} to obtain
\begin{align}
\label{e:alphas1} \talpha_{j,k}&= \left\{ \begin{array}{ll}
-\tbeta_{j,k} -{\displaystyle \sum_{m=k+2}^{j-1} \tbeta_{j,m} \talpha_{m,k}}, & \qquad \mbox{$j$ even, $k\leq j-1$, $k$ even},\\
-\tbeta_{j,k} -{\displaystyle \sum_{m=k+1}^{j-1} \tbeta_{j,m} \talpha_{m,k}}, & \qquad \mbox{$j$ even, $k\leq j-1$, $k$ odd},\\
-\tbeta_{j,k} -{\displaystyle \sum_{m=k+2}^{j-2} \tbeta_{j,m} \talpha_{m,k}}, & \qquad \mbox{$j$ odd, $k\leq j-1$, $k$ even},\\
-\tbeta_{j,k} -{\displaystyle \sum_{m=k+1}^{j-2} \tbeta_{j,m} \talpha_{m,k}}, & \qquad \mbox{$j$ odd, $k\leq j-1$, $k$ odd}.\\
\end{array}\right.
\end{align}
We assume that $\talpha_{m,k}$ is given by \eqref{e:alphas4} for all $m\leq j-1$ ($j$ even) or $m\leq j-2$ ($j$ odd), while all $\tbeta$s are given by \eqref{e:betas}. Taking $j,k$ both even (the other cases follow similarly), we substitute these known $\talpha$'s and $\tbeta$'s into the first row of \eqref{e:alphas1} to give
\begin{align}
\nonumber \talpha_{j,k} &= - \frac{{\displaystyle\Pf \bW_{k+1}^{(k\mapsto j)}}} {{\displaystyle\Pf \bW_{k+1}}} + \sum_{m=k+2 \atop m\, \mathrm{even}}^{j-2} \frac{{\displaystyle\Pf \bW_{m+1}^{(m\mapsto j)}}} {{\displaystyle\Pf \bW_{m+1}}} \frac{{\displaystyle\Pf \bW_{m-1}^{(k\mapsto m)}}} {{\displaystyle\Pf \bW_{m-1}}}+ \sum_{m=k+3 \atop m\, \mathrm{odd}}^{j-1} \frac{{\displaystyle\Pf \bW_{m}^{(m\mapsto j)}}} {{\displaystyle\Pf \bW_{m}}} \frac{{\displaystyle\Pf \bW_{m-2}^{(k\mapsto m)}}} {{\displaystyle\Pf \bW_{m-2}}}\\
\label{e:alphas2}&=- \frac{{\displaystyle\Pf \bW_{k+1}^{(k\mapsto j)}}} {{\displaystyle\Pf \bW_{k+1}}} + \sum_{m=k+2 \atop m\, \mathrm{even}}^{j-2} \frac{{\displaystyle\Pf \bW_{m+1}^{(m\mapsto j)}}} {{\displaystyle\Pf \bW_{m+1}}} \frac{{\displaystyle\Pf \bW_{m-1}^{(k\mapsto m)}}} {{\displaystyle\Pf \bW_{m-1}}}+ \frac{{\displaystyle\Pf \bW_{m+1}^{(m+1 \mapsto j)}}} {{\displaystyle\Pf \bW_{m+1}}} \frac{{\displaystyle\Pf \bW_{m-1}^{(k\mapsto m+1)}}} {{\displaystyle\Pf \bW_{m-1}}},
\end{align}
keeping in mind that we have the convention that $\Pf \bW_{-1} =1$.

We now use \cite[(5.1)]{Knuth1996} (again quoted here and rearranged for convenience)
\begin{align}
\nonumber  f[\alpha xuw] f[\alpha vyz] -f[\alpha xuv] f[\alpha wyz] &= -f[\alpha u v w] f[\alpha x y z] + f[\alpha uyz] f[\alpha xvw]\\
\label{e:Knuth2} & +f[\alpha z] f[ \alpha u v w x y] - f[\alpha y] f[\alpha uvwxz]
\end{align}
with
\begin{align}
\alpha= \{0, 1, \dots, k-1, k+1, \dots, m-1 \}, \quad &x= \{ j\}, \quad u= \{ k\}, \quad v= \{ m\}, \quad w= \{ m+1\}\\
&y=z=\emptyset \qquad \mbox{(the empty set)}.
\end{align}
Rearranging indices according to \eqref{e:KnuthPf4}, the equality \eqref{e:Knuth2} gives
\begin{align}
&\Pf \bW_{m+1}^{(m\mapsto j)} \Pf \bW_{m-1}^{(k\mapsto m)}+ \Pf \bW_{m+1}^{(m+1\mapsto j)} \Pf \bW_{m-1}^{(k\mapsto m+1)}\\
&= \Pf \bW_{m+1} \Pf \bW_{m-1}^{(k\mapsto j)}- \Pf \bW_{m-1}\Pf \bW_{m+1}^{(k\mapsto j)},
\end{align}
and substituting into \eqref{e:alphas2} we get
\begin{align}
\talpha_{j,k} &=- \frac{{\displaystyle\Pf \bW_{k+1}^{(k\mapsto j)}}} {{\displaystyle\Pf \bW_{k+1}}} + \sum_{m=k+2 \atop m\, \mathrm{even}}^{j-2} \frac{\Pf \bW_{m-1}^{(k\mapsto j)}} {{\displaystyle\Pf \bW_{m-1}}} - \frac{\Pf \bW_{m+1}^{(k\mapsto j)}} {{\displaystyle\Pf \bW_{m+1}}}.
\end{align}
This is a telescoping sum, which reduces to \eqref{e:alphas4}. The other cases in \eqref{e:alphas1} are calculated similarly.

\hfill $\Box$

\subsection{$\beta =4$ polynomials in the classical limit}
\label{a:limpolys4}

In the classical limit ($y\to \infty$) the skew inner product \eqref{d:IP4} becomes
\begin{align}
\langle f, g \rangle_4:= \frac{1}{2} \int_{- \infty}^{\infty} dx \; e^{-2 x^2} \left[ f(x) g'(x)- g(x) f'(x) \right],
\end{align}
and the associated skew-orthogonal polynomials obeying
\begin{align}
\nonumber \langle Q_{2j} , Q_{2k} \rangle_{4}&= \langle Q_{2j+1} , Q_{2k+1} \rangle_{4}= 0\\
\langle Q_{2j} , Q_{2k+1} \rangle_{4}&= -\langle Q_{2k+1} , Q_{2j} \rangle_{4} = q_j \delta_{j,k}
\end{align}
are given by \cite{NagaWada1991,AdleForrNagavanMoer2000}
\begin{align}
\nonumber Q_{2j+1}(x) &= p_{2j+1} (\sqrt{2} x), &Q_{2j}(x) &= \sum_{t=0}^{j} \left( \prod_{s=t+1}^{j} \frac{h_{2s}}{h_{2s-1}} \right) p_{2t} (\sqrt{2} x)\\
\label{e:LimPolys4} &&&= \sum_{t=0}^{j} \frac{j!}{t!}\; p_{2t} (\sqrt{2} x)
\end{align}
[up to the invariance \eqref{e:oddsymm}], where the polynomials
\begin{align}
\label{e:monHerms} p_{j} (x) = \frac{1}{2^j} H_{j} (x)
\end{align}
are the (monic, ``physicist's'') Hermite polynomials in \eqref{e:limHerms} and $h_j = h_{j} (\infty)$ from \eqref{e:InfinityNorms}. The corresponding normalizations $q_j = q_j(\infty)$ are also from \eqref{e:InfinityNorms}.

As mentioned after Proposition \ref{p:alpha4}, it can be seen that the results of that Proposition reduce to the classical polynomials \eqref{e:LimPolys4}, since in the limit $y\to \infty$ the matrix $[\Omega_{j,k}]=0$ in \eqref{e:xwxt}, and we then follow exactly the steps in \cite{AdleForrNagavanMoer2000} to obtain \eqref{e:LimPolys4}.

\section{Skew-orthogonal polynomials for $\beta=1$}
\label{a:beta1}
\setcounter{equation}{0}

We again suppress the explicit dependence on $y$ to save space, although all the quantities here depend on $y$.

We can follow the same steps as for the $\beta=4$ case in Appendix \ref{a:beta4} to find the coefficients $\alpha_{j,k}$ in \eqref{e:Rjalpha1}. With $\bp$ from \eqref{d:tQp} we first rewrite equation \eqref{e:Rjalpha1} as
\begin{align}
\bR = \bX \bp \qqimp \bp = \bX^{-1} \bR,
\end{align}
where 
\begin{align}
\bR= \left[ \begin{array}{c}
R_0 \\
R_1\\
\vdots
\end{array}
\right],
\end{align}
and $\bX$ and $\bX^{-1}$ are the same as in \eqref{d:Xmat} and \eqref{d:Xinvmat}, but without the tildes. Also define the matrices
\begin{align}
\br &:= \big[ \langle R_j, R_k \rangle_1^y \big]_{j,k = 0, 1, \dots, N-1},\\
\label{d:Bmat} \bB &:= [(p_j, A^{-1} p_k)]_{j,k = 0, 1, \dots, N-1},\\
\label{d:bPhi} \boldsymbol{\Phi} &:= [\Phi_{j,k}]_{j,k = 0, 1, \dots, N-1},
\end{align}
where $\br$ is of skew-diagonal form \eqref{d:sdiag}. Then
\begin{align}
\nonumber \br&= \Big[ \langle R_j, R_k \rangle_{1}^{y} \Big]= \left\langle \bR \bR^T \right\rangle_{1}^{y}= \left\langle \bX \bp \bp^T \bX^T \right\rangle_{1}^{y}\\
\nonumber & =\bX \left\langle \bp \bp^T \right\rangle_{1}^{y} \bX^T\\
\nonumber &= -\bX \Big( \bB + \boldsymbol{\Phi} \Big) \bX^T\\
\label{e:xvxt}&= \bX \bV \bX^T,
\end{align}
where the anti-symmetric matrix $\bV$ is defined in \eqref{d:Vm} --- we will discuss the derivation of the specific structure of the elements of $\bV$ in Appendix \ref{a:VmElts} below. (As above, the averages over matrix arguments imply that the average is applied elementwise to the matrix.)

We now follow the same steps as in \eqref{e:xinvq}--\eqref{e:betaeqns} to get
\begin{align}
\label{e:XrXinv} &\bX^{-1} \br \left( \bX^{-1} \right)^T = \bV\\
\label{e:betaeqns1} \qqimp & \sum_{m \, \mathrm{even}} r_{m/2} \left( \beta_{j,m} \beta_{k, m+1}- \beta_{j,m+1} \beta_{k, m} \right) - v_{j,k} =0
\end{align}
with $\bV = \bV_m= [v_{j,k}]_{j,k=0,\dots, m}$ from \eqref{d:Vm}. Assuming $m=2n$, taking the Pfaffian of \eqref{e:XrXinv} we have
\begin{align}
\Pf \br = \prod_{j=0}^{n-1} r_j = \Pf \bV_{2n-1}
\end{align}
and we obtain \eqref{e:rs}, with the convention \eqref{e:PfVm0}.

Then, since the equations in \eqref{e:betaeqns1} are of the same form as \eqref{e:betaeqns}, we apply the same reasoning as that in Proposition \ref{p:betas4} to obtain solutions for the $\beta_{j,k}$
\begin{align}
\beta_{j,k}&= \left\{ \begin{array}{cl}
\frac{{\displaystyle\Pf \bV_{k+1}^{(k\mapsto j)}}} {{\displaystyle\Pf \bV_{k+1}}},&\quad \mbox{$k$ even},\\
\frac{{\displaystyle\Pf \bV_{k}^{(k\mapsto j)}}} {{\displaystyle\Pf \bV_{k}}},&\quad \mbox{$k$ odd},
\end{array} \right.
\end{align}
where again, $\bV_{\mu}^{(\eta \mapsto \nu)}$ is the matrix $\bV_{\mu}$ with all occurrences of the index $\eta$ replaced by the index $\nu$. Now using the equations
\begin{align}
\bX^{-1} \bX = \bI \qqimp \alpha_{j,k} = -\sum_{m=k}^{j-1} \beta_{j,m} \alpha_{m,k},
\end{align}
we follow the same steps as in Appendix \ref{a:Proofbeta4} and we establish the remaining statements in Proposition \ref{p:alpha1}.

\subsection{Entries of the matrix $\bV_m$}
\label{a:VmElts}

For a general polynomial
\begin{align}
p_{j} (x) = c_{j,j} x^j + c_{j, j-1} x^{j-1} + \dots + c_{j,1} x + c_{j,0}
\end{align}
we use the identities (calculated via repeated integration by parts)
\begin{align}
\label{e:Ointeg} \int_{a}^{b} e^{-u^2/2} u^{2k+1} du &= (2k)!! \left( \sum_{m=0}^k \frac{e^{-a^2/2} a^{2m} - e^{-b^2/2} b^{2m}}{(2m)!!} \right)\\
\nonumber \int_{a}^{b} e^{-u^2/2} u^{2k} du &= (2k-1)!! \left( \sum_{m=1}^k \frac{e^{-a^2/2} a^{2m-1} - e^{-b^2/2} b^{2m-1}}{(2m-1)!!} \right)\\
\label{e:Einteg}&+ (2k-1)!! \sqrt{\frac{\pi}{2}} \left( \erf \left( \frac{b}{\sqrt{2}} \right)- \erf \left( \frac{a}{\sqrt{2}} \right) \right)
\end{align}
to obtain
\begin{align}
\int_{-\infty}^{\infty} e^{-z^2/2} p_k (z) dz = \sqrt{2 \pi} \sum_{t=0}^{\lfloor k/2 \rfloor} c_{k, 2t} (2t -1)!!
\end{align}
and
\begin{align}
A^{-1}p_{k} [z] &= \left( \frac{e^{x^2/2}}{2}\; \erf\left( \frac{x}{\sqrt{2}} \right) \int_{-\infty}^{\infty} e^{-z^2/2} p_{k} (z) dz \right)  - p_{k-1} (x) - (\mbox{lower order polynomials}).
\end{align}
So then, with $\bB$ defined in \eqref{d:Bmat}, we have
\begin{align}
\nonumber &\bB =
\begin{bmatrix}
-\Phi_{0,0} & -h_0 + X_{0,1} & ?? & ??&\\
X_{1,0} & -\Phi_{1,1} & -h_1 + X_{1,2} & ??& \cdots\\
X_{2,0} & X_{2,1} & -\Phi_{2,2} & \\
&\vdots &&\ddots
\end{bmatrix}\\
&= \begin{bmatrix}
-\Phi_{0,0} & -h_0 + X_{0,1} & ?? & ??&\\
h_0 -X_{0,1} -\Phi(0,1) -\Phi(1,0) & -\Phi_{1,1} & -h_1 + X_{1,2} & ??& \cdots\\
X_{2,0} & h_1 -X_{1,2} -\Phi(1,2) -\Phi(2,1) & -\Phi_{2,2} & &\\
&\vdots&&\ddots
\end{bmatrix},
\end{align}
where the $??$ represents currently unknown elements, and the second equality comes from the use of \eqref{e:AinvSymm}. Adding the matrix $\boldsymbol{\Phi}$ from \eqref{d:bPhi} gives the (negative of the) anti-symmetric matrix $\bV$ from \eqref{d:Vm}, allowing us to specify the $??$ as so
\begin{align}
\nonumber &\bB +\boldsymbol{\Phi}= -\bV =\\
&\begin{bmatrix}
0 & -h_0 + X_{0,1} +\Phi(0,1) & -X_{2,0} - \Phi(2,0) & -X_{3,0} - \Phi(3,0)& \\
h_0 -X_{0,1} -\Phi(0,1) & 0 & -h_1 + X_{1,2} +\Phi(1,2) & -X_{3,1} - \Phi(3,1)& \cdots\\
X_{2,0} + \Phi(2,0) & h_1 -X_{1,2} -\Phi(1,2) & 0 & -h_2 + X_{2,3} +\Phi(2,3)&\\
X_{3,0} + \Phi(3,0) & X_{3,1}+ \Phi(3,1) & h_2 -X_{2,3} -\Phi(2,3) & 0&\\
&\vdots &&&\ddots
\end{bmatrix}.
\end{align}

\subsection{$\beta =1$ polynomials in the classical limit}
\label{a:limpolys1}

Similar to Appendix \ref{a:limpolys4} above we have the $y \to \infty$ limit of the skew-inner product \eqref{d:IP1} as
\begin{align}
\langle f, g \rangle_1 = \frac1{2} \int_{-\infty}^{\infty} dx \; e^{-x^2/2} f(x) \int_{-\infty}^{\infty} d z \; e^{-z^2/2} g (z) \sgn (z- x),
\end{align}
with the associated skew-orthogonal polynomials obeying the equations
\begin{align}
\nonumber \langle R_{2j} , R_{2k} \rangle_{1}&= \langle R_{2j+1} , R_{2k+1} \rangle_{1}= 0\\
\langle R_{2j} , R_{2k+1} \rangle_{1}&=-\langle R_{2k+1} , R_{2j} \rangle_{1} = r_j \delta_{j,k}.
\end{align}
These polynomials are given explicitly [up to the invariance \eqref{e:oddsymm}] by \cite{NagaWada1991,AdleForrNagavanMoer2000}
\begin{align}
\nonumber R_{2j} (x) &= p_{2j} (x), &R_{2j+1} (x) &= p_{2j+1} (x) - \frac{h_{2j}} {h_{2j-1}} p_{2j-1} (x)\\
\label{e:limHerms1} &&&= p_{2j+1} (x) - j\, p_{2j-1} (x),
\end{align}
where the polynomials $p_j (x)$ are the Hermite polynomials in \eqref{e:monHerms} and $h_j = h_j (\infty)$. The normalizations $r_j = r_j (\infty)$ are from \eqref{e:InfinityNorms}.

To check coherence between \eqref{e:alphas1b} and \eqref{e:limHerms1} we can use integration by parts, the identities \eqref{e:Hermids} and
\begin{align}
\frac{d}{dx} \erf\left( \frac{x}{\sqrt{2}} \right) &= \sqrt{\frac{\pi}{2}} e^{-x^2/2}
\end{align}
to give us
\begin{align}
\int_{-\infty}^{\infty} e^{-x^2/2} H_j (x) \erf\left( \frac{x}{\sqrt{2}} \right) dx = \left\{ \begin{array}{cl}
2^{(j+2)/2} (j-1)!! , \quad& \mbox{$j$ odd},\\
0,& \mbox{$j$ even}.
\end{array}\right.
\end{align}
Substitution into \eqref{d:Xjk} yields
\begin{align}
\nonumber X_{j,k} \Big|_{y \to \infty} &=  \frac{1}{2^{j+k+1}} \left( \int_{-\infty}^{\infty} H_j (x) e^{-x^2/2}\; \erf\left( \frac{x}{\sqrt{2}} \right) dx \right) \int_{-\infty}^{\infty} e^{-z^2/2} H_k (z) dz\\
\label{e:XjkLim} &= \left\{ \begin{array}{cl}
\Gamma \left( \frac{j+1}{2} \right) \Gamma \left( \frac{k+1}{2} \right) = h_k (\infty) \frac{\Gamma \left( \frac{j+1}{2} \right)}{\Gamma \left( \frac{k+2}{2} \right)}, \quad& \mbox{$j$ odd} \wedge \mbox{$k$ even},\\
0& \mbox{otherwise},
\end{array}\right.
\end{align}
where we used \eqref{e:Hermint} for the integral over $H_k$. The second line (equalling zero) follows easily from the fact that the error function is an odd function and that $H_j(x)$ is an even or odd function depending on the parity of $j$. We will also make use of the formula
\begin{align}
\label{e:hjDupe} h_j (\infty) = \Gamma \left( \frac{j+1}{2} \right) \Gamma \left( \frac{j+2}{2} \right),
\end{align}
which can be shown via Legendre's duplication formula for gamma functions.

\begin{figure}
\begin{center}
\resizebox{0.75\textwidth}{!}{\begin{tikzpicture}[baseline=(current  bounding  box.center)] %
\draw (-0.5,0)--(8.5,0); %
\node[below] at (0,0) {$0$}; %
\node[above] at (0.5,0.5) {$h_0$};
\node[below] at (1,0) {$1$}; %
\node[below] at (2,0) {$2$}; %
\node[above] at (2.5,0.5) {$h_2$};
\node[below] at (3,0) {$3$}; %
\node[below] at (4,0.5) {\Large $\dots$}; %
\node[below] at (5,0) {$2j\!\!-\!\!4$}; %
\node[above] at (5.5,0.5) {$h_{2j-4}$};
\node[below] at (6,0) {$2j\!\!-\!\!3$}; %
\node[below] at (7,0) {$2j\!\!-\!\!2$}; %
\node[above] at (7.5,0.5) {$h_{2j-2}$};
\node[below] at (8,0) {$2j\!\!-\!\!1$}; %
\draw[smooth] (0,0) to[out=60,in=180] (0.5,0.33) to[out=0,in=120] (1,0); %
\draw[smooth] (2,0) to[out=60,in=180] (2.5,0.33) to[out=0,in=120] (3,0); %
\draw[smooth] (5,0) to[out=60,in=180] (5.5,0.33) to[out=0,in=120] (6,0); %
\draw[smooth] (7,0) to[out=60,in=180] (7.5,0.33) to[out=0,in=120] (8,0); %
\end{tikzpicture}
}
\end{center}
\caption{\label{f:denomLinks} The only possible link diagram satisfying the conditions that every even site connects to the right and every odd site connects to the left is the identity diagram, where every link is a little link. The corresponding matrix entry is written above each link.}
\end{figure}
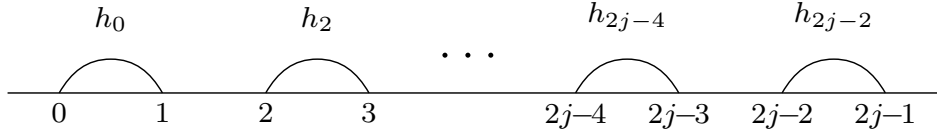

In the case that $y= \infty$ then from \eqref{d:Phijk} the function $\Phi_{j,k}=0$ and we also use \eqref{e:XjkLim} to find that the matrix $\bV_m$ in \eqref{d:Vm} has entries
\begin{align}
\label{e:VmLim} \bV_m= 
\begin{bmatrix}
0&h_0 &0 &X_{3,0}&0&X_{5,0}&\\
-h_0& 0& 0& 0& 0&0& \\
0&0&0&h_2 & 0& X_{5,2}& \cdots\\
-X_{3,0}&0 &-h_2& 0& 0&0&\\
0&0&0&0&0& h_4\\
-X_{5,0}& 0& -X_{5,2}& 0& -h_4& 0\\
&&\vdots &&&&\ddots
\end{bmatrix}
\end{align}
meaning
\begin{align}
v_{j,k} = \left\{ \begin{array}{cl}
h_j, &\quad \mbox{$j$ even $\wedge$ $k=j+1$},\\
0,&\quad \mbox{$j$ odd $\vee$ $k$ even},\\
X_{k,j}, &\quad \mbox{$j$ even $\wedge$ $k$ odd $\wedge$ $j<k-1$},
\end{array} \right.
\end{align}
with the anti-symmetry condition
\begin{align}
\label{e:VmLimAS} v_{j,k}= -v_{k,j}.
\end{align}
So $\bV_m$ is a sparse chequerboard matrix (as in \cite[Eqn. (6.4)]{AdlevanMoer2002}), and in particular, the second row of $\bV_m$ has the structure
\begin{align}
[-h_0\; 0\; 0 \;\dots \;0 ].
\end{align}
This latter fact tells us that if we perform a Pfaffian Laplace expansion (as discussed in Appendix \ref{a:Pfaffs} above) along the first row (with $j=0$), then the Pfaffian minors $M_{1,k}$ have a first row entirely of zeros, except when $k=2$. Since $v_{j,k}=X_{k,j} =0$ for all odd $j$ (with $j>k$), this patterns repeats for all the Pfaffian sub-minors and so
\begin{align}
\label{e:PfVm1} \Pf \bV_{2j-1} = h_0 h_2 \cdots h_{2j-2},
\end{align}
which gives us the denominator of $\alpha_{j,k}$ in \eqref{e:alphas1b}. We can also understand this via the definition in terms of perfect matchings in \eqref{e:PfPerfMatch}: the structure of the upper triangle of the matrix in \eqref{e:VmLim} tells us that $v_{j,k}=0$ unless $j$ is even and $k$ is odd, meaning that all even sites in the link diagram connect to the right and all odd sites connect to the left. The only possible diagram satisfying this condition is the identity link pattern in Figure \ref{f:denomLinks}, which corresponds to the product in \eqref{e:PfVm1}.

For the numerator of $\alpha_{j,k}$ we have four cases to consider, being the four possibilities given by the parities of $j$ and $k$.

\noindent\underline{$\alpha_{2j, 2k}$}:

\medskip In the $2k$-th column we have the matrix entries
\begin{align}
v_{s, 2k} \mapsto v_{s, 2j} = X_{2j, s} =0 \qquad (s< 2k)
\end{align}
while in the $2k$-th row we have
\begin{align}
v_{2k, t} \mapsto v_{2j, t} = -X_{2j, t} =0 \qquad (2k<t)
\end{align}
so we have zeros above and to the right of the $(2k, 2k)$ entry (in the same column and row), which gives us
\begin{align}
\label{e:PfLim1} \Pf \bV_{2j-1}^{(2k\mapsto 2j)} =0,
\end{align}
since at least one of these zero factors must appear in each term of the Pfaffian.

\noindent\underline{$\alpha_{2j, 2k+1}$}:

\medskip Similar to the above, we have
\begin{align}
v_{s, 2k+1} \mapsto v_{s, 2j} = X_{2j, s} =0 \qquad (s< 2k+1)
\end{align}
and
\begin{align}
v_{2k+1, t} \mapsto v_{2j, t} = -X_{2j, t} =0 \qquad (2k+1 < t).
\end{align}
So now we have zeros above and to the right of the $(2k+1, 2k+1)$ entry, which gives us
\begin{align}
\label{e:PfLim2} \Pf \bV_{2j-1}^{(2k+1 \mapsto 2j)} =0.
\end{align}

\noindent\underline{$\alpha_{2j+1, 2k}$}:

\medskip Now we have
\begin{align}
v_{s, 2k} \mapsto v_{s, 2j+1} = X_{2j+1, s} =0 \qquad (s< 2k \wedge \mbox{$s$ odd})
\end{align}
so we still have every odd row containing only zeros (in the upper triangle). Thus, as in \eqref{e:PfVm1}, the only term in the Laplace expansion that could be non-zero is $h_0 h_2 \cdots h_{2j-3}$. However,
\begin{align}
h_{2k} = v_{2k, 2k+1} \mapsto v_{2j+1, 2k+1} = -X_{2j+1, 2k+1} =0,
\end{align}
and so
\begin{align}
\label{e:PfLim3} \Pf \bV_{2j-1}^{(2k \mapsto 2j+1)} =0.
\end{align}

\noindent\underline{$\alpha_{2j+1, 2k+1}$}:

\medskip Using the expressions \eqref{e:XjkLim} and \eqref{e:hjDupe} we have the identity
\begin{align}
\label{e:Xhid} X_{2m+1, 2t} X_{2t+1, 2n} = h_{2t} X_{2m+1, 2n}, \qquad (m> t > n),
\end{align}
which will make use of below. First we recall from \eqref{e:VmLim} that in the upper triangle $v_{j,k} \neq 0$ only when $j$ is even and when $k$ is odd, which implies that all the even sites in the corresponding diagram connect to the right, and all the odd sites connect to the left. However, we will have an exception to this when we make the replacement $2k+1 \mapsto 2j+1$. Specifically, in terms of link diagrams there are two possibilities for the links involving site $2j+1$: either $(2s, 2j+1)$ or $(2j+1, 2t)$ (so $2j+1$ is either the right or left vertex of the link). We note that the other vertex must be even, since any odd-odd or even-even link results in $X_{\mathrm{odd}, \mathrm{odd}} = 0 = X_{\mathrm{even}, \mathrm{even}}$. It is easiest to consider the two cases separately:
\begin{itemize}
\item[(i)]{Assume $2j+1$ connects to the left, that is we have a link $(2s, 2j+1)$. Since all other odd sites connect left and all other even sites connect right, this must be the identity link diagram, similar to Figure \ref{f:denomLinks}, so $s=j$.}

\item[(ii)]{Assume $2j+1$ connects to the right, that is we have a link $(2j+1, 2t)$, then we must have identity links at sites to the left of $2k$ and to the right of $2t+1$, as depicted in Figure \ref{f:linksii1}. [The left-pointing arrow on the edge $(2j+1, 2t)$ indicates that the left vertex is greater than the right vertex, which is the opposite convention to all the other links, and this introduces a negative sign from \eqref{e:VmLimAS}.] In this case, we see from the diagram that there are 2 possible connections for $2t-2$, and then another 2 possible connections for $2t-4$, and so on. Thus there are $2^{t-k-1}$ link diagrams corresponding to Figure \ref{f:linksii1}}.
\end{itemize}
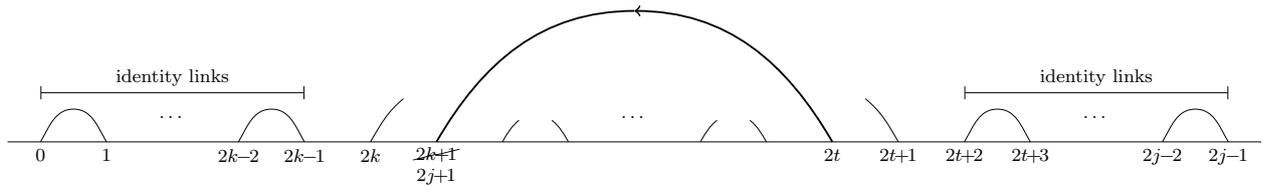
\begin{figure}
\begin{center}
\resizebox{\textwidth}{!}{\begin{tikzpicture}[baseline=(current  bounding  box.center)] %
\draw (-0.5,0)--(18.5,0); %
\node[below] at (0,0) {$0$}; %
\node[below] at (1,0) {$1$}; %
\node[below] at (2,0.5) {$\dots$}; %
\node[below] at (3,0) {$2k\!\!-\!\!2$}; %
\node[below] at (4,0) {$2k\!\!-\!\!1$}; %
\node[below] at (5,0) {$2k$}; %
\node[below] at (6,0) {$\cancel{2k\!\!+\!\!1}$}; %
\node[below] at (6,-0.3) {$2j\!\!+\!\!1$}; %
\node[below] at (9,0.5) {$\dots$}; %
\node[below] at (12,0) {$2t$}; %
\node[below] at (13,0) {$2t\!\!+\!\!1$}; %
\node[below] at (14,0) {$2t\!\!+\!\!2$}; %
\node[below] at (15,0) {$2t\!\!+\!\!3$}; %
\node[below] at (16,0.5) {$\dots$}; %
\node[below] at (17,0) {$2j\!\!-\!\!2$}; %
\node[below] at (18,0) {$2j\!\!-\!\!1$}; %
\draw[smooth] (0,0) to[out=60,in=180] (0.5,0.5) to[out=0,in=120] (1,0); %
\draw[smooth] (3,0) to[out=60,in=180] (3.5,0.5) to[out=0,in=120] (4,0); %
\node[above] at (2,0.75) {identity links};
\draw[|-|] (0,0.75) -- (4,0.75);
\draw[smooth] (5,0) to[out=60,in=220] (5.5,0.66); %
\draw[smooth,thick] (6,0) to[out=60,in=180] (9,2); %
\draw[smooth,thick,<-] (9,2) to[out=0,in=120] (12,0); %
\draw[smooth] (7,0) to[out=60,in=220] (7.25,0.33);
\draw[smooth] (7.75,0.33) to[out=320,in=120] (8,0);
\draw[smooth] (10,0) to[out=60,in=220] (10.25,0.33);
\draw[smooth] (10.75,0.33) to[out=320,in=120] (11,0);
\draw[smooth] (12.5,0.66) to[out=320,in=120] (13,0); %
\draw[smooth] (14,0) to[out=60,in=180] (14.5,0.5) to[out=0,in=120] (15,0); %
\draw[smooth] (17,0) to[out=60,in=180] (17.5,0.5) to[out=0,in=120] (18,0); %
\node[above] at (16,0.75) {identity links};
\draw[|-|] (14,0.75) -- (18,0.75);
\end{tikzpicture}}
\end{center}
\caption{\label{f:linksii1}A general link diagram in the case that the vertex $2j+1$ connects to the right. The left-pointing arrow on this link indicates that the corresponding matrix entry has row index larger than the column index (which is different to the convention on all other links). From \eqref{e:VmLimAS} we see that this left-pointing arrow will introduce a negative sign.}
\end{figure}
Summing over the possible values of $t=k+1, \dots, j-1$ in (ii), and adding the identity link pattern from (i), we have the number of valid link patterns on $N$ sites $L(N)$ given by
\begin{align}
\label{e:NumLinks} L(N)= 1+ \sum_{t=k+1}^{j-1} 2^{t-k-1} = 2^{j-k-1}.
\end{align}
So for $2k+1< 2j-1$ we have an even number of terms in the Pfaffian, and it turns out that they all cancel.

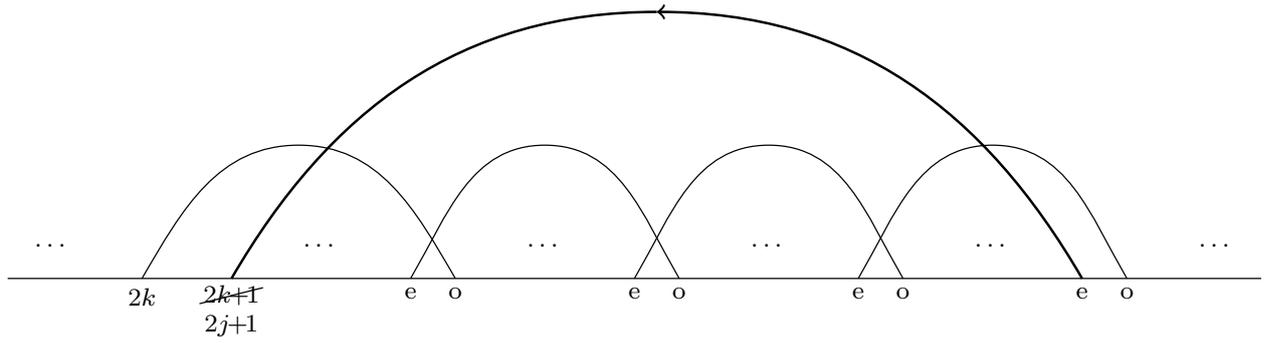
\begin{figure}
\begin{center}
\resizebox{\textwidth}{!}{\begin{tikzpicture}[baseline=(current  bounding  box.center)] %
\draw (-0.5,0)--(13.5,0); %
\node[below] at (0,0.5) {$\dots$}; %
\node[below] at (1,0) {$2k$}; %
\node[below] at (2,0) {$\cancel{2k\!\!+\!\!1}$}; %
\node[below] at (2,-0.3) {$2j\!\!+\!\!1$}; %
\node[below] at (3,0.5) {$\dots$}; %
\node[below] at (4,0) {e}; %
\node[below] at (4.5,0) {o}; %
\node[below] at (5.5,0.5) {$\dots$}; %
\node[below] at (6.5,0) {e}; %
\node[below] at (7,0) {o}; %
\node[below] at (8,0.5) {$\dots$}; %
\node[below] at (9,0) {e}; %
\node[below] at (9.5,0) {o}; %
\node[below] at (10.5,0.5) {$\dots$}; %
\node[below] at (11.5,0) {e}; %
\node[below] at (12,0) {o}; %
\node[below] at (13,0.5) {$\dots$}; %
\draw[smooth,thick] (2,0) to[out=60,in=180] (6.75,3); %
\draw[smooth,thick,<-] (6.75,3) to[out=0,in=120] (11.5,0); %
\draw[smooth] (1,0) to[out=60,in=180] (2.75,1.5) to[out=0,in=120] (4.5,0); %
\draw[smooth] (4,0) to[out=60,in=180] (5.5,1.5) to[out=0,in=120] (7,0); %
\draw[smooth] (6.5,0) to[out=60,in=180] (8,1.5) to[out=0,in=120] (9.5,0); %
\draw[smooth] (9,0) to[out=60,in=180] (10.5,1.5) to[out=0,in=120] (12,0); %
\end{tikzpicture}}
\end{center}
\caption{\label{f:GenLinks0} An example of the type of link diagrams possible with the restrictions in Figure \ref{f:linksii1}. The ellipses ``$\dots$'' denote identity links. The labels ``e'' and ``o'' denote generic even and odd vertices respectively.}
\end{figure}

To show this, note that the restriction that all odd vertices connect to the left and all even vertices connect to the right (except for $2j+1$ and $2t$) means that a general link diagram must look like that in Figure \ref{f:GenLinks0}. That is, big interconnected links, with a large rainbow link $(2j+1, \mathrm{even})$, and interspersed with little links. The big links must interconnect at neighbouring sites, since otherwise we would have two neighbouring vertices pointing in the same direction, violating the even/right--odd/left rule.
We can construct every diagram of the type in Figure \ref{f:GenLinks0} by application of the equality \eqref{e:Xhid}, by recasting that equation into the link diagram equalities in Figure \ref{f:linkIdentity2}, for the particular case when $m=j$. In Figure \ref{f:linkIdentity2}~(a) note the link diagram on the right has a left-pointing arrow (implying that the row index is larger than the column index), and so from \eqref{e:VmLimAS} we introduce a negative sign on the corresponding matrix entry. In Figure \ref{f:linkIdentity2}~(b) we have left-pointing arrows on both sides of the equality, but we have an additional sign introduced since the diagrams differ by an odd number of crossings.
\begin{figure}
\begin{center}
\subfloat[{}]{
\resizebox{0.75\textwidth}{!}{
\begin{tikzpicture} 
\draw (5.125,0)--(9,0); %
\node[below] at (5.5,0) {\scriptsize $2n$}; %
\node[above] at (6,0) {\dots}; %
\node[below] at (6.5,0) {\scriptsize $\cancel{2k\!\!+\!\!1}$}; %
\node[below] at (6.5,-0.3) {\scriptsize $2j\!\!+\!\!1$}; %
\node[above] at (7.25,0) {\dots}; %
\node[below] at (8,0) {\scriptsize $2t$}; %
\node[below] at (8.5,0) {\scriptsize $2t\!\!+\!\!1$}; %
\draw[smooth] (5.5,0) to[out=60,in=180] (6,0.5) to[out=0,in=120] (6.5,0); %
\node[above] at (6,0.6) {\scriptsize $X_{2j\!+\!1\!, 2n}$}; %
\draw[smooth] (8,0) to[out=60,in=180] (8.25,0.25) to[out=0,in=120] (8.5,0); %
\node[above] at (8.25,0.4) {\scriptsize $h_{2t}$}; %
\end{tikzpicture}
\begin{tikzpicture} %
\node[below] at (-0.3,0.5) {\scriptsize $= (-1)$}; %
\draw (0,0)--(4,0); %
\node[below] at (0.5,0) {\scriptsize $2n$}; %
\node[above] at (1.125,0) {\dots}; %
\node[below] at (1.5,0) {\scriptsize $\cancel{2k\!\!+\!\!1}$}; %
\node[below] at (1.5,-0.3) {\scriptsize $2j\!\!+\!\!1$}; %
\node[above] at (2.25,0) {\dots}; %
\node[below] at (3,0) {\scriptsize $2t$}; %
\node[below] at (3.5,0) {\scriptsize $2t\!\!+\!\!1$}; %
\draw[smooth] (0.5,0) to[out=60,in=180] (2,1.25) to[out=0,in=120] (3.5,0); %
\node[above] at (2,1.25) {\scriptsize $X_{2t\!+\!1\!, 2n}$}; %
\draw[smooth] (1.5,0) to[out=60,in=180] (2.25,0.45); %
\draw[smooth,<-] (2.25,0.45) to[out=0,in=120] (3,0); %
\node[above] at (2.25,0.45) {\scriptsize $-X_{2j\!+\!1\!, 2t}$}; %
\end{tikzpicture}
}}

\subfloat[{}]{
\resizebox{0.75\textwidth}{!}{
\begin{tikzpicture} %
\draw (5.2,0)--(9,0); %
\node[below] at (5.5,0) {\scriptsize $\cancel{2k\!\!+\!\!1}$}; %
\node[below] at (5.5,-0.3) {\scriptsize $2j\!\!+\!\!1$}; %
\node[above] at (6,0) {\dots}; %
\node[below] at (6.5,0) {\scriptsize $2n$}; %
\node[above] at (7.25,0) {\dots}; %
\node[below] at (8,0) {\scriptsize $2t$}; %
\node[below] at (8.5,0) {\scriptsize $2t\!\!+\!\!1$}; %
\draw[smooth] (5.5,0) to[out=60,in=180] (6,0.5);
\draw[smooth,<-] (6,0.5) to[out=0,in=120] (6.5,0); %
\node[above] at (6,0.6) {\scriptsize $-X_{2j\!+\!1\!, 2n}$}; %
\draw[smooth] (8,0) to[out=60,in=180] (8.25,0.25) to[out=0,in=120] (8.5,0); %
\node[above] at (8.25,0.4) {\scriptsize $h_{2t}$}; %
\end{tikzpicture}
\begin{tikzpicture} %
\node[below] at (-0.2,0.5) {\scriptsize $= (-1)$}; %
\draw (0,0)--(4,0); %
\node[below] at (0.5,0) {\scriptsize $\cancel{2k\!\!+\!\!1}$}; %
\node[below] at (0.5,-0.3) {\scriptsize $2j\!\!+\!\!1$}; %
\node[above] at (1.125,0) {\dots}; %
\node[below] at (1.5,0) {\scriptsize $2n$}; %
\node[above] at (2.25,0) {\dots}; %
\node[below] at (3,0) {\scriptsize $2t$}; %
\node[below] at (3.5,0) {\scriptsize $2t\!\!+\!\!1$}; %
\draw[smooth] (0.5,0) to[out=60,in=180] (1.75,1); %
\draw[smooth,<-] (1.75,1) to[out=0,in=120] (3,0); %
\node[above] at (1.25,1) {\scriptsize $-X_{2j\!+\!1\!, 2n}$}; %
\draw[smooth] (1.5,0) to[out=60,in=180] (2.5,1) to[out=0,in=120] (3.5,0); %
\node[above] at (3,1) {\scriptsize $X_{2t\!+\!1\!, 2n}$}; %
\draw[red] (2.175,0.92) circle (0.1cm);
\end{tikzpicture}
}}
\end{center}
\caption{\label{f:linkIdentity2}Two possible link diagrams corresponding to equation \eqref{e:Xhid} when $2k+1 \mapsto 2j+1$. A left-pointing arrow on a link indicates that the left vertex of the edge is greater than the right vertex, and so from \eqref{e:VmLimAS} we pick up a negative sign (since this corresponds to an element in the lower triangle of the anti-symmetric matrix). Use of diagram (b) introduces/removes an odd number of crossings, and so there is also a factor of $(-1)$ in this equality.}
\end{figure}

\begin{figure}
\begin{center}
\resizebox{\textwidth}{!}{\begin{tikzpicture}[baseline=(current  bounding  box.center)] %
\node[above] at (-0.5,0) {$(+1)$};
\draw (0,0)--(13,0); %
\node[below] at (0.5,0.33) {$\dots$}; %
\node[below] at (1,0) {$2k$}; %
\node[below] at (2,0) {$\cancel{2k\!\!+\!\!1}$}; %
\node[below] at (2,-0.3) {$2j\!\!+\!\!1$}; %
\node[below] at (2.75,0.33) {$\dots$}; %
\node[below] at (3.5,0) {e}; %
\node[below] at (4.5,0) {o}; %
\node[below] at (5.25,0.33) {$\dots$}; %
\node[below] at (6,0) {e}; %
\node[below] at (7,0) {o}; %
\node[below] at (7.75,0.33) {$\dots$}; %
\node[below] at (8.5,0) {e}; %
\node[below] at (9.5,0) {o}; %
\node[below] at (10.25,0.33) {$\dots$}; %
\node[below] at (11,0) {e}; %
\node[below] at (12,0) {o}; %
\node[below] at (12.5,0.33) {$\dots$}; %
\draw[smooth] (1,0) to[out=60,in=180] (1.5,0.33) to[out=0,in=120] (2,0); %
\draw[smooth] (3.5,0) to[out=60,in=180] (4,0.33) to[out=0,in=120] (4.5,0); %
\draw[smooth] (6,0) to[out=60,in=180] (6.5,0.33) to[out=0,in=120] (7,0); %
\draw[smooth] (8.5,0) to[out=60,in=180] (9,0.33) to[out=0,in=120] (9.5,0); %
\draw[smooth] (11,0) to[out=60,in=180] (11.5,0.33) to[out=0,in=120] (12,0); %
\end{tikzpicture}
}

\resizebox{\textwidth}{!}{\begin{tikzpicture}[baseline=(current  bounding  box.center)] %
\draw[->] (-0.5,2) to (-0.5,1);
\node[above] at (0,1.25) {(a)};
\node[above] at (-0.5,0) {$(-1)$};
\draw (0,0)--(13,0); %
\node[below] at (0.5,0.33) {$\dots$}; %
\node[below] at (1,0) {$2k$}; %
\node[below] at (2,0) {$\cancel{2k\!\!+\!\!1}$}; %
\node[below] at (2,-0.3) {$2j\!\!+\!\!1$}; %
\node[below] at (2.75,0.33) {$\dots$}; %
\node[below] at (3.5,0) {e}; %
\node[below] at (4.5,0) {o}; %
\node[below] at (5.25,0.33) {$\dots$}; %
\node[below] at (6,0) {e}; %
\node[below] at (7,0) {o}; %
\node[below] at (7.75,0.33) {$\dots$}; %
\node[below] at (8.5,0) {e}; %
\node[below] at (9.5,0) {o}; %
\node[below] at (10.25,0.33) {$\dots$}; %
\node[below] at (11,0) {e}; %
\node[below] at (12,0) {o}; %
\node[below] at (12.5,0.33) {$\dots$}; %
\draw[smooth] (1,0) to[out=60,in=180] (2.75,1) to[out=0,in=120] (4.5,0); %
\draw[smooth] (2,0) to[out=60,in=180] (2.75,0.5); %
\draw[smooth,<-] (2.75,0.5) to[out=0,in=120] (3.5,0); %
\draw[smooth] (6,0) to[out=60,in=180] (6.5,0.33) to[out=0,in=120] (7,0); %
\draw[smooth] (8.5,0) to[out=60,in=180] (9,0.33) to[out=0,in=120] (9.5,0); %
\draw[smooth] (11,0) to[out=60,in=180] (11.5,0.33) to[out=0,in=120] (12,0); %
\end{tikzpicture}
}

\resizebox{\textwidth}{!}{\begin{tikzpicture}[baseline=(current  bounding  box.center)] %
\draw[->] (-0.5,2) to (-0.5,1);
\node[above] at (0,1.25) {(b)};
\node[above] at (-0.5,0) {$(+1)$};
\draw (0,0)--(13,0); %
\node[below] at (0.5,0.33) {$\dots$}; %
\node[below] at (1,0) {$2k$}; %
\node[below] at (2,0) {$\cancel{2k\!\!+\!\!1}$}; %
\node[below] at (2,-0.3) {$2j\!\!+\!\!1$}; %
\node[below] at (2.75,0.33) {$\dots$}; %
\node[below] at (3.5,0) {e}; %
\node[below] at (4.5,0) {o}; %
\node[below] at (5.25,0.33) {$\dots$}; %
\node[below] at (6,0) {e}; %
\node[below] at (7,0) {o}; %
\node[below] at (7.75,0.33) {$\dots$}; %
\node[below] at (8.5,0) {e}; %
\node[below] at (9.5,0) {o}; %
\node[below] at (10.25,0.33) {$\dots$}; %
\node[below] at (11,0) {e}; %
\node[below] at (12,0) {o}; %
\node[below] at (12.5,0.33) {$\dots$}; %
\draw[smooth] (1,0) to[out=60,in=180] (2.75,0.75) to[out=0,in=120] (4.5,0); %
\draw[smooth] (2,0) to[out=60,in=180] (4,1); %
\draw[smooth,<-] (4,1) to[out=0,in=120] (6,0); %
\draw[smooth] (3.5,0) to[out=60,in=180] (5.25,0.75) to[out=0,in=120] (7,0); %
\draw[smooth] (8.5,0) to[out=60,in=180] (9,0.33) to[out=0,in=120] (9.5,0); %
\draw[smooth] (11,0) to[out=60,in=180] (11.5,0.33) to[out=0,in=120] (12,0); %
\end{tikzpicture}
}

\resizebox{\textwidth}{!}{\begin{tikzpicture}[baseline=(current  bounding  box.center)] %
\draw[->] (-0.5,2) to (-0.5,1);
\node[above] at (0,1.25) {(b)};
\node[above] at (-0.5,0) {$(-1)$};
\draw (0,0)--(13,0); %
\node[below] at (0.5,0.33) {$\dots$}; %
\node[below] at (1,0) {$2k$}; %
\node[below] at (2,0) {$\cancel{2k\!\!+\!\!1}$}; %
\node[below] at (2,-0.3) {$2j\!\!+\!\!1$}; %
\node[below] at (2.75,0.33) {$\dots$}; %
\node[below] at (3.5,0) {e}; %
\node[below] at (4.5,0) {o}; %
\node[below] at (5.25,0.33) {$\dots$}; %
\node[below] at (6,0) {e}; %
\node[below] at (7,0) {o}; %
\node[below] at (7.75,0.33) {$\dots$}; %
\node[below] at (8.5,0) {e}; %
\node[below] at (9.5,0) {o}; %
\node[below] at (10.25,0.33) {$\dots$}; %
\node[below] at (11,0) {e}; %
\node[below] at (12,0) {o}; %
\node[below] at (12.5,0.33) {$\dots$}; %
\draw[smooth] (1,0) to[out=60,in=180] (2.75,0.75) to[out=0,in=120] (4.5,0); %
\draw[smooth] (2,0) to[out=60,in=180] (5.25,1.25); %
\draw[smooth,<-] (5.25,1.25) to[out=0,in=120] (8.5,0); %
\draw[smooth] (3.5,0) to[out=60,in=180] (5.25,0.75) to[out=0,in=120] (7,0); %
\draw[smooth] (6,0) to[out=60,in=180] (7.75,0.75) to[out=0,in=120] (9.5,0); %
\draw[smooth] (11,0) to[out=60,in=180] (11.5,0.33) to[out=0,in=120] (12,0); %
\end{tikzpicture}
}

\resizebox{\textwidth}{!}{\begin{tikzpicture}[baseline=(current  bounding  box.center)] %
\draw[->] (-0.5,2) to (-0.5,1);
\node[above] at (0,1.25) {(b)};
\node[above] at (-0.5,0) {$(+1)$};
\draw (0,0)--(13,0); %
\node[below] at (0.5,0.33) {$\dots$}; %
\node[below] at (1,0) {$2k$}; %
\node[below] at (2,0) {$\cancel{2k\!\!+\!\!1}$}; %
\node[below] at (2,-0.3) {$2j\!\!+\!\!1$}; %
\node[below] at (2.75,0.33) {$\dots$}; %
\node[below] at (3.5,0) {e}; %
\node[below] at (4.5,0) {o}; %
\node[below] at (5.25,0.33) {$\dots$}; %
\node[below] at (6,0) {e}; %
\node[below] at (7,0) {o}; %
\node[below] at (7.75,0.33) {$\dots$}; %
\node[below] at (8.5,0) {e}; %
\node[below] at (9.5,0) {o}; %
\node[below] at (10.25,0.33) {$\dots$}; %
\node[below] at (11,0) {e}; %
\node[below] at (12,0) {o}; %
\node[below] at (12.5,0.33) {$\dots$}; %
\draw[smooth] (1,0) to[out=60,in=180] (2.75,0.75) to[out=0,in=120] (4.5,0); %
\draw[smooth] (2,0) to[out=60,in=180] (6.5,1.5); %
\draw[smooth,<-] (6.5,1.5) to[out=0,in=120] (11,0); %
\draw[smooth] (3.5,0) to[out=60,in=180] (5.25,0.75) to[out=0,in=120] (7,0); %
\draw[smooth] (6,0) to[out=60,in=180] (7.75,0.75) to[out=0,in=120] (9.5,0); %
\draw[smooth] (8.5,0) to[out=60,in=180] (10.25,0.75) to[out=0,in=120] (12,0); %
\end{tikzpicture}
}
\end{center}
\caption{Constructing the link diagram in Figure \ref{f:GenLinks0} using the diagram equalities in Figure \ref{f:linkIdentity2}.  The labels to the left of each link diagram refer to which of the equalities in Figure \ref{f:linkIdentity2} was applied, and the sign of the corresponding term in the Pfaffian.\label{f:GenLinks2}.}
\end{figure}
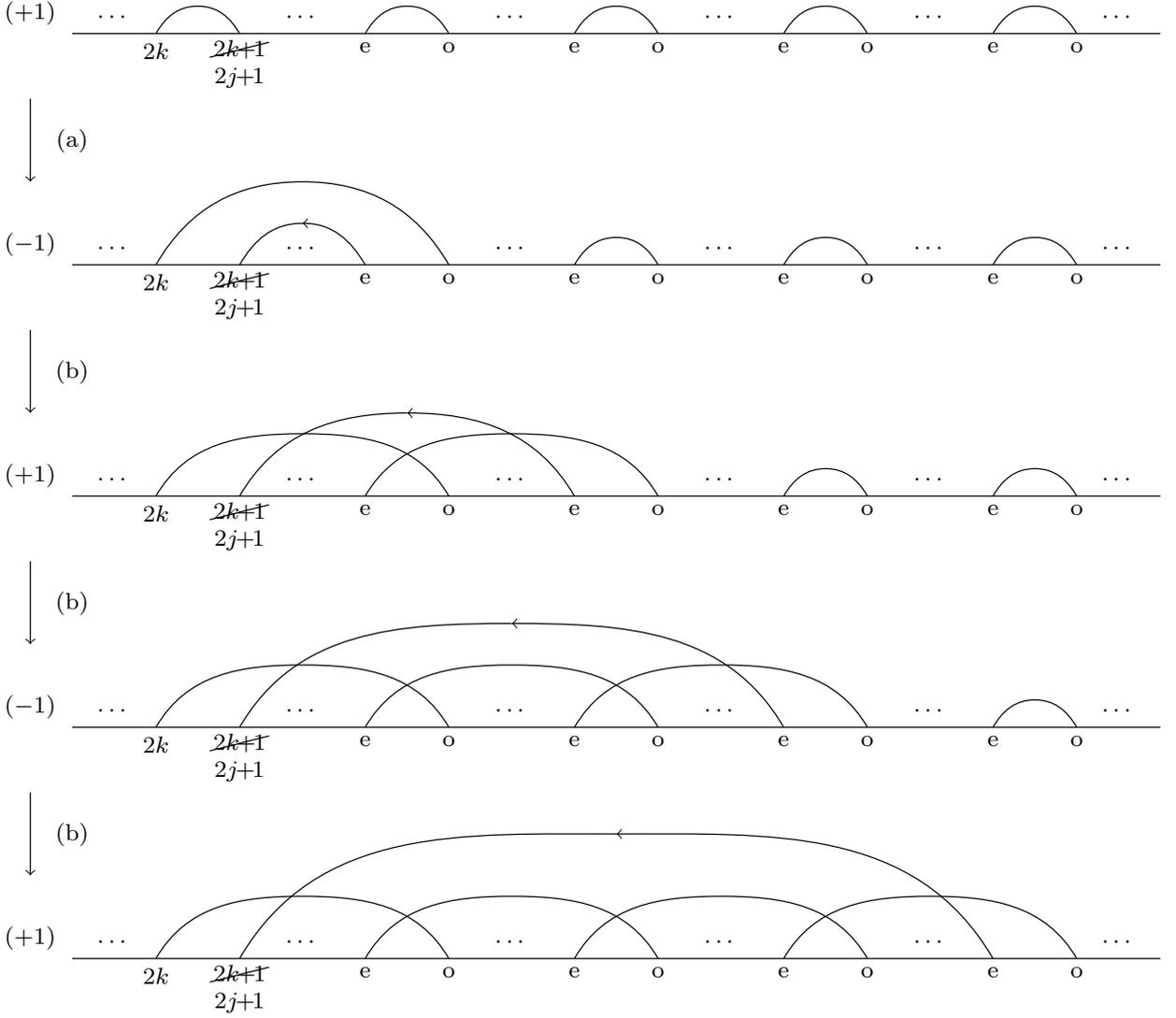

In Figure \ref{f:GenLinks2} we give the example of constructing the link diagram in Figure \ref{f:GenLinks0} from the identity diagram by repeated application of equalities in Figure \ref{f:linkIdentity2} --- starting from the left at the link $(2k, 2j+1)$ we first apply equality (a), and then, moving to the right, we repeatedly apply (b) until we have the final diagram. Each application of the equalities (a) and (b) introduces a negative sign.

In the identity diagram there are $j-k-1$ little links to the right of site $2k+1$, so there are ${j-k-1 \choose p}$ link diagrams obtained from $p$ uses of the equalities in Figure \ref{f:linkIdentity2}, which gives us that
\begin{align}
\nonumber \Pf \bV_{2j-1}^{(2k+1 \mapsto 2j+1)} &= (h_0 h_2 \cdots h_{2k-2}) v_{2k,2j+1} (h_{2k+2} \cdots h_{2j-2}) \sum_{p=0}^{j-k-1} (-1)^p {j-k-1 \choose p}\\
\label{e:altBinom} &= 0 \qquad \mbox{(for $k< j-1$)},
\end{align}
where $(h_0 h_2 \cdots h_{2k-2}) v_{2k,2j+1} (h_{2k+2} \cdots h_{2j-2})$ is the term from the identity link diagram (i.e. the top diagram in Figure \ref{f:GenLinks2}). The second equality follows since the sum of alternating binomial coefficients is equal to zero, which can be seen from the binomial expansion of $(x-y)^{j-k-1}$, with $x, y\to 1$. Thus $\alpha_{2j+1, 2k+1}=0$ when $k<j-1$.

From \eqref{e:NumLinks} we see the only scenario where we do not have an even number of cancelling link diagrams is when $k = j-1$, and we have only the identity link pattern. In this case, equation \eqref{e:altBinom} becomes
\begin{align}
\label{e:PfLim} \Pf \bV_{2j-1}^{(2j-1 \mapsto 2j+1)} = h_0 h_2 \cdots h_{2j-4} X_{2j+1, 2j-2}
\end{align}
since $v_{2j-2, 2j-1}\mapsto v_{2j-2, 2j+1}= X_{2j+1, 2j-2}$. Substituting \eqref{e:PfLim} and \eqref{e:PfVm1} (with $m=2j-1$) into \eqref{e:alphas1b} we have
\begin{align}
\alpha_{2j+1, 2j-1} = - \frac{X_{2j+1, 2j-2}} {h_{2j-2}} = - \frac{h_{2j}} {h_{2j-1}} = -\frac{\Gamma(j+1)}{\Gamma(j)} = -j,
\end{align}
where we used \eqref{e:XjkLim} for the second equality. Combining this result with \eqref{e:PfLim1}, \eqref{e:PfLim2}, \eqref{e:PfLim3} and \eqref{e:altBinom} we recover \eqref{e:limHerms1}.

\end{appendices}

\clearpage

\bibliographystyle{spmpsci}      
\bibliography{MPS2020}   

\begin{thebibliography}{10}
\providecommand{\url}[1]{{#1}}
\providecommand{\urlprefix}{URL }
\expandafter\ifx\csname urlstyle\endcsname\relax
  \providecommand{\doi}[1]{DOI~\discretionary{}{}{}#1}\else
  \providecommand{\doi}{DOI~\discretionary{}{}{}\begingroup
  \urlstyle{rm}\Url}\fi

\bibitem{AbraSteg1972}
Abramowitz, M., Stegun, I.A. (eds.): Handbook of mathematical functions, 10
  edn.
\newblock United States Department of Commerce, Washington D.C. (1972)

\bibitem{AdleForrNagavanMoer2000}
Adler, M., Forrester, P., Nagao, T., {van Moerbeke}, P.: Classical skew
  orthogonal polynomials and random matrices.
\newblock Journal of Statistical Physics \textbf{99}(1--2), 141--170 (2000)

\bibitem{AdlevanMoer2002}
Adler, M., van Moerbeke, P.: {T}oda versus {P}faff lattice and related
  polynomials.
\newblock Duke Math. J. \textbf{112}(1), 1--58 (2002)

\bibitem{amir2011probability}
Amir, G., Corwin, I., Quastel, J.: Probability distribution of the free energy
  of the continuum directed random polymer in 1+1 dimensions.
\newblock Communications on pure and applied mathematics \textbf{64}(4),
  466--537 (2011)

\bibitem{baik2018pfaffian}
Baik, J., Barraquand, G., Corwin, I., Suidan, T., et~al.: {P}faffian {S}chur
  processes and last passage percolation in a half-quadrant.
\newblock The Annals of Probability \textbf{46}(6), 3015--3089 (2018)

\bibitem{baik1999distribution}
Baik, J., Deift, P., Johansson, K.: On the distribution of the length of the
  longest increasing subsequence of random permutations.
\newblock Journal of the American Mathematical Society \textbf{12}(4),
  1119--1178 (1999)

\bibitem{baik2000limiting}
Baik, J., Rains, E.M.: Limiting distributions for a polynuclear growth model
  with external sources.
\newblock Journal of Statistical Physics \textbf{100}(3/4), 523--541 (2000)

\bibitem{barraquand2020half}
Barraquand, G., Krajenbrink, A., Doussal, P.L.: Half-space stationary
  {K}ardar-{P}arisi-{Z}hang equation.
\newblock arXiv preprint arXiv:2003.03809  (2020)

\bibitem{biroli2007top}
Biroli, G., Bouchaud, J.P., Potters, M.: On the top eigenvalue of heavy-tailed
  random matrices.
\newblock EPL (Europhysics Letters) \textbf{78}(1), 10001 (2007)

\bibitem{bloemendal2013limits}
Bloemendal, A., Vir{\'a}g, B.: Limits of spiked random matrices {I}.
\newblock Probability Theory and Related Fields \textbf{156}(3), 795--825
  (2013)

\bibitem{BoroSosh2003}
Borodin, A., Soshnikov, A.: Janossy densities. {I}. {D}eterminantal ensembles.
\newblock Journal of Statistical Physics \textbf{113}(3), 595--610 (2003)

\bibitem{borot2012right}
Borot, G., Nadal, C.: Right tail asymptotic expansion of {T}racy--{W}idom beta
  laws.
\newblock Random Matrices: Theory and Applications \textbf{1}(03), 1250006
  (2012)

\bibitem{calabrese2010free}
Calabrese, P., Le~Doussal, P., Rosso, A.: Free-energy distribution of the
  directed polymer at high temperature.
\newblock EPL (Europhysics Letters) \textbf{90}(2), 20002 (2010)

\bibitem{cao2014continuous}
Cao, M., Chen, Y., Griffin, J.: Continuous and discrete {P}ainlev{\'e}
  equations arising from the gap probability distribution of the finite $n$
  {G}aussian unitary ensembles.
\newblock Journal of Statistical Physics \textbf{157}(2), 363--375 (2014)

\bibitem{deBruijn1955}
{de Bruijn}, N.: On some multiple integrals involving determinants.
\newblock Journal of the Indian Mathematical Society. New Series \textbf{19},
  133--151 (1955)

\bibitem{dean2015finite}
Dean, D.S., Le~Doussal, P., Majumdar, S.N., Schehr, G.: Finite-temperature free
  fermions and the {K}ardar-{P}arisi-{Z}hang equation at finite time.
\newblock Physical Review Letters \textbf{114}(11), 110402 (2015)

\bibitem{dean2016noninteracting}
Dean, D.S., Le~Doussal, P., Majumdar, S.N., Schehr, G.: Noninteracting fermions
  at finite temperature in a $d$-dimensional trap: Universal correlations.
\newblock Physical Review A \textbf{94}(6), 063622 (2016)

\bibitem{dean2019noninteracting}
Dean, D.S., Le~Doussal, P., Majumdar, S.N., Schehr, G.: Noninteracting fermions
  in a trap and random matrix theory.
\newblock Journal of Physics A: Mathematical and Theoretical \textbf{52}(14),
  144006 (2019)

\bibitem{dotsenko2010bethe}
Dotsenko, V.: {B}ethe ansatz derivation of the {T}racy-{W}idom distribution for
  one-dimensional directed polymers.
\newblock EPL (Europhysics Letters) \textbf{90}(2), 20003 (2010)

\bibitem{dyson1962statistical}
Dyson, F.J.: Statistical theory of the energy levels of complex systems. {I}.
\newblock Journal of Mathematical Physics \textbf{3}(1), 140--156 (1962)

\bibitem{ForrMays2009}
Forrester, P., Mays, A.: A method to calculate correlation functions for
  $\beta=1$ random matrices of odd size.
\newblock Journal of Statistical Physics \textbf{134}(3), 443--462 (2009)

\bibitem{forrester1993spectrum}
Forrester, P.J.: The spectrum edge of random matrix ensembles.
\newblock Nuclear Physics B \textbf{402}(3), 709--728 (1993)

\bibitem{Forrester2010}
Forrester, P.J.: Log-gases and random matrices, \emph{London Mathematical
  Society Monographs}, vol.~34.
\newblock Princeton University Press, Princeton (2010)

\bibitem{forrester2011non}
Forrester, P.J., Majumdar, S.N., Schehr, G.: Non-intersecting {B}rownian
  walkers and {Y}ang--{M}ills theory on the sphere.
\newblock Nuclear Physics B \textbf{844}(3), 500--526 (2011)

\bibitem{fyodorov2011level}
Fyodorov, Y.V.: Level curvature distribution: From bulk to the soft edge of
  random {H}ermitian matrices.
\newblock Acta Physica Polonica A \textbf{120}(6A) (2011)

\bibitem{fyodorov2015large}
Fyodorov, Y.V., Perret, A., Schehr, G.: Large time zero temperature dynamics of
  the spherical $p= 2$-spin glass model of finite size.
\newblock Journal of Statistical Mechanics: Theory and Experiment
  \textbf{2015}(11), P11017 (2015)

\bibitem{gueudre2012directed}
Gueudr{\'e}, T., Le~Doussal, P.: Directed polymer near a hard wall and {KPZ}
  equation in the half-space.
\newblock EPL (Europhysics Letters) \textbf{100}(2), 26006 (2012)

\bibitem{imamura2004fluctuations}
Imamura, T., Sasamoto, T.: Fluctuations of the one-dimensional polynuclear
  growth model with external sources.
\newblock Nuclear Physics B \textbf{699}(3), 503--544 (2004)

\bibitem{Janossy1950}
Janossy, L.: On the absorption of a nucleon cascade.
\newblock Proceedings of the Royal Irish Academy. Section A: Mathematical and
  Physical Sciences \textbf{53}, 181--188 (1950)

\bibitem{Knuth1996}
Knuth, D.E.: Overlapping {P}faffians.
\newblock Electronic Journal of Combinatorics \textbf{3} (1996)

\bibitem{le2012kpz}
Le~Doussal, P., Calabrese, P.: The {KPZ} equation with flat initial condition
  and the directed polymer with one free end.
\newblock Journal of Statistical Mechanics: Theory and Experiment
  \textbf{2012}(06), P06001 (2012)

\bibitem{liechty2012nonintersecting}
Liechty, K.: Nonintersecting {B}rownian motions on the half-line and discrete
  {G}aussian orthogonal polynomials.
\newblock Journal of Statistical Physics \textbf{147}(3), 582--622 (2012)

\bibitem{majumdar2007course}
Majumdar, S.N.: Course 4 random matrices, the {U}lam problem, directed polymers
  \& growth models, and sequence matching.
\newblock Les Houches \textbf{85}, 179--216 (2007)

\bibitem{majumdar2004anisotropic}
Majumdar, S.N., Nechaev, S.: Anisotropic ballistic deposition model with links
  to the {U}lam problem and the {T}racy-{W}idom distribution.
\newblock Physical Review E \textbf{69}(1), 011103 (2004)

\bibitem{majumdar2005exact}
Majumdar, S.N., Nechaev, S.: Exact asymptotic results for the {B}ernoulli
  matching model of sequence alignment.
\newblock Physical Review E \textbf{72}(2), 020901 (2005)

\bibitem{majumdar2020extreme}
Majumdar, S.N., Pal, A., Schehr, G.: Extreme value statistics of correlated
  random variables: A pedagogical review.
\newblock Physics Reports \textbf{840}, 1--32 (2020)

\bibitem{majumdar2014top}
Majumdar, S.N., Schehr, G.: Top eigenvalue of a random matrix: Large deviations
  and third order phase transition.
\newblock Journal of Statistical Mechanics: Theory and Experiment
  \textbf{2014}(1), P01012 (2014)

\bibitem{makey2020universality}
Makey, G., Galioglu, S., Ghaffari, R., Engin, E.D., Y{\i}ld{\i}r{\i}m, G.,
  Yavuz, {\"O}., Bekta{\c{s}}, O., Nizam, {\"U}.S., Akbulut, {\"O}.,
  {\c{S}}ahin, {\"O}., et~al.: Universality of dissipative self-assembly from
  quantum dots to human cells.
\newblock Nature Physics \textbf{16}(7), 795--801 (2020)

\bibitem{Mays2011thesis}
Mays, A.: A geometrical triumvirate of real random matrices.
\newblock Ph.D. thesis, The University of Melbourne, Parkville (2011)

\bibitem{mays2020prep}
Mays, A., Ponsaing, A., Schehr, G.: In preparation  (2020)

\bibitem{Mehta2004}
Mehta, M.L.: Random matrices, vol. 142, 3rd edn.
\newblock Academic Press, Boston (2004)

\bibitem{monthus2013typical}
Monthus, C., Garel, T.: Typical versus averaged overlap distribution in spin
  glasses: Evidence for droplet scaling theory.
\newblock Physical Review B \textbf{88}(13), 134204 (2013)

\bibitem{nadal2011matrices}
Nadal, C.: Matrices al{\'e}atoires et leurs applications {\`a} la physique
  statistique et quantique.
\newblock Ph.D. thesis, Paris 11 (2011)

\bibitem{nadal2009nonintersecting}
Nadal, C., Majumdar, S.N.: Nonintersecting {B}rownian interfaces and {W}ishart
  random matrices.
\newblock Physical Review E \textbf{79}(6), 061117 (2009)

\bibitem{NadaMaju2011}
Nadal, C., Majumdar, S.N.: A simple derivation of the
  {T}racy{\textendash}{W}idom distribution of the maximal eigenvalue of a
  {G}aussian unitary random matrix.
\newblock Journal of Statistical Mechanics: Theory and Experiment
  \textbf{2011}(04), P04001 (2011)

\bibitem{NagaWada1991}
Nagao, T., Wadati, M.: Correlation functions of random matrix ensembles related
  to classical orthogonal polynomials.
\newblock Journal of The Physical Society of Japan \textbf{60}(10), 3298--3322
  (1991)

\bibitem{nguyen2017non}
Nguyen, G.B., Remenik, D.: Non-intersecting {B}rownian bridges and the
  {L}aguerre orthogonal ensemble.
\newblock Annales de l'Institut Henri Poincar{\'e}, Probabilit{\'e}s et
  Statistiques \textbf{53}(4), 2005--2029 (2017)

\bibitem{PerrSche2014}
Perret, A., Schehr, G.: Near-extreme eigenvalues and the first gap of
  {H}ermitian random matrices.
\newblock Journal of Statistical Physics \textbf{156}(5), 843--876 (2014)

\bibitem{perret2015density}
Perret, A., Schehr, G.: The density of eigenvalues seen from the soft edge of
  random matrices in the {G}aussian $\beta$-ensembles.
\newblock Acta Physica Polonica B \textbf{46}(9), 1693 (2015)

\bibitem{prahofer2000universal}
Pr{\"a}hofer, M., Spohn, H.: Universal distributions for growth processes in
  1+1 dimensions and random matrices.
\newblock Physical Review Letters \textbf{84}(21), 4882 (2000)

\bibitem{Rote2001}
Rote, G.: Division-free algorithms for the determinant and the {P}faffian:
  Algebraic and combinatorial approaches.
\newblock In: H.~Alt (ed.) Computational Discrete Mathematics: Advanced
  Lectures, pp. 119--135. Springer Berlin Heidelberg, Berlin, Heidelberg (2001)

\bibitem{sasamoto2010one}
Sasamoto, T., Spohn, H.: One-dimensional {K}ardar-{P}arisi-{Z}hang equation: An
  exact solution and its universality.
\newblock Physical review letters \textbf{104}(23), 230602 (2010)

\bibitem{Soshnikov2003}
Soshnikov, A.: Janossy densities. {II}. {P}faffian ensembles.
\newblock Journal of Statistical Physics \textbf{113}(3), 611--622 (2003)

\bibitem{Soshnikov2004}
Soshnikov, A.: Janossy densities of coupled random matrices.
\newblock Communications in Mathematical Physics \textbf{251}(3), 447--471
  (2004)

\bibitem{stephan2019free}
St{\'e}phan, J.M.: Free fermions at the edge of interacting systems.
\newblock SciPost Physics \textbf{6}, 057 (2019)

\bibitem{takeuchi2010universal}
Takeuchi, K.A., Sano, M.: Universal fluctuations of growing interfaces:
  Evidence in turbulent liquid crystals.
\newblock Physical Review Letters (23), 230601 (2010)

\bibitem{takeuchi2011growing}
Takeuchi, K.A., Sano, M., Sasamoto, T., Spohn, H.: Growing interfaces uncover
  universal fluctuations behind scale invariance.
\newblock Scientific reports \textbf{1}, 34 (2011)

\bibitem{tracy1994level}
Tracy, C.A., Widom, H.: Level-spacing distributions and the {A}iry kernel.
\newblock Communications in Mathematical Physics \textbf{159}(1), 151--174
  (1994)

\bibitem{TracWido1996}
Tracy, C.A., Widom, H.: On orthogonal and symplectic matrix ensembles.
\newblock Communications in Mathematical Physics \textbf{177}(3), 727--754
  (1996)

\bibitem{TracWido1998}
Tracy, C.A., Widom, H.: Correlation functions, cluster functions, and spacing
  distributions for random matrices.
\newblock Journal of Statistical Physics \textbf{92}(5), 809--835 (1998)

\bibitem{witte2013joint}
Witte, N., Bornemann, F., Forrester, P.: Joint distribution of the first and
  second eigenvalues at the soft edge of unitary ensembles.
\newblock Nonlinearity \textbf{26}(6), 1799 (2013)

\bibitem{witte2012variance}
Witte, N., Forrester, P.: On the variance of the index for the {G}aussian
  unitary ensemble.
\newblock Random Matrices: Theory and Applications \textbf{1}(04), 1250010
  (2012)

\end{thebibliography}

%
%

\end{document}